\documentclass[runningheads]{llncs}
\usepackage{graphicx}
\usepackage{comment}
\usepackage{amsmath,amssymb} % define this before the line numbering.
\usepackage{color}
\usepackage{enumerate}
\usepackage[width=122mm,left=12mm,paperwidth=146mm,height=193mm,top=12mm,paperheight=217mm]
{geometry}
\usepackage[letterpaper=true,colorlinks,urlcolor=blue,citecolor=blue,linkcolor=red,bookmarks=false]{hyperref}
\usepackage{booktabs}
\usepackage{multirow}
\usepackage{subfigure}
\begin{document}

% \linenumbers

\pagestyle{headings}
\mainmatter
\def\ECCVSubNumber{1520}  % Insert your submission number here

%\title{AT$^2$SR: Attention in Attention Network for Single Image Super-Resolution} % Replace with your title
\title{ Single Image Super-Resolution via \\ a Holistic Attention Network}
%\begin{comment}
%\titlerunning{ECCV-20 submission ID \ECCVSubNumber} 
%\authorrunning{ECCV-20 submission ID \ECCVSubNumber} 
%\author{Anonymous ECCV submission}
%\institute{Paper ID \ECCVSubNumber}
%\end{comment}

%\begin{comment}
\titlerunning{Single Image SR via a Holistic Attention Network}

%\author{Ben Niu\inst{1,\footnotemark[1]} \and
%Weilei Wen\inst{2,\footnotemark[1]} \and
%%Weilei Wen$^{2,3}$$^$ {Equal contributions} \and
%Wenqi Ren$^{3}$ \and
%Xiangde Zhang$^{1}$ \and
%%Lianping Yang$^{1}$\thanks{Corresponding author} \and
%%Lianping Yang$^{1,\footnotemark[4]}$ \and
%Lianping Yang\inst{1,\footnotemark[4]} \and
%Shuzhen Wang$^{2}$ \and
%Kaihao Zhang$^{5}$ \and
%Xiaochun Cao$^{3,4}$ \and
%Haifeng Shen$^{6}$
%}

\author{Ben Niu\inst{1,\footnotemark[1]} \and
	Weilei Wen\inst{2,3,\footnotemark[1]} \and
	Wenqi Ren\inst{3} \and
	Xiangde Zhang\inst{1} \and
	Lianping Yang\inst{1,\footnotemark[4]} \and
	Shuzhen Wang\inst{2} \and
	Kaihao Zhang\inst{5} \and
	Xiaochun Cao\inst{3,4} \and
	Haifeng Shen\inst{6}}

\authorrunning{B. Niu et al.}
\institute{$^{1}$Northeastern University ~
$^{2}$Xidian University ~
$^{3}$SKLOIS, IIE, CAS ~ \\
$^{4}$ Peng Cheng Laboratory, Cyberspace Security Research Center, China \\
$^{5}$ANU ~ $^{6}$AI Labs, Didi Chuxing, China}

\renewcommand{\thefootnote}{\fnsymbol{footnote}}
\footnotetext[1]{Equal contribution}
\footnotetext[4]{Corresponding author}
\maketitle

\begin{abstract}
	Informative features play a crucial role in the single image super-resolution task. %However, we observe that previous work did not make efficient use of the convolutional features of each layer. 
	Channel attention has been demonstrated to be effective for preserving information-rich features in each layer.
	However, channel attention treats each convolution layer as a separate process that misses the correlation among different layers. 
	To address this problem, we propose a new holistic attention network (HAN), which consists of a layer attention module (LAM) and a channel-spatial attention module (CSAM), to model the holistic interdependencies among layers, channels, and positions.
	%learn the holistic interdependencies of both layers and channels. 
	Specifically, the proposed LAM adaptively emphasizes hierarchical features by considering correlations among layers.
	%	 learns the correlation among hierarchical features. Then apply different weights on hierarchical features. The weight depends on the contribution of hierarchical features. 
	Meanwhile, CSAM learns the confidence at all the positions of each channel to selectively capture more informative features.
	%for adaptive feature-modulation.
	%
	%
	%In the inner of two modules, we apply multiple skip connections to make the backbone of our model focus on high-frequency information. 
	%
	%The proposed HAN achieves the goal of rescale feature-wise attention adaptively by the correlations of different-aspect features. So that the detail information can be extracted better by optimizing the weight parameters of each feature map. 
	Extensive experiments demonstrate that the proposed HAN performs favorably against the state-of-the-art single image super-resolution approaches.
	% and achieves more detailed textures.
	
\end{abstract}

\section{Introduction}

Single image super-resolution (SISR) is an important task in computer vision and image processing. Given a low-resolution image, the goal of super-resolution (SR) is to generate a high-resolution (HR) image with necessary edge structures and texture details. 
%SISR is thus a desirable technique for both computational photography and high-level computer vision tasks, 
The advance of SISR will immediately benefit many application fields, such as video surveillance and pedestrian detection. 
%
%With the deep convolutional neural networks (CNNs) development, many image processing tasks including, SISR, image deblurring~\cite{tao2018scale}, image denoising~\cite{zhang2017beyond}, have got significant improvements.
%SISR is a typical low-level computer vision task, which generates high-resolution (HR) images from low-resolution counterparts. It is significant for many high-level computer vision tasks. Such as pedestrian detection ~\cite{liu2019high}.\\

%In the past decade, numerous work have been proposed to address the image SR problem by learning the mapping functions from LR input to HR output. 
%
SRCNN~\cite{dong2014learning} is an unprecedented work to tackle the SR problem by learning the mapping function from LR input to HR output using convolutional neural networks (CNNs). Afterwards, numerous deep CNN-based methods~\cite{ren2019face,ren2018deep} have been proposed in recent years and generate a significant progress.
The superior reconstruction performance of CNNs based methods are
mainly from deep architecture and residual learning \cite{he2016deep}.
Networks with very deep layers have larger receptive fields and are able to provide a powerful capability to learn a complicated mapping between the LR input and the HR counterpart.
%
%Kim et al. \cite{kim2016accurate} proposed the VDSR by increasing the network to 20 layers. Compared to SRCNN, VDSR achieved better results due to network deepening.
% 
%To encourage the model convergence, VDSR learned a residual mapping between the HR and LR image. 
%
%Tai et al.~\cite{tai2017image}introduced DRRN which increase the depth of network to 52 layers. Due to the proposed of residual net (ResNet) by He at al.~\cite{he2016deep}, the CNN model can be deeper. With the addition of residual structure, the SR network can become deeper and the SR results can be improved effectively. 
%
%Lim et al.~\cite{lim2017enhanced} modified the residual structure to remove Batch Normalization~\cite{ioffe2015batch} layer and proposed EDSR network, which further deepened the SR network.
%\hl{high light}
%\underline{he introduction of residual network} 
Due to the residual learning, the depth of the SR networks are going to deeper since residual learning could  efficiently alleviate the gradient vanishing and exploding problems.

Though significant progress have been made, we note that the texture details of the LR image often tend to be smoothed in the super-resolved result since most existing CNN-based SR methods neglect the feature correlation of intermediate layers.
Therefore, generating detailed textures is still a non-trivial problem in the SR task.
Although the results obtained by using channel attention \cite{zhang2018image,dai2019second} retain some detailed information, these channel attention-based approaches struggle in preserving informative textures
%neglecting  the feature correlation of intermediate  layers from all the hierarchical features and hl{
and restoring natural details since they treat the feature maps at different layers equally and result in lossing some detail parts in the reconstructed image. 

To address these problems, we present a novel approach termed as holistic attention network (HAN) that is capable of exploring the correlations among hierarchical layers, channels of each layer, and all positions of each channel. Therefore, HAN is able to stimulate the representational power of CNNs.
Specifically, we propose a layer attention module (LAM) and a channel-spatial attention module (CSAM) in the HAN for more powerful feature expression and correlation learning. 
These two sub-attention modules are inspired by channel attention \cite{zhang2018image} which weighs the internal features of each layer to make the network pay more attention to information-rich feature channels.
% and explores the interdependencies between channels in each layer. 
%
However, we notice that channel attention cannot weight the features from multi-scale layers. Especially the long-term information from the shallow layers are easily weakened. Although the shallow features can be recycled via skip connections, they are treated equally with deep features across layers after long skip connection, hence hindering the representational ability of CNNs.
To solve this problem, we consider exploring the interrelationship among features at hierarchical levels, and propose a layer attention module (LAM). 
On the other hand, channel attention neglects that the importance of different positions in each feature map varies significantly. Therefore, we also propose a channel-spatial attention module (CSAM) to collaboratively improve the discrimination ability of the proposed SR network.

Our contributions in this paper are summarized as follows:

\begin{enumerate}[$\bullet$]

	\item We propose a novel super-resolution algorithm named Holistic Attention Network (HAN), which enhances the representational ability of feature representations for super-resolution.
	%makes the network employ the features from different layers and channels more reasonably and efficiently. %so that the detailed information and textures can be recovered better.

	\item We introduce a layer attention module (LAM) to learn the weights for hierarchical features by considering correlations of multi-scale layers.
	Meanwhile, a channel-spatial attention module (CSAM) is presented to learn the channel and spatial interdependencies of features in each layer. 

	%\item  We present a channel-spatial attention module (CSAM), which can not only rescale features among channels of each layer, but also capture more informative positions in each channel. so that the detailed information and textures can be recovered better.
	\item The proposed two attention modules collaboratively improve the SR results by modeling informative features among hierarchical layers, channels, and positions. Extensive experiments demonstrate that our algorithm performs favorably against the state-of-the-art SISR approaches.
	% Such LAM further enhances the representational ability of the our network.
	
\end{enumerate}

\section{Related Work}
Numerous algorithms and models have been proposed to solve the problem of image SR, which can be roughly divided into two categories. One is the traditional algorithm~\cite {yang2008image,huang2018robust,huang2015single}, the other one is the deep learning model based on neural network \cite{kim2016accurate,dong2016accelerating,lai2017deep,lim2017enhanced,zhang2018residual,kim2016deeply,tai2017image,tai2017memnet}. Due to the limitation of space, we only introduce the SR algorithms based on deep CNN.\par   
%{\flushleft \bf Deep CNN for super-resolution.}
%\subsubsection{Deep CNN for super-resolution.}
%
\textbf{Deep CNN for super-resolution.} Dong et al.~\cite{dong2014learning} proposed a CNN architecture named SRCNN, which was the pioneering work to apply deep learning to single image super-resolution. 
%This work achieves the seminal success by using only convolutional layers for SR. SRCNN network is simple and straightforward. They used three convolutional layers and two ReLU layers to extract feature map and get the approximate non-linear mapping between LR images and paired HR images. 
Since SRCNN successfully applied deep learning network to SR task, various efficient and deeper architectures have been proposed for SR.
%
%Kim et al.~\cite{kim2016accurate} proposed a very deep network (VDSR) which increases the depth of network from 3 to 20 layers and shows significant improvement over SRCNN. To facilitate training the deeper model with a fast convergence speed, VDSR learns a residual mapping between the HR and LR image rather than the actual pixel values.
%trains the network to predict the residuals rather the actual pixel values. they learned a residual mapping between the HR and LR image.
%
Wang et al. \cite{wang2015deep}combined the domain knowledge of sparse coding with a
deep CNN and trained a cascade network to recover images progressively.
To alleviate the phenomenon of gradient explosion and reduce the complexity of the model, DRCN~\cite{kim2016deeply} and DRRN~\cite{tai2017image} were proposed by using a recursive convolutional network.
Lai et al.~\cite{lai2017deep} proposed a LapSR network which employs a pyramidal framework to progressively generate $\times 8$ images by three sub-networks. 
Lim et al.~\cite{lim2017enhanced} modified the ResNet~\cite{he2016deep} by removing batch normalization (BN) layers, which greatly improves the SR effect.

In addition to above MSE minimizing based methods, perceptual constraints are proposed to achieve better visual quality \cite{sajjadi2017enhancenet}. 
SRGAN~\cite{ledig2017photo} uses a generative adversarial networks (GAN) to predict high-resolution outputs by introducing a multi-task loss including a MSE loss, a perceptual loss~\cite{johnson2016perceptual}, and an adversarial loss~\cite{NIPS2014_5423}.
%
%Sajjadi et al. \cite{sajjadi2017enhancenet} incorporated the texture matching loss to improve textures in the reconstructed SR result.
%
Zhang et al. \cite{zhang2019image} further transferred textures from reference images according to the textural similarity to enhance textures.
However, the aforementioned models either result in the loss of detailed textures in intermediate features due to the very deep depth, or produce some unpleasing artifacts or inauthentic textures. 
In contrast, we propose a holistic attention network consists of a layer attention and a channel-spatial attention to investigate the interaction of different layers, channels, and positions.
%the characteristics of different network depths should be weighted differently, so we design two attention mechanisms, namely local attention module and global attention module.

%{\flushleft \bf Attention mechanism.} 
%Attention mechanism is widely used in natural language processing, but seldom used in image enhancement. 
%\noindent \textbf{Attention mechanism.}
\textbf{Attention mechanism.} Attention mechanisms direct the operational focus of deep neural networks to areas where there is more information. In short, they help the network ignore irrelevant information and focus on important information \cite{hu2018squeeze,hu2019channel}. 
Recently, attention mechanism has been successfully applied into deep CNN based image enhancement methods. 
%Wang et al.~\cite{wang2018non} introduced non-local neural network in video classification which contains non-local operations for spatial attention. In image classification task, Hu et al.~\cite{hu2018squeeze} proposed SENet to exploit channel-wise relationships to achieve noteworthy performance gain. 
%
Zhang et al.~\cite{zhang2018image} proposed a residual channel attention network (RCAN) in which residual channel attention blocks (RCAB) allow the network to focus on the more informative channels. 
%The  in  is equal to the squeeze-and-excitation (SE) block in SENet. 
Woo et al.~\cite{woo2018cbam} proposed channel attention (CA) and spatial attention (SA) modules to exploit both inter-channel and inter-spatial relationship of feature maps. 
%CSAR block~ also proposed CA and SA modules, but it remove squeeze process and combined two modules in a parallel manner. 
%
Kim et al.~\cite{kim2018ram} introduced a residual attention module for SR which is composed of residual blocks and spatial channel attention for learning the inter-channel and intra-channel correlations. More recently, Dai et al.~\cite{dai2019second} presented a second-order channel attention (SOCA) module to adaptively refine features using second-order feature statistics.
%
%All of these aforementioned methods put forward relatively efficient channel attention mechanism and spatial attention mechanism. 
%
\begin{figure}[t]\footnotesize
	\begin{center}
		\includegraphics[width = 0.8\textwidth]{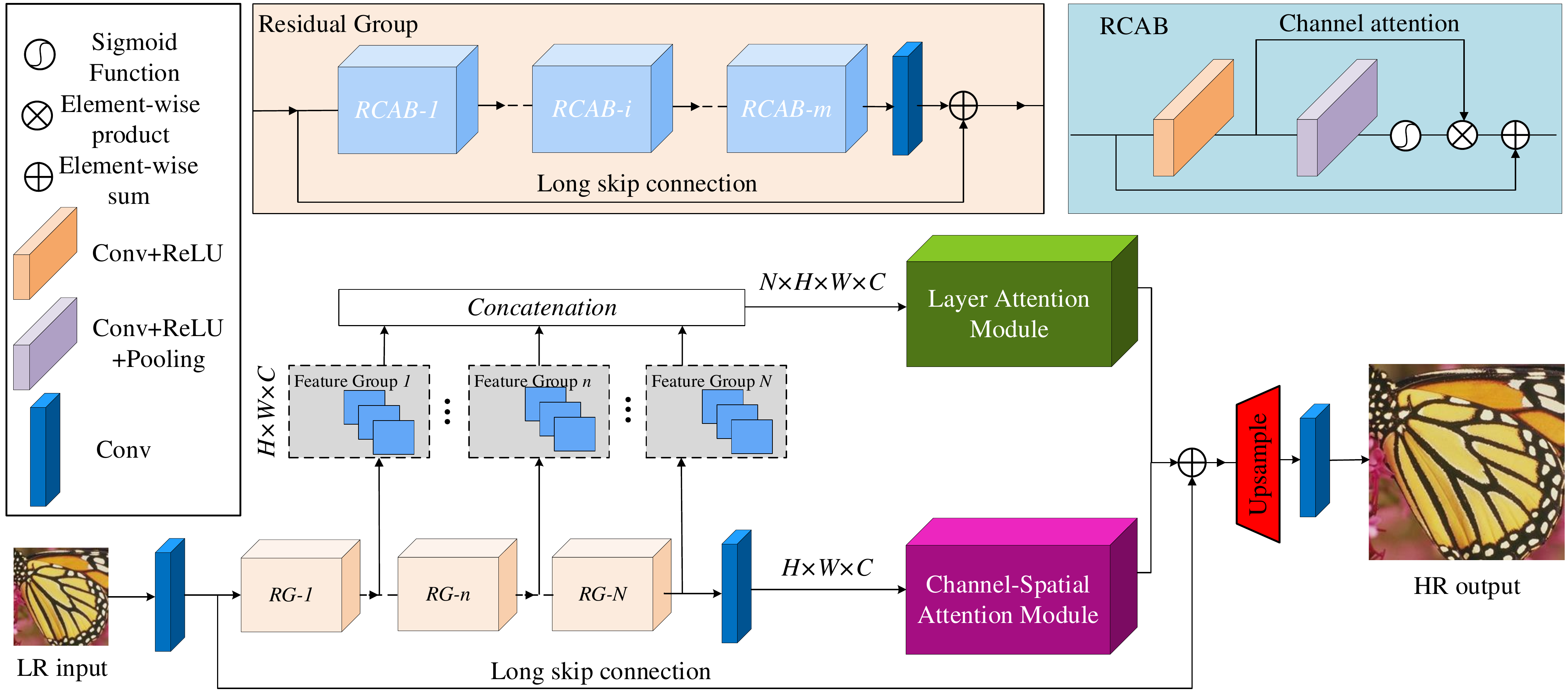}
	\end{center}

	\caption{Network architecture of the proposed holistic attention network(HAN). Given a low-resolution image, the first convolutional layer of the HAN extracts a set of shallow feature maps. Then a series of residual groups further extract deeper feature representations of the low-resolution input. We propose a layer attention module (LAM) to learn the correlations of each output from RGs and a channel-spatial attention module (CSAM) to investigate the interdependencies between channels and pixels. Finally, an upsampling block produces the high-resolution image
	}

	\label{fig-Holistic_Attention_Network}
\end{figure}

However, these attention based methods only consider the channel and spatial correlations while ignore the interdependencies between multi-scale layers. To solve this problem, we propose a layer attention module (LAM) to
exploit the nonlinear feature interactions among hierarchical layers.  
% for features of different depths, they only used spatial attention mechanism to extract information roughly, instead of hierarchical processing of features of different depths. We propose a global attention and a local attention mechanism, in which different depth features are weighted to represent the different contributions to the SR results.

\section{Holistic Attention Network (HAN) for SR}
\label{sec-att}
In this section, we first present the overview of HAN network for SISR. Then we give the detailed configurations of the proposed layer attention module (LAM) and channel-spatial attention module (CSAM).

\subsection{Network Architecture}
As shown in Figure \ref{fig-Holistic_Attention_Network}, our proposed HAN consists of four parts: feature extraction, layer attention module, channel-spatial attention module, and the final reconstruction block.
%\noindent \textbf{Features extraction.}
%

\textbf{Features extraction.} Given a LR input $I_{LR}$, a convolutional layer is used to extract the shallow feature $F_0$ of the LR input
\begin{equation}
\mathit{F_0} = \text{Conv}(I_{LR}) .
\end{equation}
Then we use the backbone of the RCAN \cite{zhang2018image} to extract the intermediate features $F_i$ of the LR input
%
%where $RB$ represents the backbone of the RCAN from \cite{bibid}. 
%%
%RCAN extracts feature maps $F$ which are used for the subsequent layer attention module. The input of the layer attention module is the feature maps of the convolutional layer output at different depths. $F$ can be expressed as $F_{1}$,$F_{2}$,...,$F_{N}$ each component is represented as follows:
%
\begin{equation}
\mathit{F_{i}} = H_{RB_i}(F_{i-1}) , ~~i = 1,2,...,N ,
\end{equation}
where $H_{RB_i}$ represents the $i$-th residual group (RG) in the RCAN, $N$ is the number of the residual groups. Therefore, except $F_{N}$ is the final output of RCAN network backbone, all other feature maps are intermediate outputs.
%Specific residual group selection will be given in the experimental Settings section.

%
%The process of layer attention is as follows:
\textbf{Holistic attention.} After extracting hierarchical features $F_i$ by a set of residual groups, we further conduct a holistic feature weighting, which includes: \textit{i}) layer attention of hierarchical features, and \textit{ii}) channel-spatial attention of the last layer of RCAN. 

The proposed layer attention makes full use of features from all the preceding layers and can be represented as
\begin{equation}
\mathit{F_{L}} = H_{LA}(\text{concatenate}(F_1, F_2, ..., F_N)) ,
\end{equation}
%
%{\color{red}}
where $ H_{LA} $ represents the LAM which learns the feature correlation matrix of all the features from RGs' output and then weights the fused intermediate features $F_i$ capitalized on the correlation matrix (see Section \ref{sec-lam}). As a results, LAM enables the high contribution feature layers to be enhanced and the redundant ones to be suppressed.
% where $ F_{L} $ denotes fusion feature of layer attention. $F^{l} $ is the input of LAM which can be expressed as the result of parallel operation after extracting feature F dimension expansion by RCAN. 
%Therefore, the acquisition method of $F^{l} $ is formulated as follows:
%%
%\begin{equation}
%\mathit{F^l}=concat([unsqueeze(F_{1}),unsqueeze(F_{2}),…,unsqueeze(F_{N})],1)
%\end{equation}
%%
%where concat denotes parallel operation, and unsqueeze means adding one dimension,that is change $ H $$\times$$ W $$\times$$ C $ into  $ H $$\times$$ W $$\times$$ C $$\times$$ N $.

	In addition, channel-spatial attention aims to modulate features for adaptively capturing more important information of inter-channel and intra-channel for the final reconstruction, which can be written as
	\begin{equation}
	\mathit{F_{CS}}=H_{CSA}(F_N) ,
	\end{equation}
	\noindent where $ H_{CSA} $ represents the CSAM to produce channel-spatial attention for discriminately abtaining feature information, $F_{CS}$ denotes the filtered features after channel-spatial attention (details can be found in Section \ref{sec-csam}). 
	Although we can filter all the intermediate features of $F_i$ using CSAM, we only modulate the last feature layer of $F_N$ as a trade-off between accuracy and speed.
%
%$F^{cs} $is the input of channel-spatial attention module which can be obtained by parallel and series acquisition two methods. For parallel acquisition,$F^{cs} $ denotes the last layer of RCAN output as input, i.e.
%
%\begin{equation}
%\mathit{F^{cs}}=F_{N}
%\end{equation}
%
%For serial acquisition, local attention takes the outputs of LAM as the input, i.e. 

%\begin{equation}
%\mathit{F^{cs}}=conv2d(F_{L})
%\end{equation}

%\noindent where parallel and series means that global attention and local attention are combined in parallel. 

%
\textbf{Image reconstruction.} After obtaining features from both LAM and CSAM, we integrate the layer attention and channel-spatial attention units by element-wise summation.
%We did not use cascading operations in any attention module which can effectively reduce GPU memory consumption. In addition, our attention model is very lightweight, and the implementation of attention does not use any 2D convolution. Therefore, our attention module can be extended to more tasks as a feature enhancement.
%
Then, we employ the sub-pixel convolution \cite{shi2016real} as the last upsampling module,
%The last upsampling module is selected as the sub-pixel convolution \cite{bibid},
which converts the scale sampling with a given magnification factor
%, and then obtains the up-sampled result 
by pixel translation.  We perform the sub-pixel convolution operation to aggregate low-resolution feature maps and simultaneously impose projection to high dimensional space to reconstruct the HR image. We formulate the process as follows
\begin{equation}
\mathit{I_{SR}}=U_{\uparrow}(F_0 + F_L + F_{CS}) ,
\end{equation}
where $ U_{\uparrow} $ represents the operation of sub-pixel convolution, and $ I_{SR} $ is the reconstructed SR result.
The long skip connection is introduced in HAN to stabilize the training of the proposed deep network, \textit{i.e.,} the sub-pixel upsampling block takes $F_0 + F_L + F_{CS}$ as input.
% $ F^{u} $ is the input of the up-sampling module. 
%For the serial case, the sub-pixel up-sampling module directly takes the result of global attention as the input:
%
%\begin{equation}
%\mathit{F^{u}}=F_{L}
%\end{equation}

%As shown in Figure \ref{fig-Holistic_Attention_Network} , the 
%\begin{equation}
%	\mathit{F^{u}}=fusion(F_{L},F_{CS},F_{0})
%\end{equation}
%where $ fusion(\cdot) $ donets the way of fusion method. In this paper, we choose element-wise sum as the fusion function.

\begin{figure*}[t]\footnotesize
	\begin{center}
		\includegraphics[width = 1\textwidth]{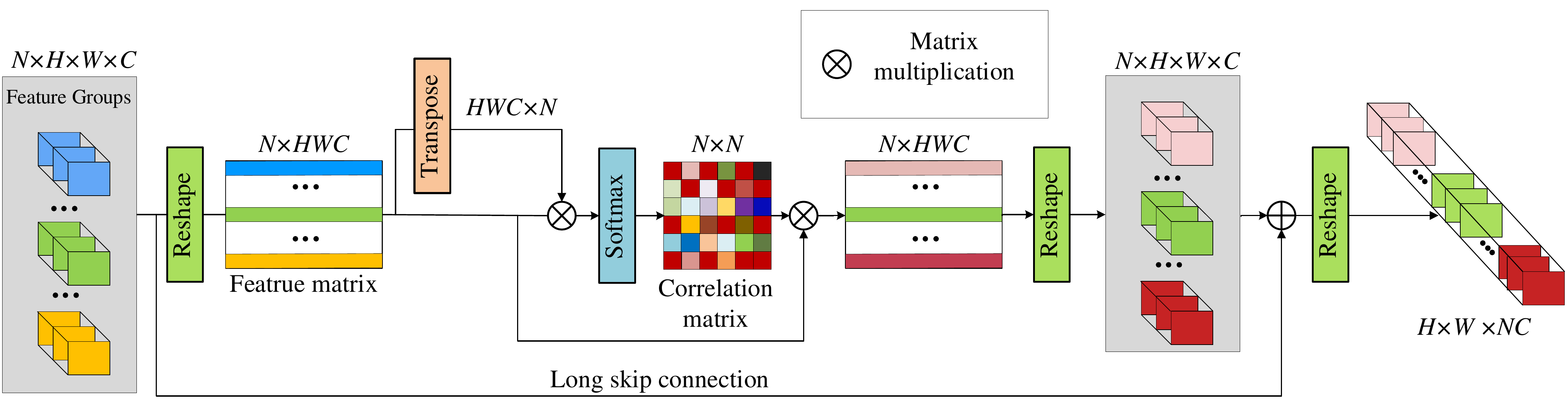}
	\end{center}

	\caption{Architecture of the proposed layer attention module
	}

	\label{fig-Layer_attention}
\end{figure*}

%Several loss functions are applied to SISR reconstruction task, including $ L_{1} $ loss,$ L_{2} $ loss, $ perception  $ loss, $ adversarial $ loss, etc. 
%
\textbf{Loss function.} Since we employ the RCAN network as the backbone of the proposed method, only $ L_{1} $ distance is selected as our loss function as in \cite{zhang2018image} for a fair comparison
	%
	%baseline of this paper is selected as RCAN network, so in order to be consistent with it, only $ L_{1} $ loss is selected as the target loss function. We represent the input sample pairs as$ \{I^{i}_{LR}, I^{i}_{HR}\}^{m}_{i=1} $,where $ m $ denotes the number of sample pairs. The final loss function of our proposed network is:
	%
	\begin{equation}
	\mathit{L(\Theta)} = \dfrac{1}{m}\sum_{i=1}^{m}\big\|H_{HAN}(I^{i}_{LR})-I^{i}_{HR}\big\|_{1}= \dfrac{1}{m}\sum_{i=1}^{m}\big\|I^{i}_{SR}-I^{i}_{HR}\big\|_{1} ,
	\end{equation}
	where $H_{HAN}$, $ \Theta $, and $m$ denote the function of the proposed HAN, the learned parameter of the HAN, and the number of training pairs, respectively.
	Note that we do not use other sophisticated loss functions such as adversarial loss \cite{NIPS2014_5423} and perceptual loss~\cite{johnson2016perceptual}. We show that simply using the
	naive image intensity loss $L(\Theta)$ can already achieve
	competitive results as demonstrated in Section \ref{sec-exp}.

\subsection{Layer Attention Module}
\label{sec-lam}
Although dense connections~\cite{huang2017densely} and skip connections \cite{he2016deep} allow shallow information to be bypassed to deep layers, 
%
%feature reuse and bypass setup, dense connection network~\cite{huang2017densely}makes the fusion of different depth features possible. The network combines information flow and feature reuse, but 
these operations do not exploit interdependencies between the different layers. 
%
%chooses the simple parallel mode for information reuse. 
In contrast, we treat the feature maps from each layer as a response to a specific class, and the responses from different layers are related to each other. 
By obtaining the dependencies between features of different depths, the network can allocate different attention weights to features of different depths and automatically improve the representation ability of extracted features. Therefore, we propose an innovative LAM that learns the relationship between features of different depths, which automatically improve the feature representation ability.

The structure of the proposed layer attention is shown in Figure \ref{fig-Layer_attention}. The input of the module is the extracted intermediate feature groups $FG$s, with the dimension of $ N $$\times$$ H $$\times$$ W $$\times$$ C $, from $N$ residual groups. 
%
%Different from the general input of attention module, we first preprocessed the input by adding one dimension.
%
%Compared with the parallel connection of the channel dimension in the general dense network. We add another dimension to the channel dimension, that is, change $ H $$ \times $$ W $$ \times $$ C $ to  $ N $$\times$$ H $$\times$$ W $$\times$$ C $, which can make the backbone feature depth independent of the network channel, so that different weights can be applied to features at different levels according to the depth. This is the design idea of the LAM.
%
%It should be pointed out that 
%
%the feature tensor dimension of general network is three-dimension (excluding batch dimension), however, 
%
%the input of the LAM is a 4D feature ($ N $$\times$$ H $$\times$$ W $$\times$$ C $) by adding a depth dimension. 
%
%Then, we calculate the correlation matrix of feature groups.
%at different depths with the help of the preprocessed feature groups whose dimention is $ N $$\times$$ H $$\times$$ W $$\times$$ C $. 
Then, we reshape the feature groups $FG$s into a 2D matrix with the dimension of $ N\times HWC$, and apply matrix multiplication with the corresponding transpose
%to $ FG $ and the transpose $ FG^\top$ 
to calculate the correlation $W_{la} = w_{i,j=1}^N$ between different layers
\begin{equation}
w_{j,i}= \delta(\varphi(FG)_i \cdot (\varphi(FG))_j^{\mathrm{T}}) ,  ~~i,j = 1,2,...,N ,
\end{equation}
%
%In particular,we first reshape the input into $ N\times HWC$, denotes as $ FG $. Then, apply matrix multiplication to $ FG $ and the transpose $ FG^\top$. Finally, we applied softmax function to obtain the correlation matrix
%$ X\in R^{N\times N} $.
%
%\begin{equation}
%x_{i, j}=\frac{\exp \left(FG_{i} \cdot FG^{\top}_{j}\right)}{\sum_{i=1}^{N}\sum_{j=1}^{N} \exp \left(FG_{i} \cdot FG^{\top}_{j}\right)},
%\end{equation}
%
\begin{figure*}[t]\footnotesize
	\begin{center}
		\includegraphics[width = 0.8\textwidth]{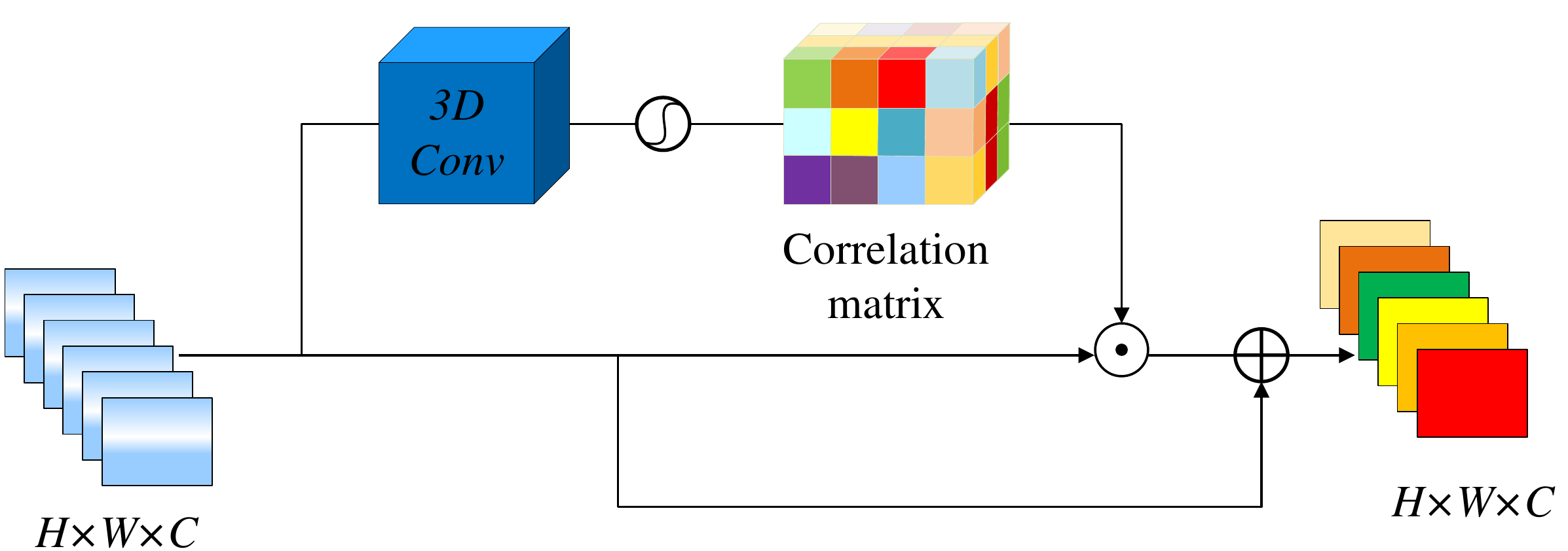}
	\end{center}

	\caption{Architecture of the proposed channel-spatial attention module}

	\label{fig-CSAM}
\end{figure*}
where $\delta(\cdot)$ and $\varphi(\cdot)$ denote the softmax and reshape operations, $ x_{i,j} $ represents the correlation index between $i$-th and $j$-th feature groups. 
%
%In the experiments, since exponential computing has high complexity, we use the square function to substitute 
%
%consumes a lot of computing power. We find that the square function will have the same effect and save abundant of computing power. We adopt squaremax activation function to obtain the relational matrix:
%\begin{equation}
%\operatorname { squaremax }(\mathrm{x})=\frac{x_{i}^{2}}{\sum_{i} x_{i}^{2}}
%\end{equation}
%To obtain weighted result, 
Finally, we multiply the reshaped feature groups $FG$s by the predicted correlation matrix with a scale factor $ \alpha $, and add the input features $FGs$
%performs multiplication with the reshaped input feature groups $FG$. Then weighted result is multiplied by a scale factor $ \alpha $, which were added to the input feature by pixel to obtain the weighted result $ F_{L} $.
%
\begin{equation}
F_{L_j}=\alpha \sum_{i=1}^{N} w_{i,j} FG_{i}+FG_{j} ,
\end{equation}
where $ \alpha $ is initialized to $0$ and is automatically assigned by the network in the following epochs. 
As a result, the weighted sum of features allow the main parts of network to focus on more informative layers of the intermediate LR features.

%\url{www.pamitc.org/documents/mermin.pdf}.

%\section{Policies}
\subsection{Channel-Spatial Attention}
\label{sec-csam}
The existing spatial attention mechanisms~\cite{woo2018cbam,kim2018ram} mainly focuse on the scale dimension of the feature, with little uptake of channel dimension information, while the recent channel attention mechanisms~\cite{zhang2018image,zhang2018residual,dai2019second} ignore the scale information. 
To solve this problem, we propose a novel channel-spatial attention mechanism (CSAM) that contains responses from all dimensions of the feature maps. Note that although we can perform the CSAM for all the feature groups $FG$s extracted from RCAN, we only modulate the last feature group of $F_N$ for a trade-off between accuracy and speed as shown in Figure \ref{fig-Holistic_Attention_Network}.

%The design of CSAM has two pointcuts: relational matrix or convolution method. If we proposed a scheme similar to the LAM to learn relational matrix, instead of the LAM learning the $ N $$\times$$ N $ matrix, we need to learn the $HWC$$ \times $$HWC$ matrix. Such matrix relationships are not suitable for deep networks because of limited computing and storage capacity. The feature of weight sharing in the convolutional layer can greatly reduce the calculation. But 2D convolution is no longer applicable for feature acquisition in 3 dimensions. Therefore, we design a CSAM using 3D convolution.

The architecture of the proposed CSAM is shown in Figure \ref{fig-CSAM}. 
Given the last layer feature maps $ F_N \in R^{H\times W \times C} $, 
%we apply unsqueeeze processing to $ F $. And the reason for adding another dimension here is to cater to the need of 3D convolution.
we feed $F_N$ to a 3D convolution layer~\cite{ji20123d} to generate attention map by capturing joint
channel and spatial features.
We operate the 3D convolution via convolving 3D kernels with the cube constructed from multiple neighboring channels of $F_N$. Specifically, we perform 3D convolutions with kernel size of $ 3\times 3\times 3 $ with step size of 1 (\textit{i.e.,} three groups of consecutive channels are convolved with a set of 3D kernels respectively), resulting
in three groups of channel-spatial attention maps $W_{csa}$.
By doing so, our CSAM can extract powerful representations to describe inter-channel and intra-channel information in continuous channels.

In addition, we perform element-wise multiplication with the attention map $ W_{csa} $ and the input feature $ F_N $. Finally, multiply the weighted result by a scale factor $ \beta $, and then add the input feature $F_N$ to obtain the weighted features
%
%F_{CS}=\beta \sigma(W_{csa}) \cdot F_N + F_N,
\begin{equation}
F_{CS}=\beta \sigma(W_{csa}) \odot F_N + F_N ,
\end{equation}
where $\sigma(\cdot)$ is the sigmoid function, $ \odot $ is the element-wise product, the scale factor $ \beta $ is initialized as 0 and progressively improved in the follow iterations. 
As a results, $F_{CS}$ is the weighted sum of all channel-spatial position features as well as the original features. Compared with conventional spatial attention and channel attention, our CSAM adaptively learns the inter-channel and intra-channel feature responses by explicitly modelling channel-wise and spatial feature interdependencies.
% information and can selectively obtain features according to the fusion information.

%Compared with 2D convolution in computer vision tasks, 3D convolution~\cite{ji20123d} has fewer parameters and greater receptive field. Due to the limited space, we're not going to do a theoretical expansion, 

%we set the convolution kernel as $ 3\times 3\times 3 $ as the size of all 3D Conv layers and set 1 as step size, zero-padding strategy is used to maintain size fixed. 
%
%local attention map needs to be activated by sigmoid, and the values range of the activated attention map are between 0 and 1. The reason for choosing this activation function is to normalize feature.
%

%\begin{equation}
%Y=\frac{1}{1+\exp (-\operatorname{conv3d}(\operatorname{unsqueeze}(F)))}
%\end{equation}

%
\begin{figure}[t]\tiny
	\begin{center}
		\tabcolsep 1pt
		\begin{tabular}{@{}ccccccccc@{}}
			HR&
			Bicubic & 
			VDSR\cite{kim2016accurate} & 
			EDSR\cite{lim2017enhanced} & 
			RDN\cite{zhang2018residual} & 
			RCAN\cite{zhang2018image}  &
			SRFBN\cite{li2019feedback} &
			SAN\cite{dai2019second} &
			HAN(our) \\

			\includegraphics[width = 0.1\textwidth]{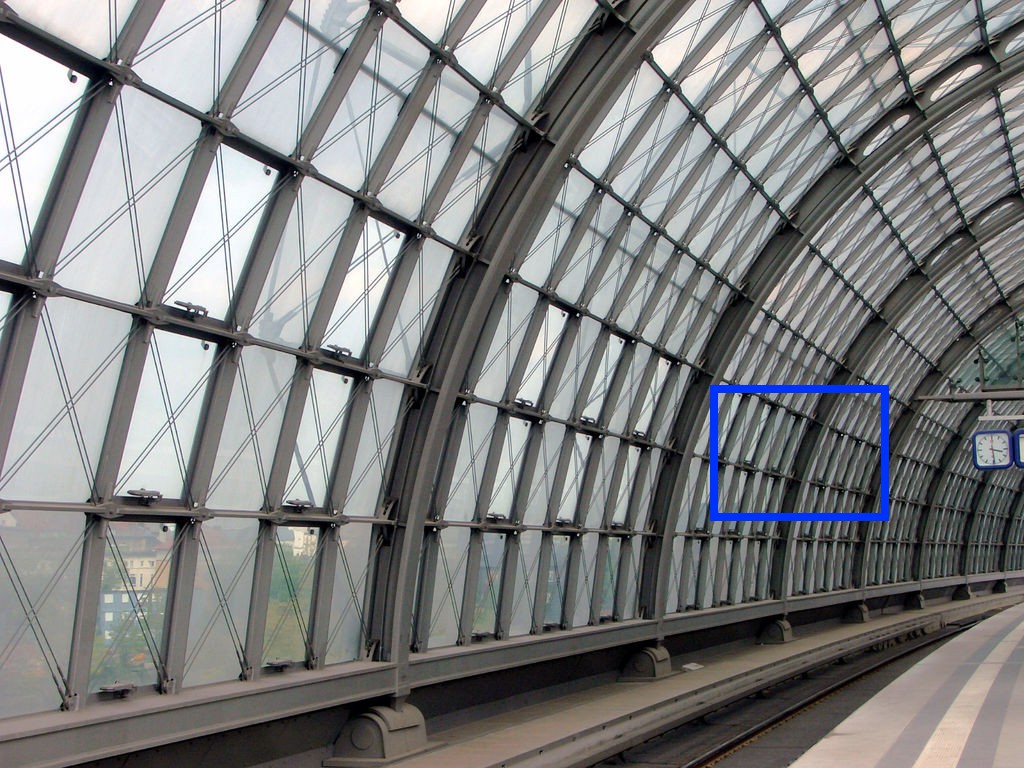}&
			\includegraphics[width = 0.1\textwidth]{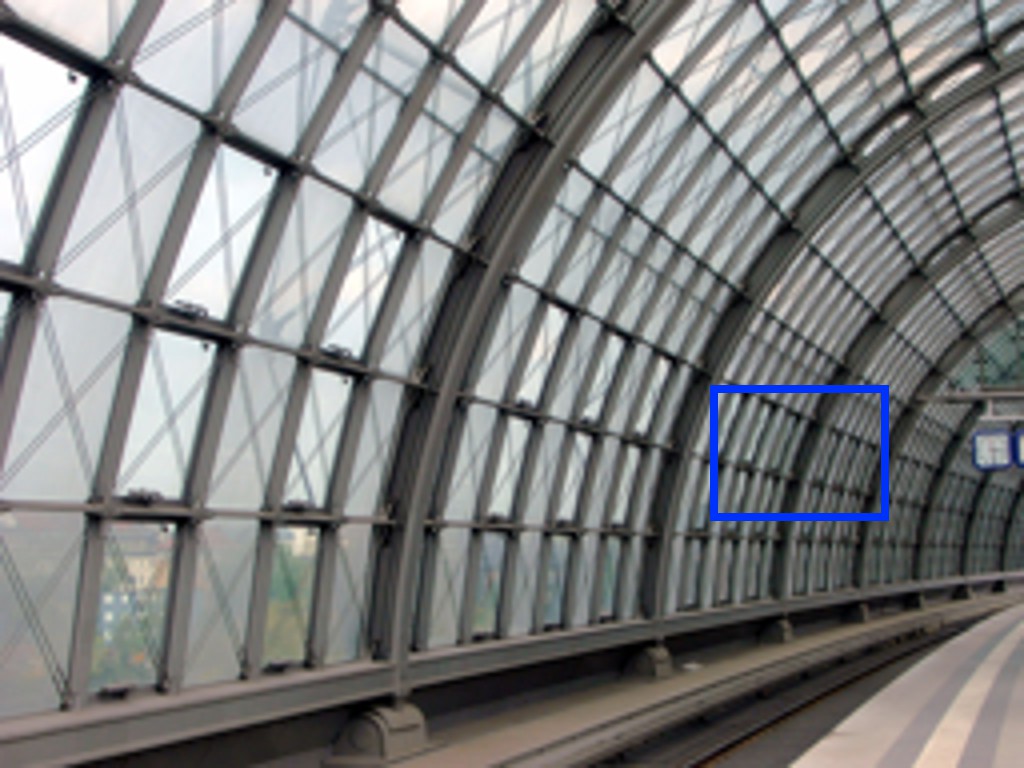} & 
			\includegraphics[width = 0.1\textwidth]{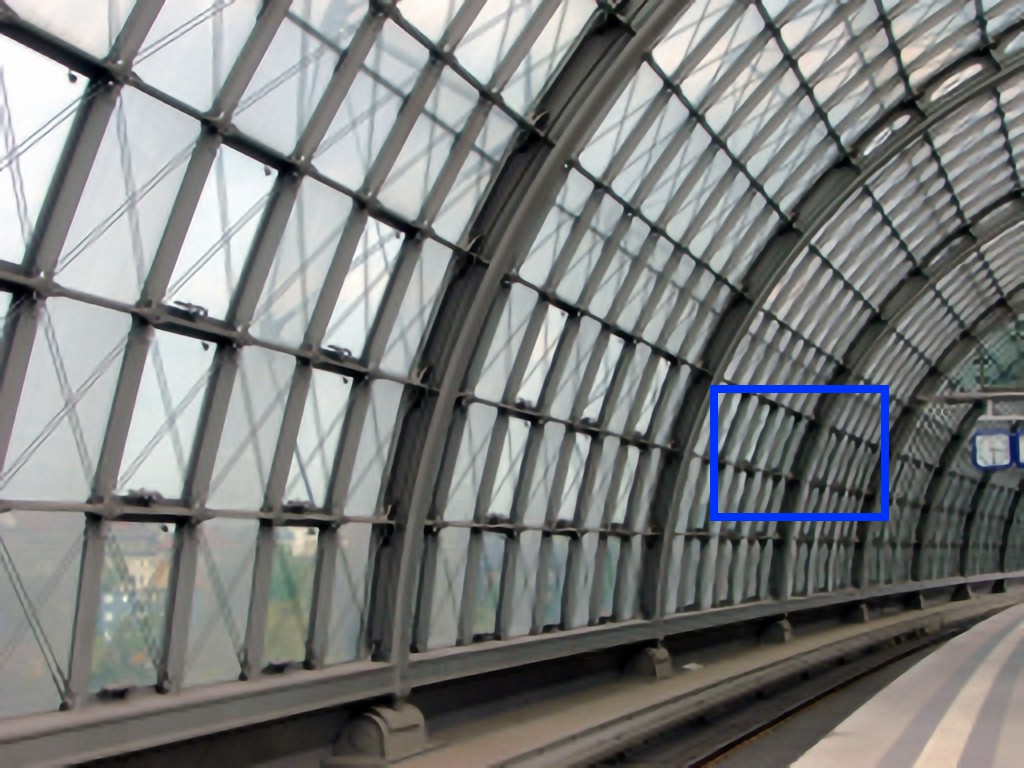} &
			\includegraphics[width = 0.1\textwidth]{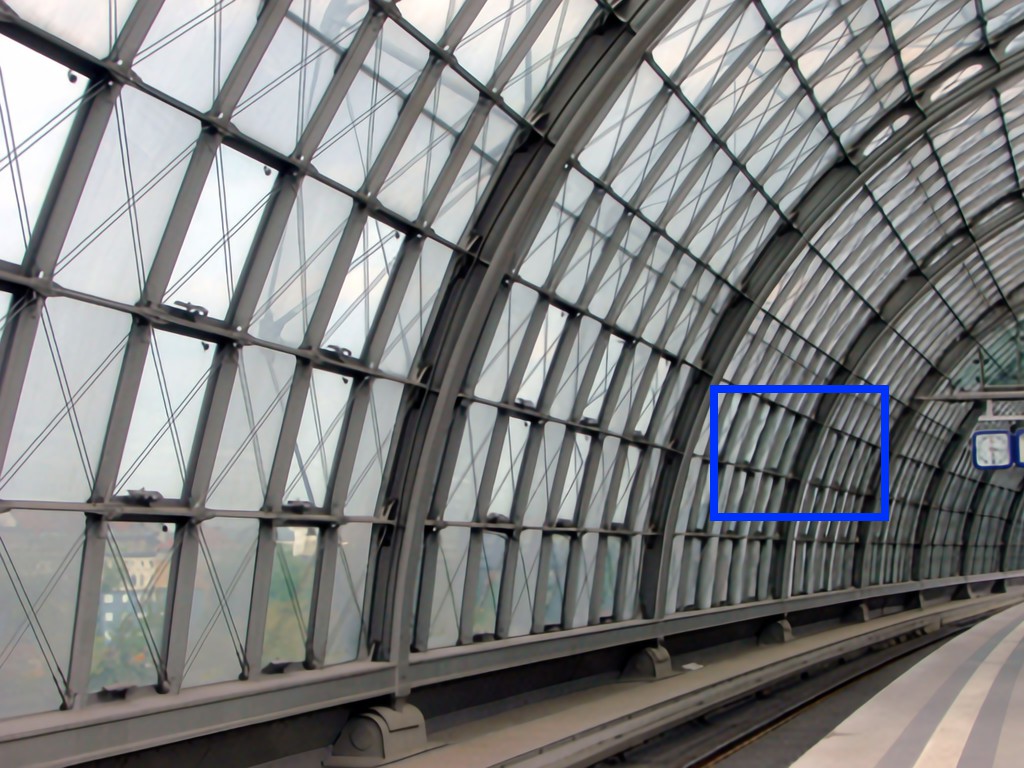} & 
			\includegraphics[width = 0.1\textwidth]{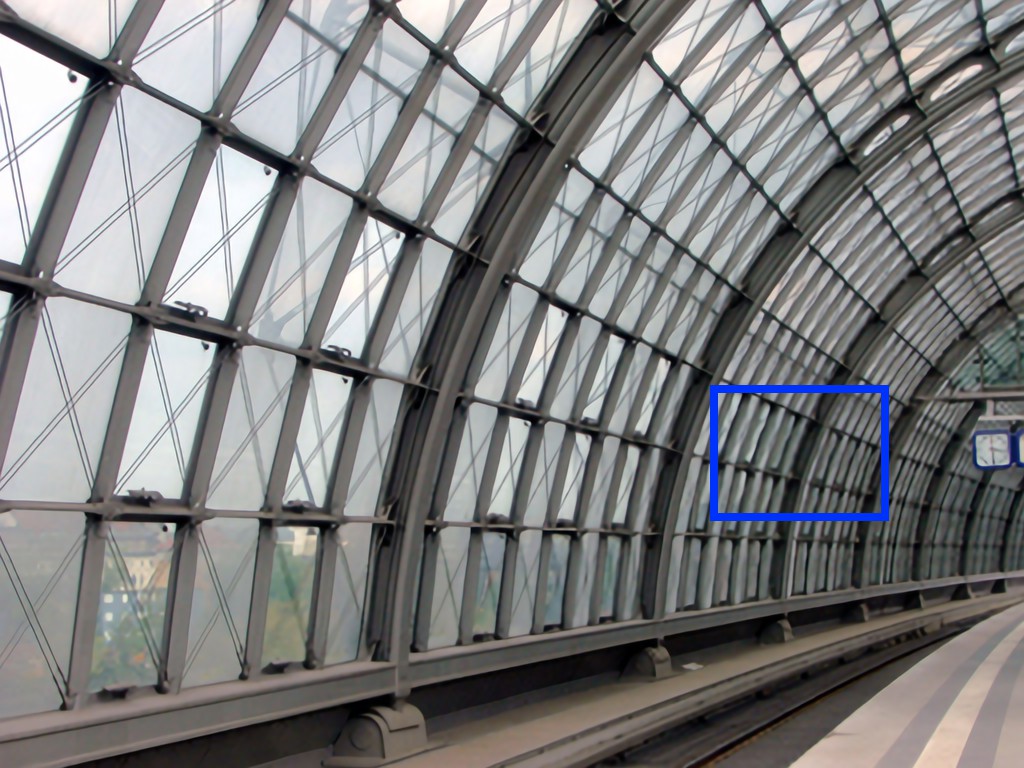} & 
			\includegraphics[width = 0.1\textwidth]{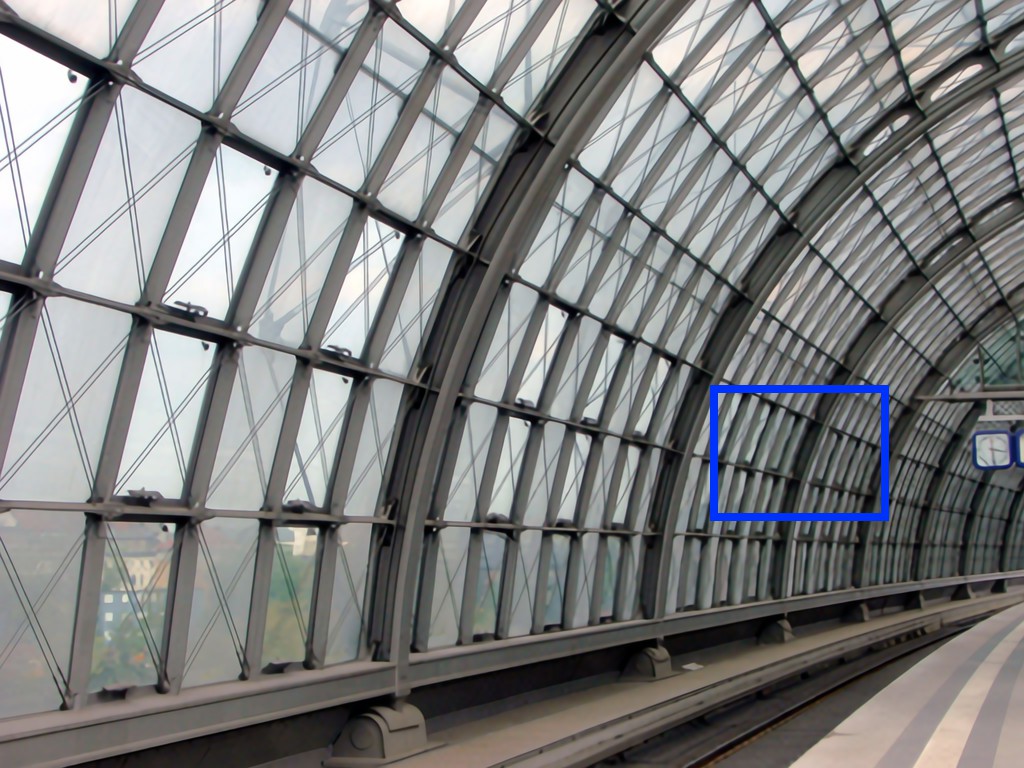} & 
			\includegraphics[width = 0.1\textwidth]{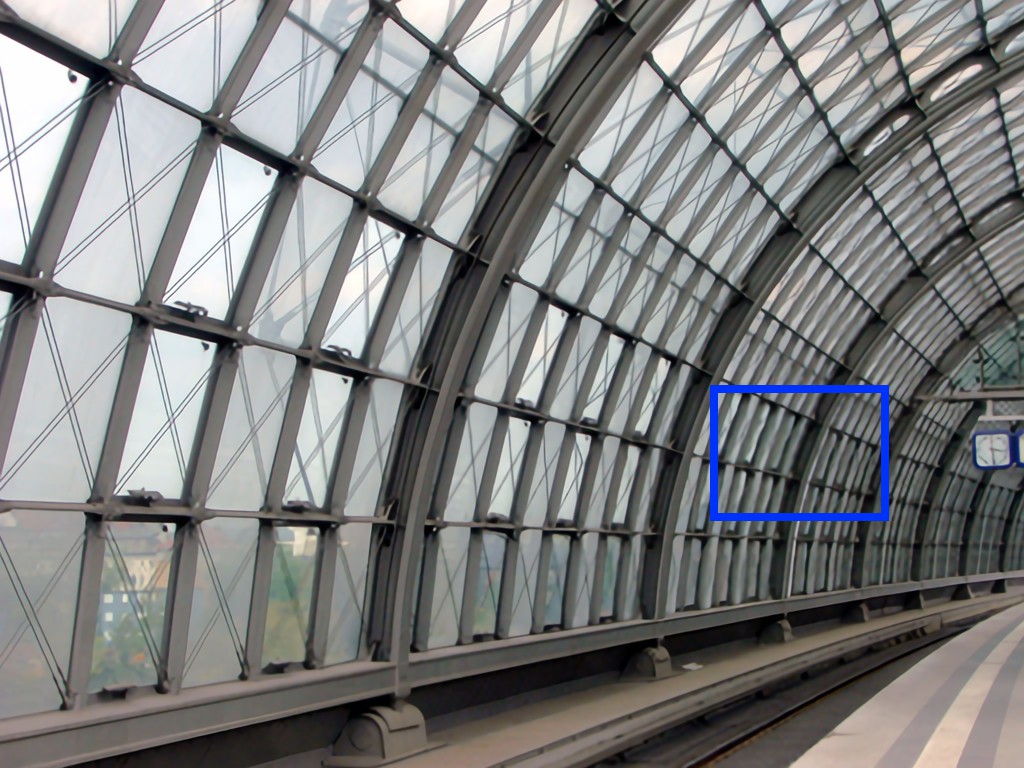}& 
			\includegraphics[width = 0.1\textwidth]{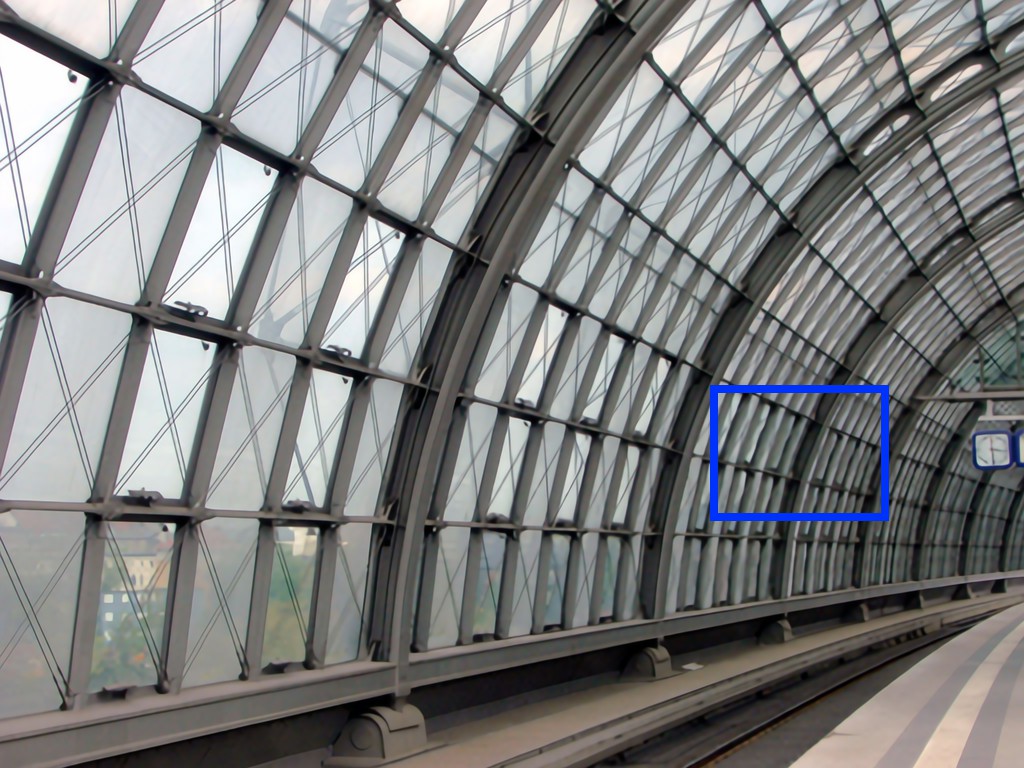} & 
			\includegraphics[width = 0.1\textwidth]{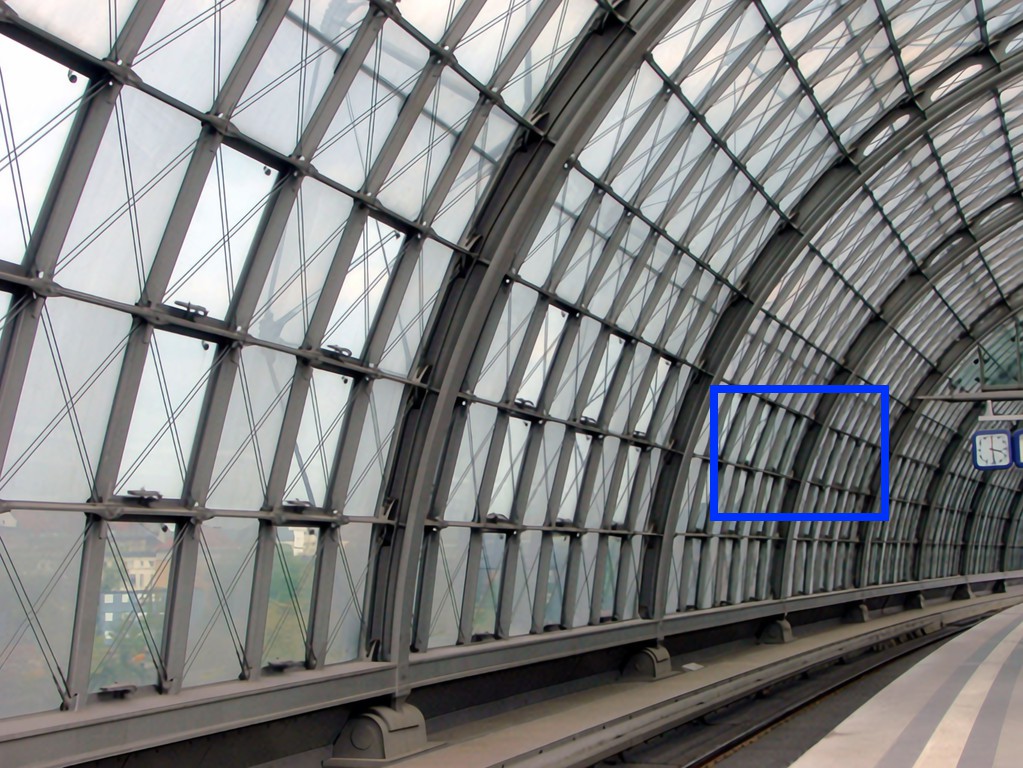} \\ 

			\includegraphics[width = 0.1\textwidth]{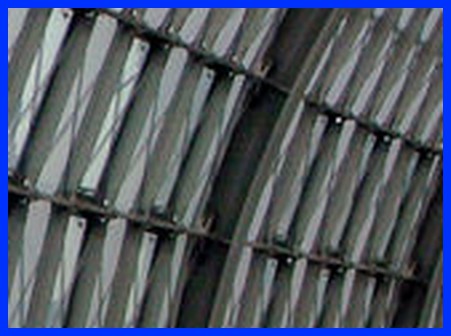}&
			\includegraphics[width = 0.1\textwidth]{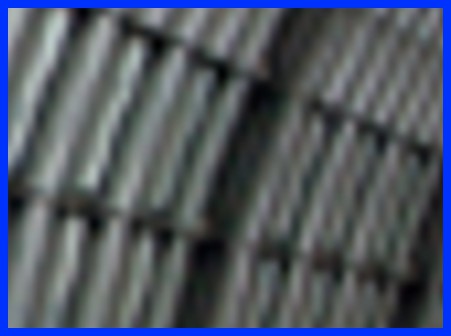} & 
			\includegraphics[width = 0.1\textwidth]{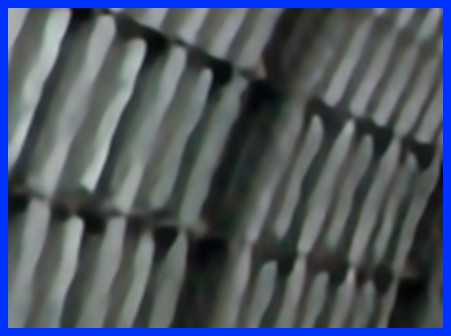} &
			\includegraphics[width = 0.1\textwidth]{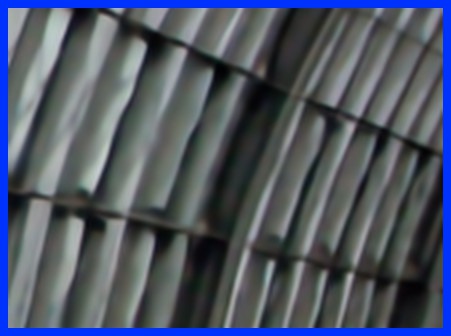} & 
			\includegraphics[width = 0.1\textwidth]{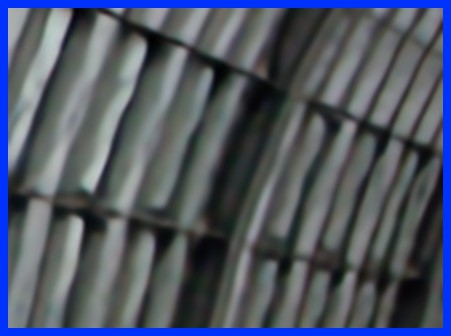} & 
			\includegraphics[width = 0.1\textwidth]{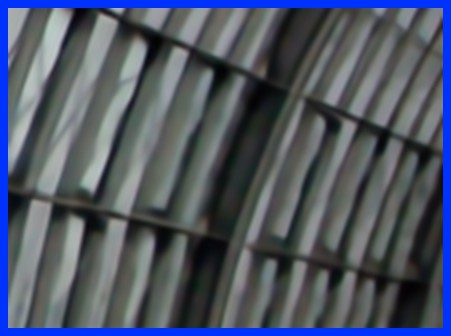} & 
			\includegraphics[width = 0.1\textwidth]{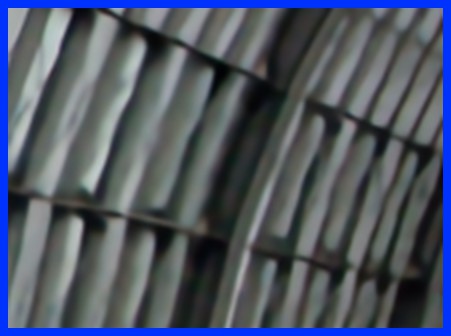}& 
			\includegraphics[width = 0.1\textwidth]{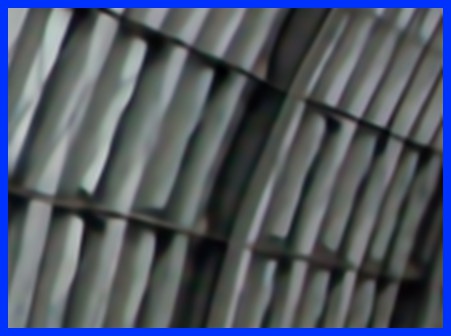} & 
			\includegraphics[width = 0.1\textwidth]{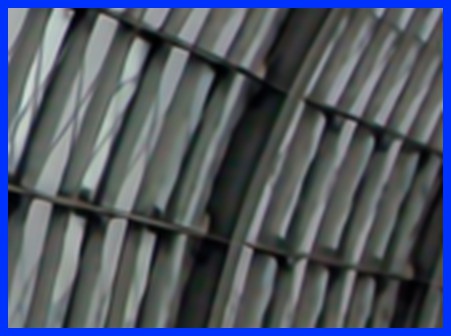} \\

			PSNR/SSIM & 24.18/0.678  & 25.63/0.763 & 27.66/0.849 & 27.12/0.832 & 27.95/0.857 & 27.43/0.843 & 27.99/0.857 & \textbf{28.05}/\textbf{0.859} \\	
			%	PSNR/SSIM & 24.18/0.6783  & 25.63/0.7632 & 27.66/0.8486 & 27.12/0.8318 & 27.95/0.8570 & 27.43/0.8427 & 27.99/0.8566 & 28.05/0.8590 \\ 
			
			\includegraphics[width = 0.1\textwidth]{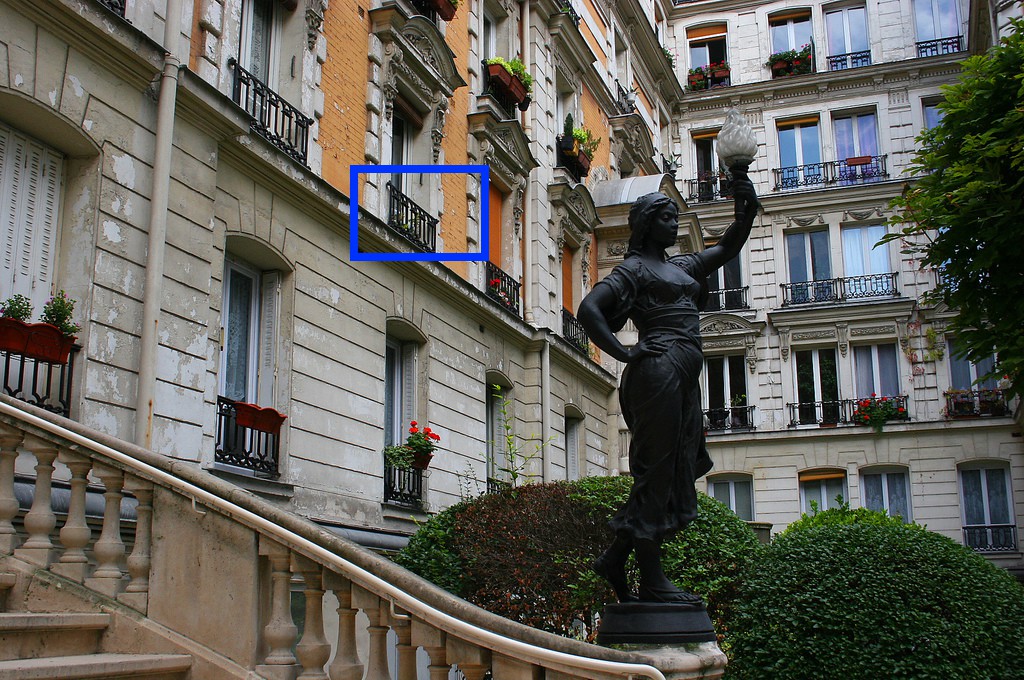}&
			\includegraphics[width = 0.1\textwidth]{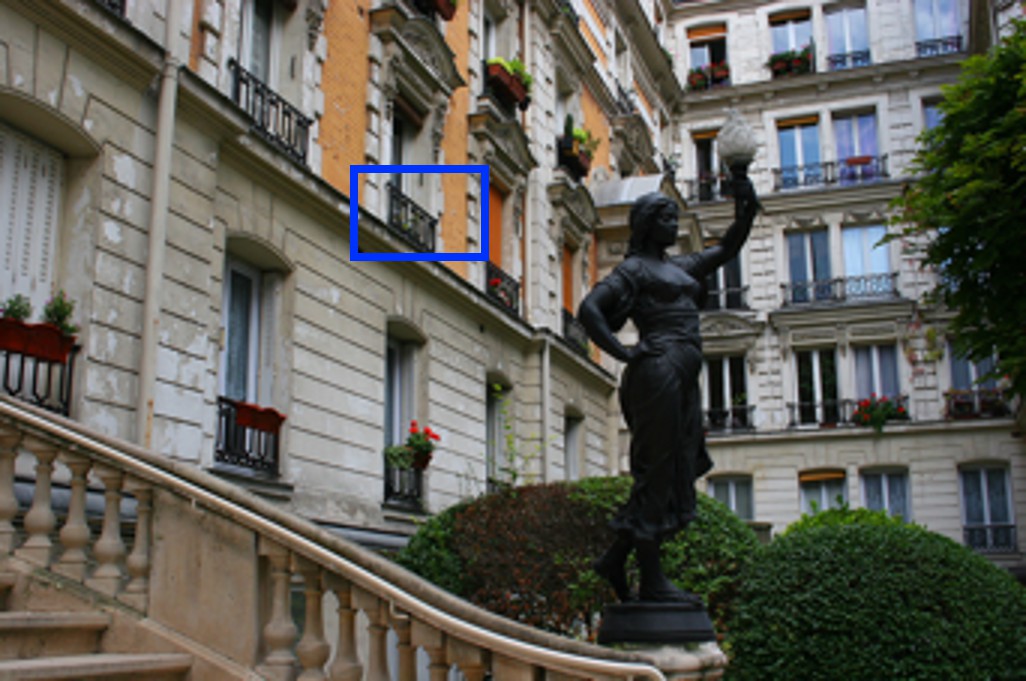} & 
			\includegraphics[width = 0.1\textwidth]{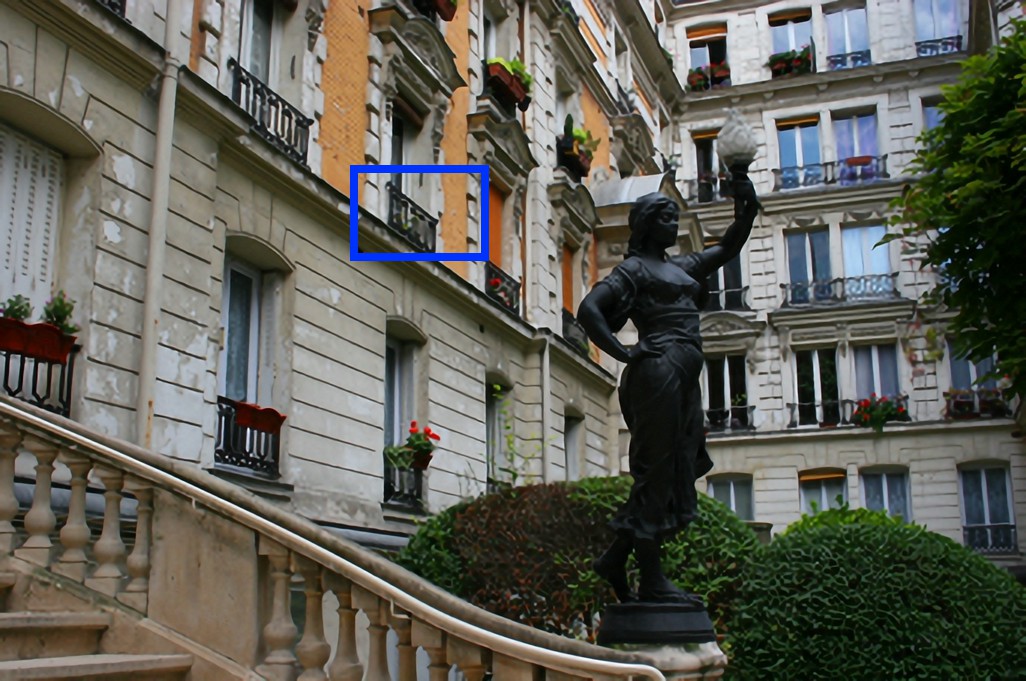} &
			\includegraphics[width = 0.1\textwidth]{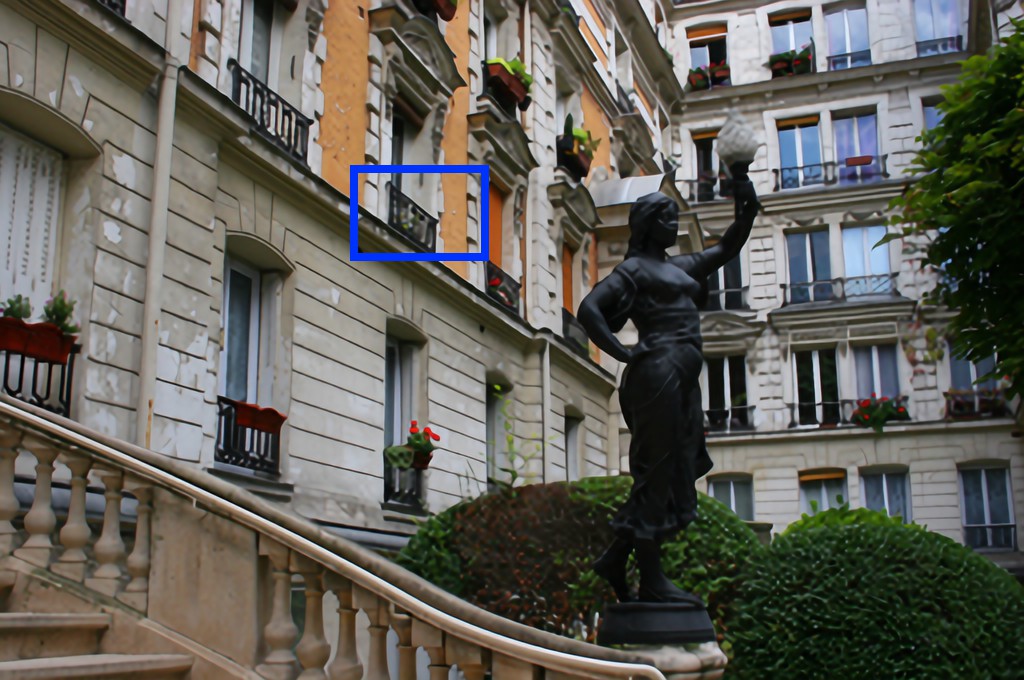} & 
			\includegraphics[width = 0.1\textwidth]{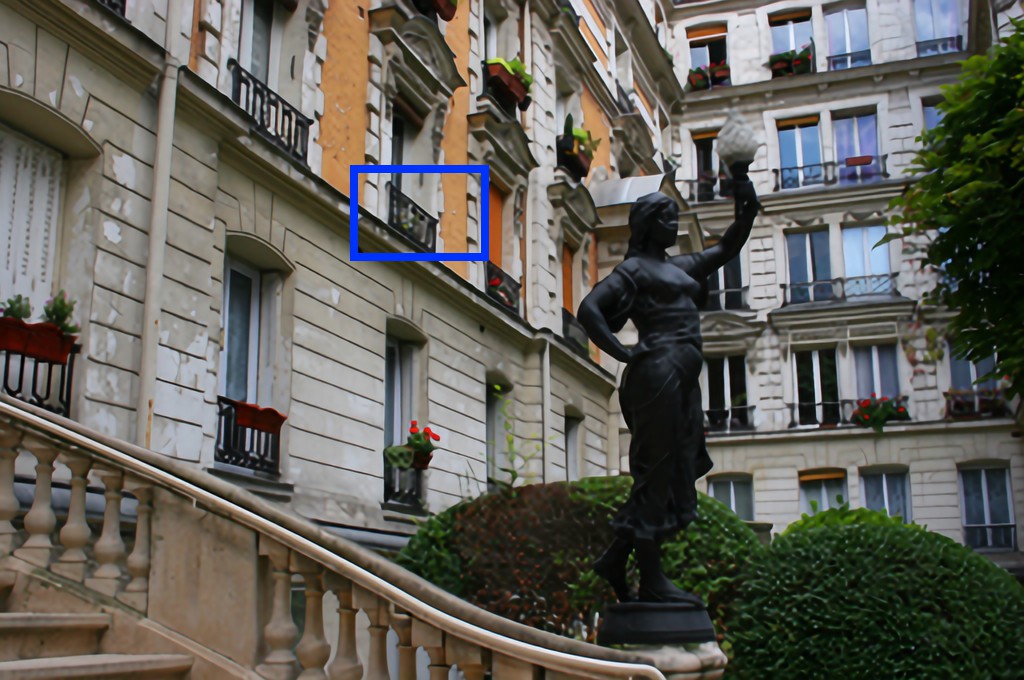} & 
			\includegraphics[width = 0.1\textwidth]{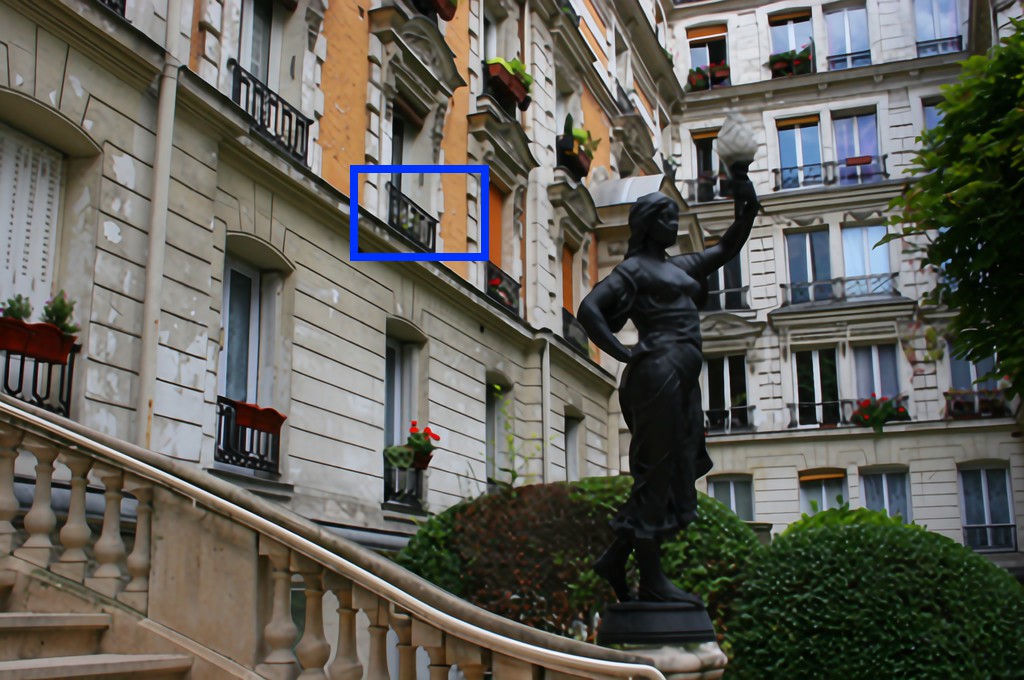} & 
			\includegraphics[width = 0.1\textwidth]{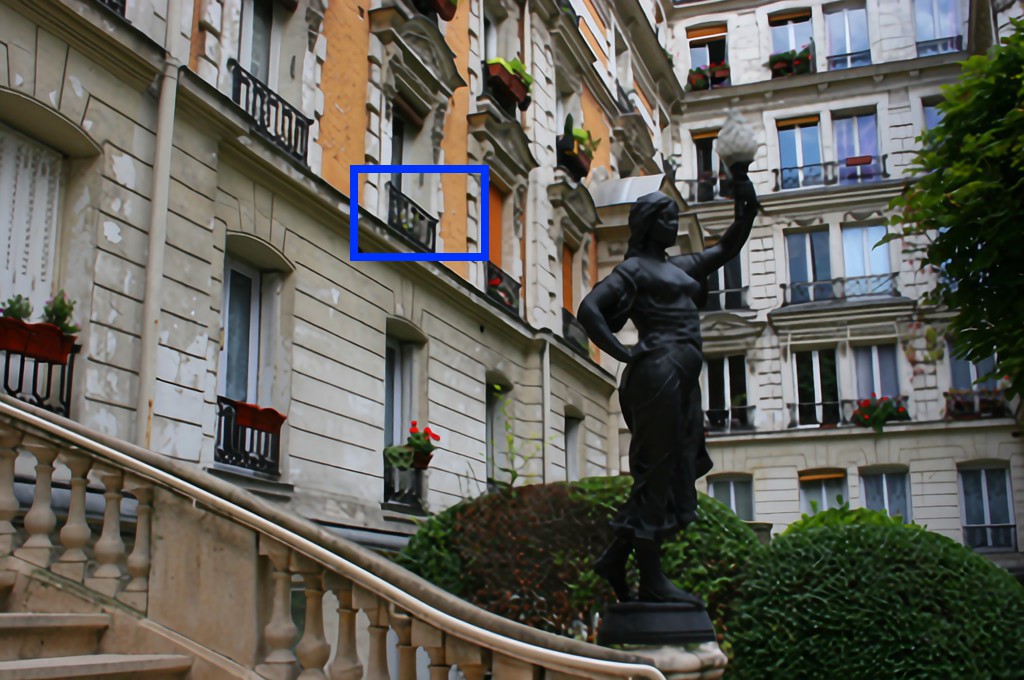}& 
			\includegraphics[width = 0.1\textwidth]{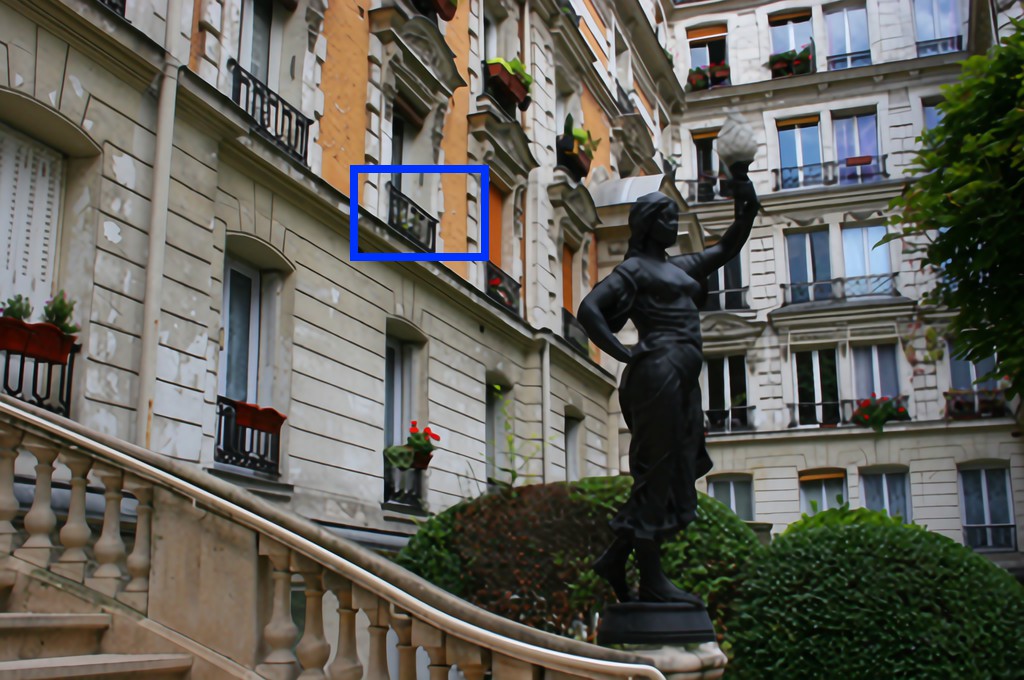} & 
			\includegraphics[width = 0.1\textwidth]{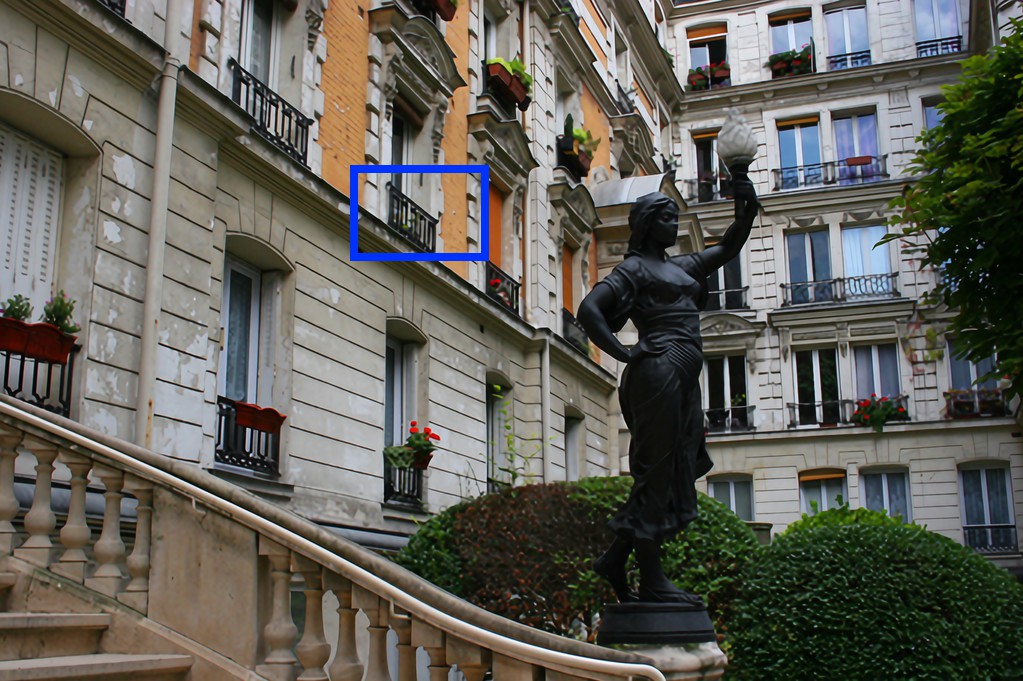} \\ 

			\includegraphics[width = 0.1\textwidth]{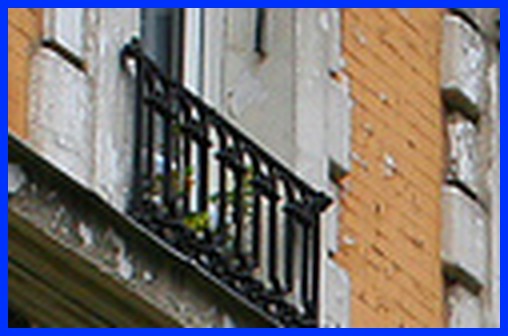}&
			\includegraphics[width = 0.1\textwidth]{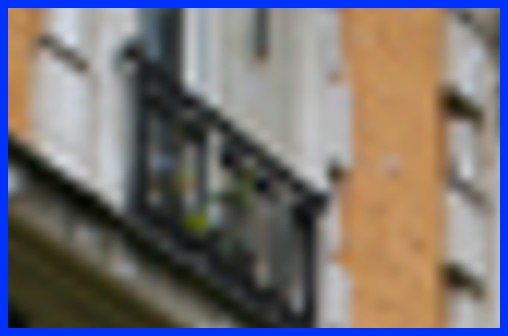} & 
			\includegraphics[width = 0.1\textwidth]{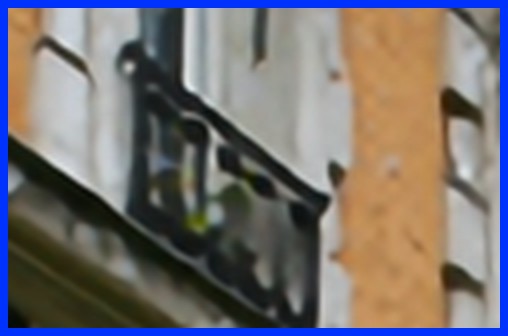} &
			\includegraphics[width = 0.1\textwidth]{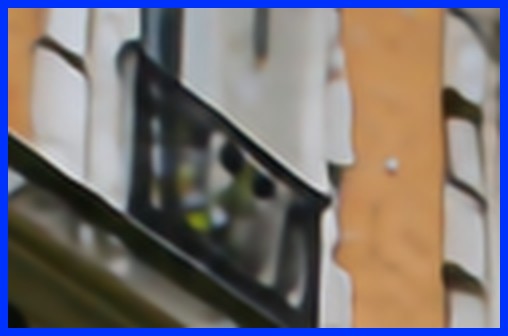} & 
			\includegraphics[width = 0.1\textwidth]{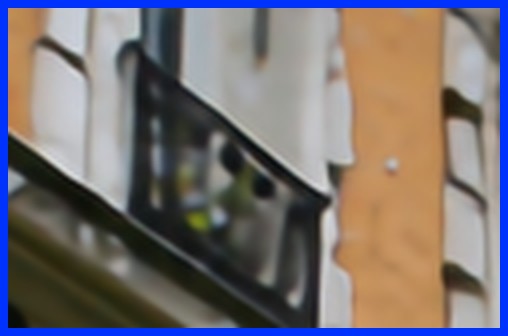} & 
			\includegraphics[width = 0.1\textwidth]{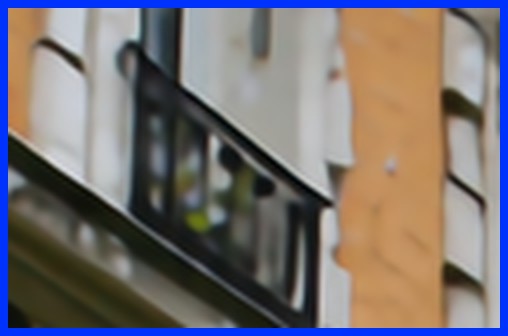} & 
			\includegraphics[width = 0.1\textwidth]{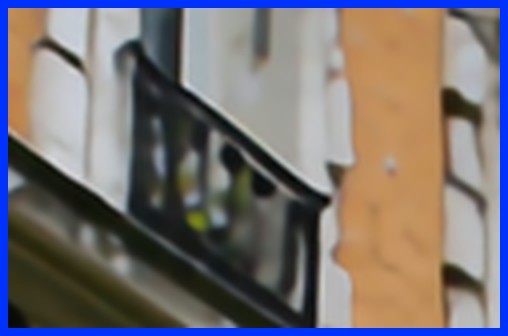}& 
			\includegraphics[width = 0.1\textwidth]{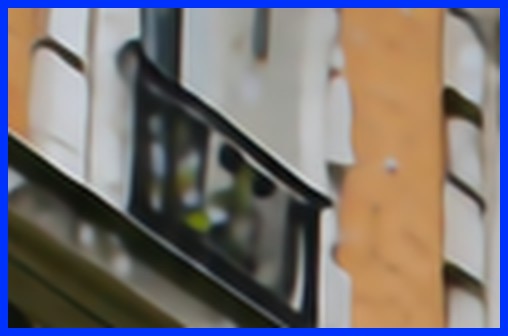} & 
			\includegraphics[width = 0.1\textwidth]{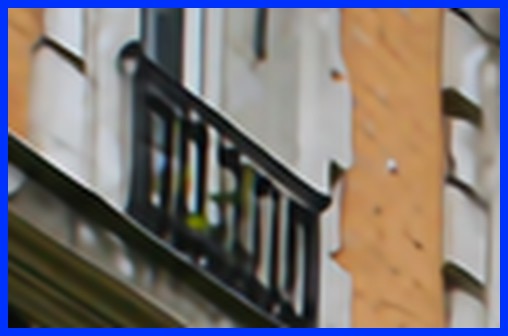} \\ 

			PSNR/SSIM & 22.97/0.636 &24.59/0.741 & 24.00/0.695 &24.00/0.698 &24.26/ \textbf{0.711} &24.13/0.706 &24.20/0.709 & \textbf{24.87}/ 0.710 \\

			\includegraphics[width = 0.1\textwidth]{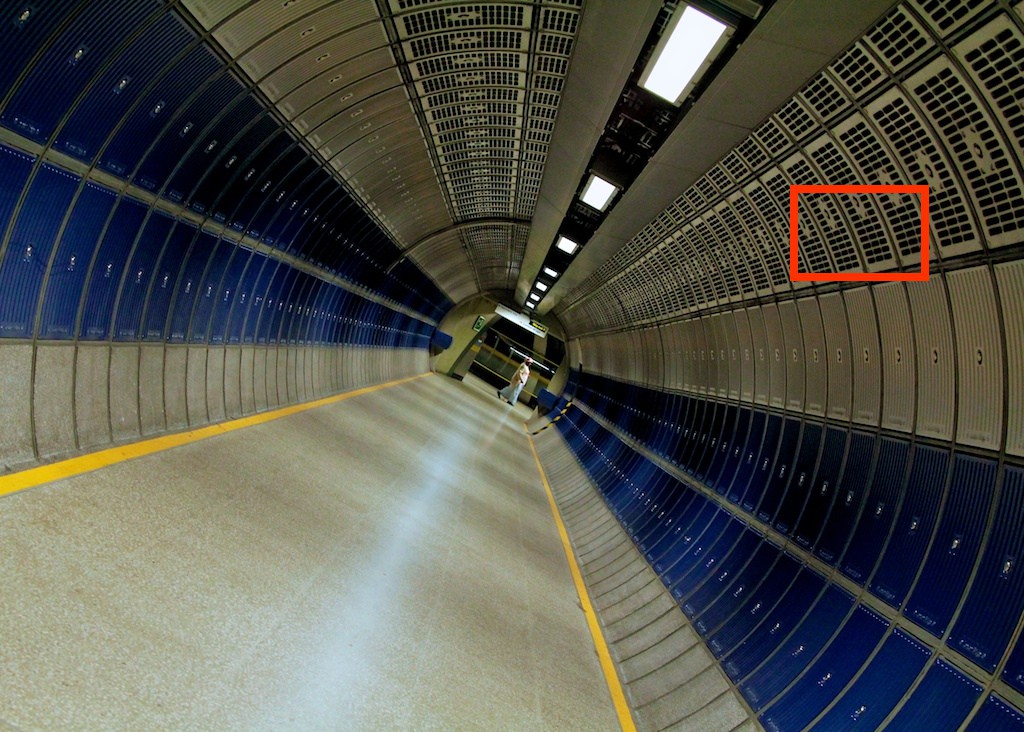}&
			\includegraphics[width = 0.1\textwidth]{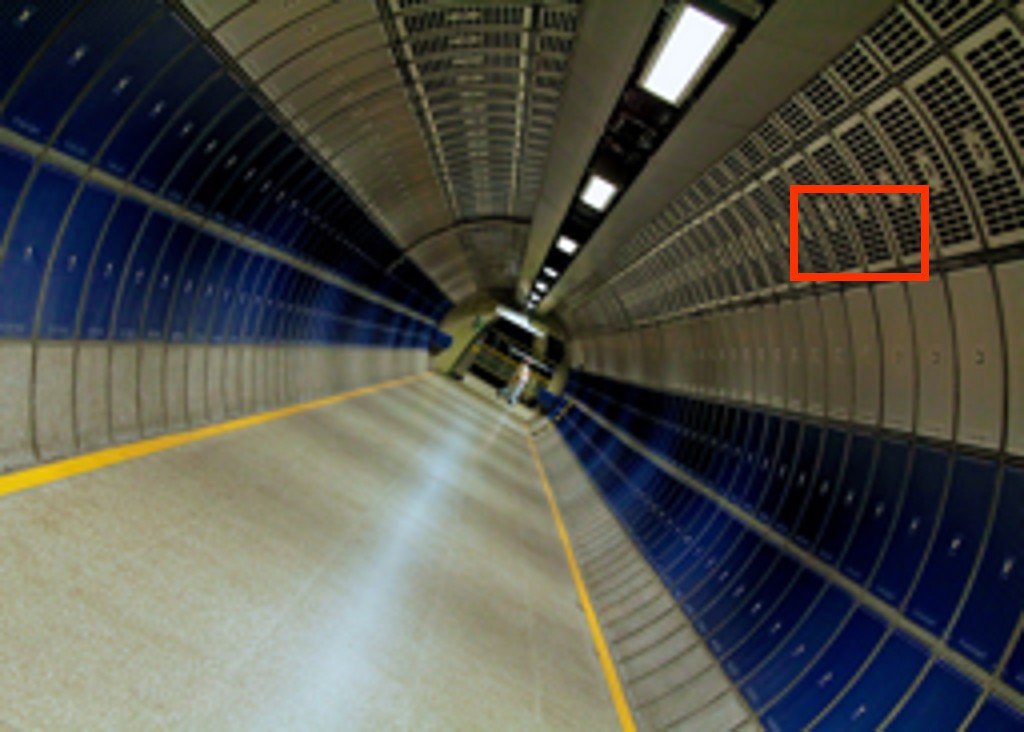} & 
			\includegraphics[width = 0.1\textwidth]{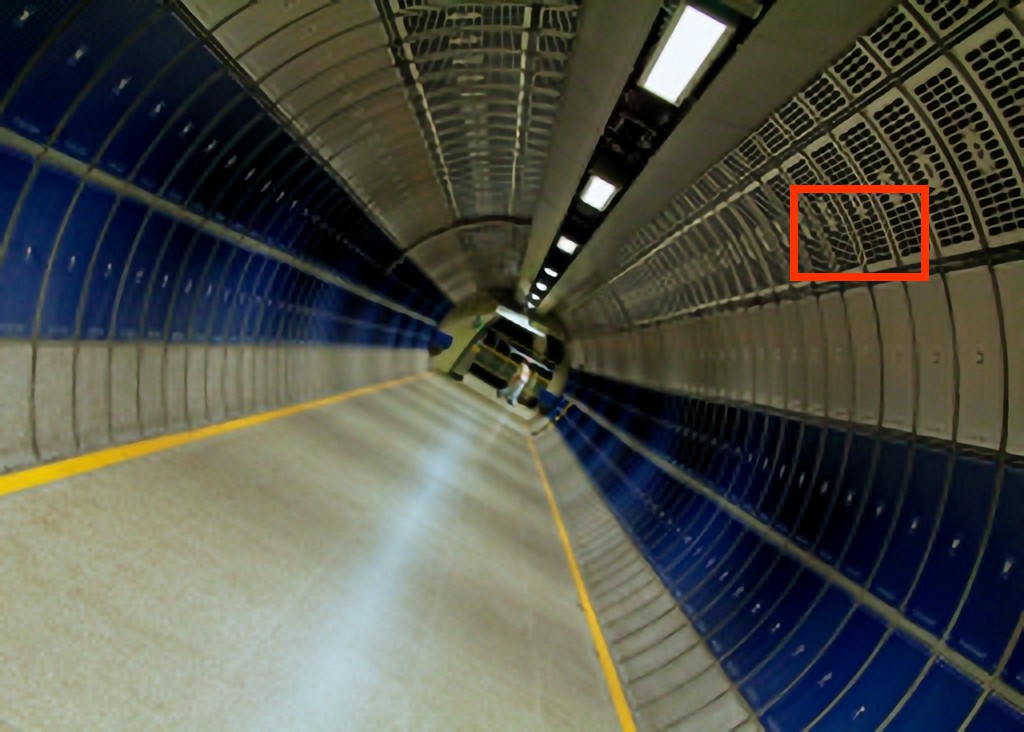} &
			\includegraphics[width = 0.1\textwidth]{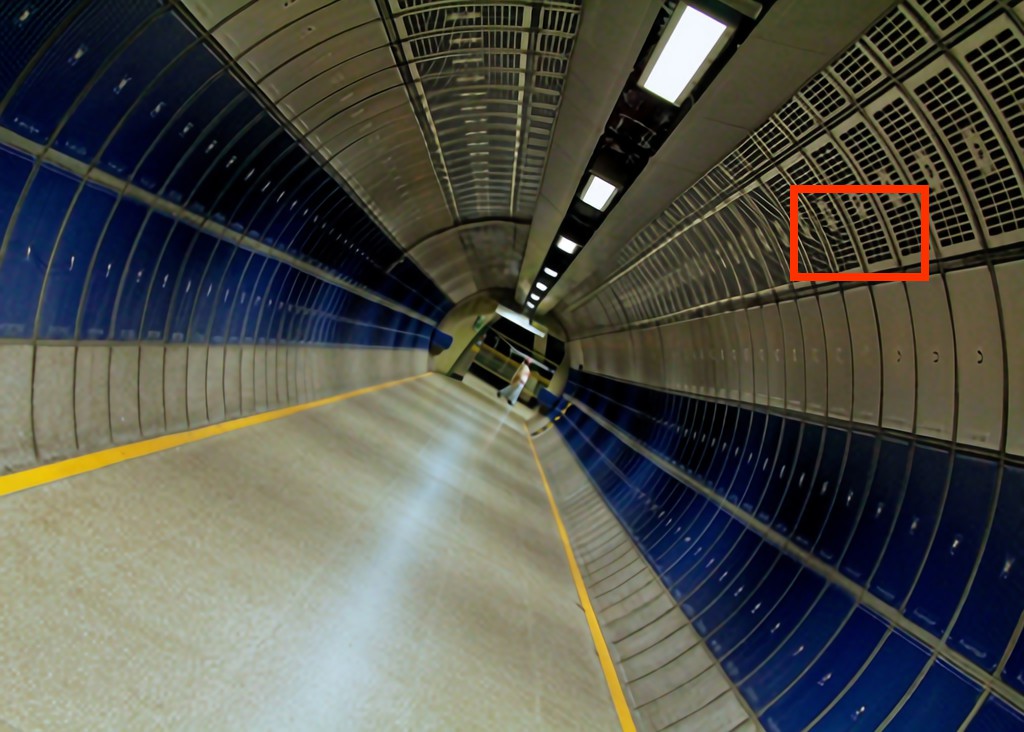} & 
			\includegraphics[width = 0.1\textwidth]{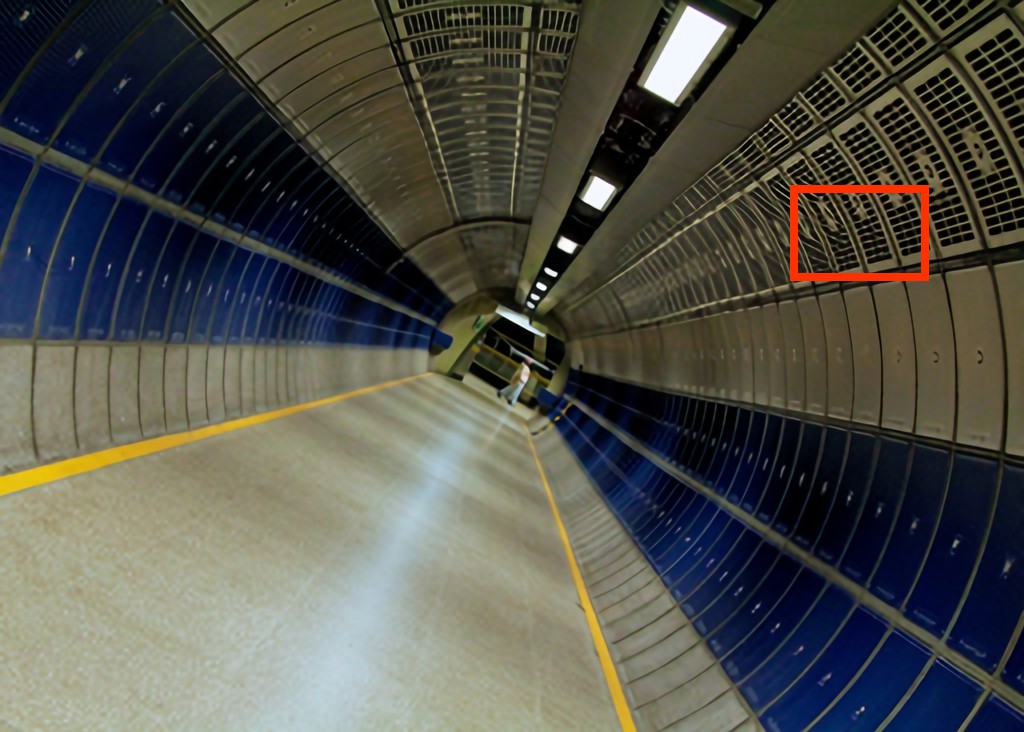} & 
			\includegraphics[width = 0.1\textwidth]{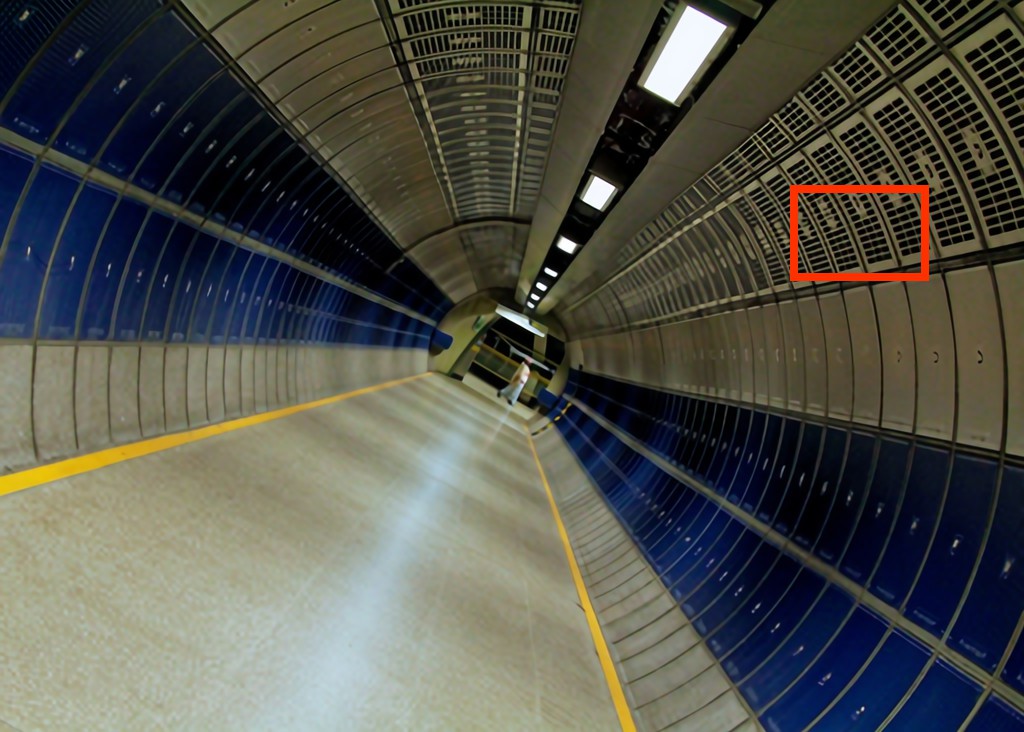} & 
			\includegraphics[width = 0.1\textwidth]{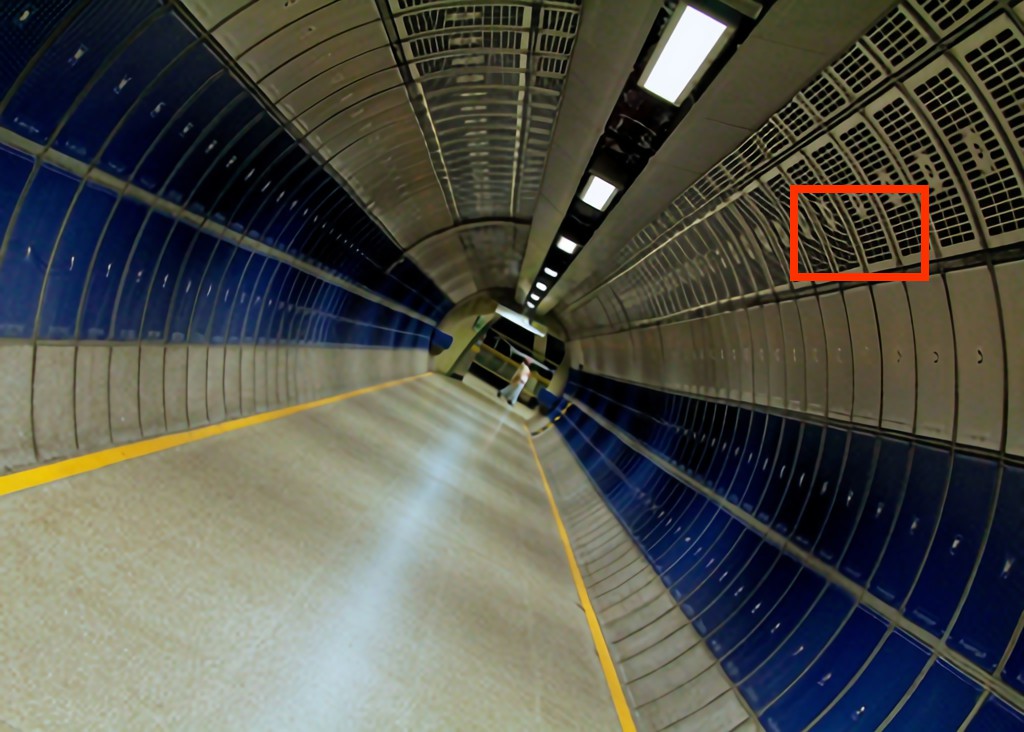}& 
			\includegraphics[width = 0.1\textwidth]{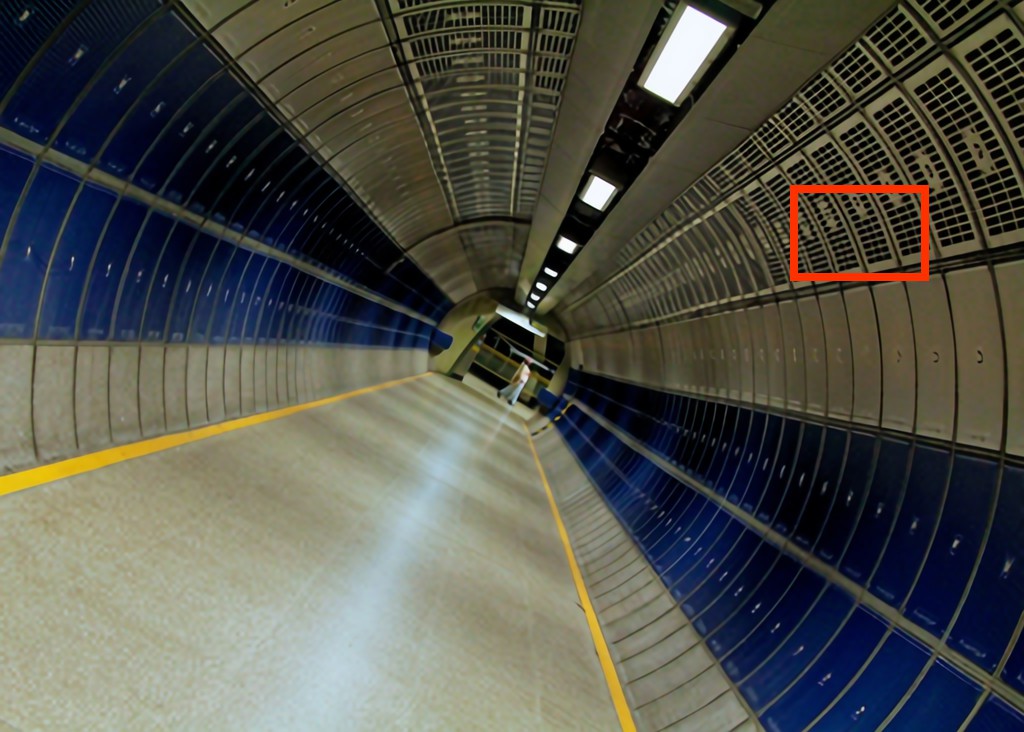} & 
			\includegraphics[width = 0.1\textwidth]{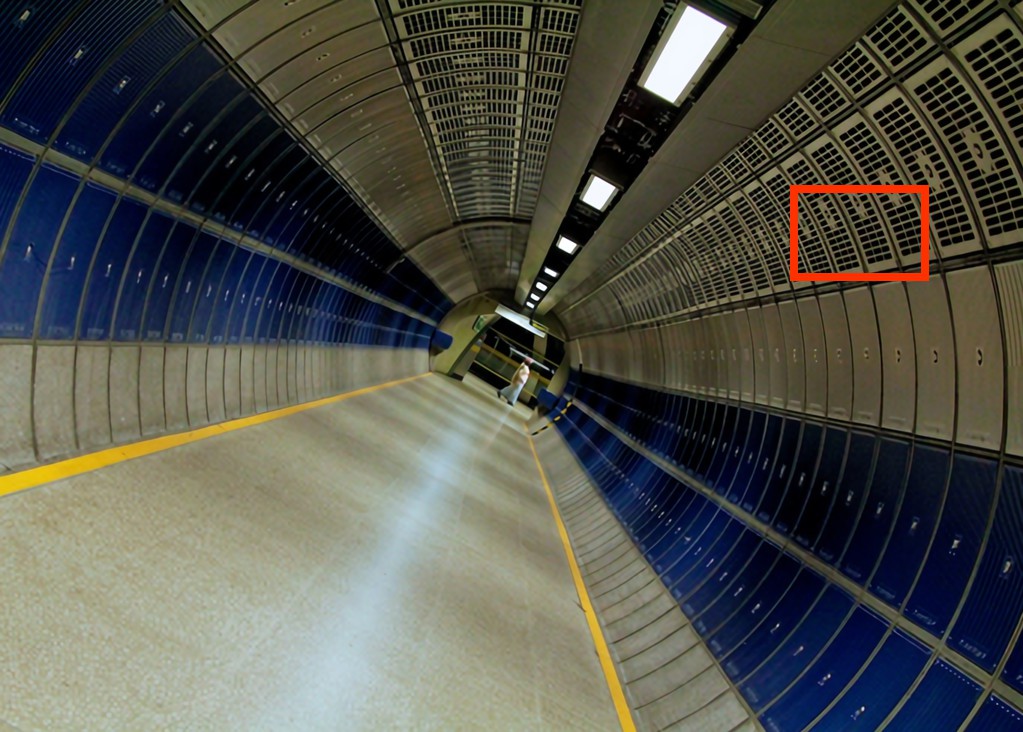}\\ 

			\includegraphics[width = 0.1\textwidth]{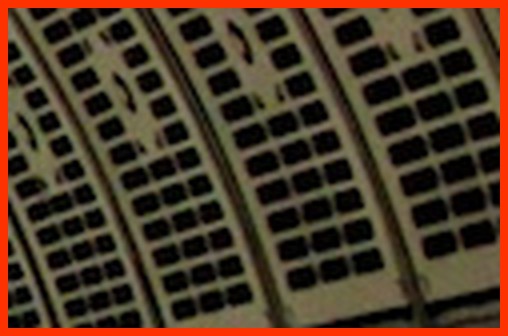}&
			\includegraphics[width = 0.1\textwidth]{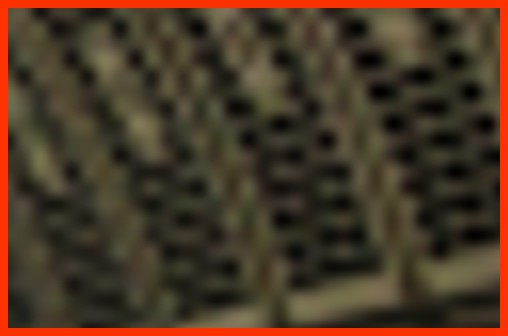} & 
			\includegraphics[width = 0.1\textwidth]{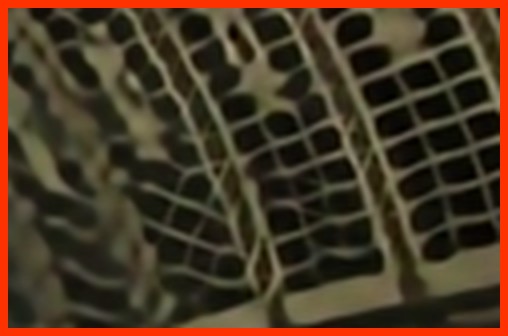} &
			\includegraphics[width = 0.1\textwidth]{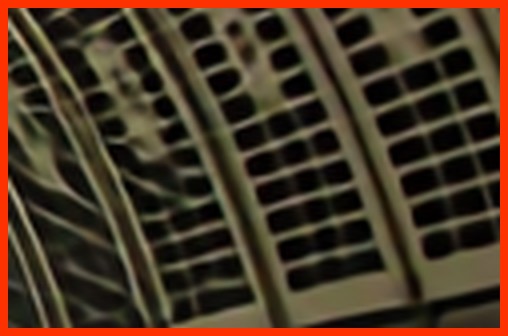} & 
			\includegraphics[width = 0.1\textwidth]{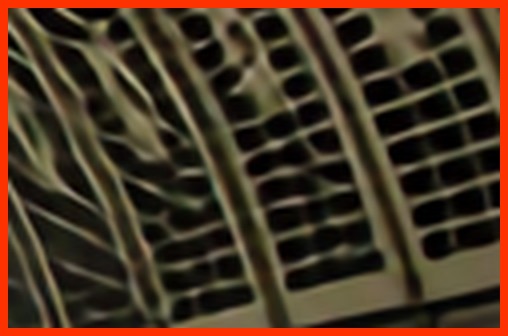} & 
			\includegraphics[width = 0.1\textwidth]{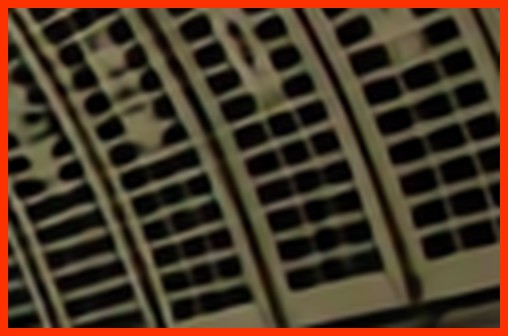} & 
			\includegraphics[width = 0.1\textwidth]{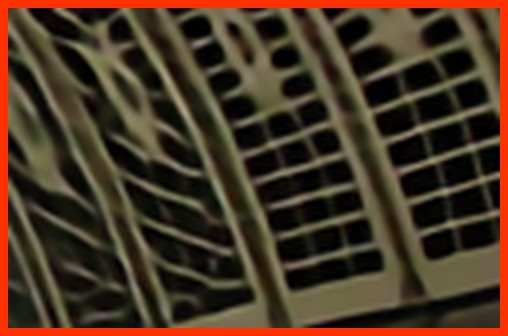}& 
			\includegraphics[width = 0.1\textwidth]{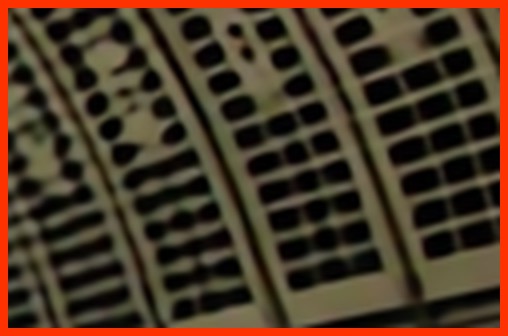} & 
			\includegraphics[width = 0.1\textwidth]{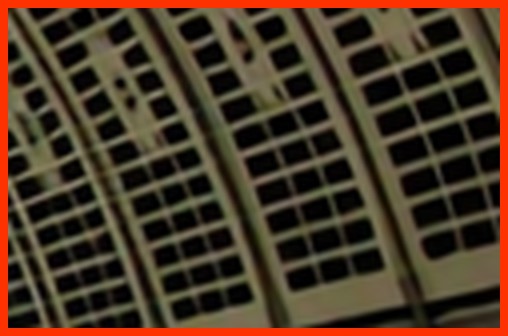} \\ 
			PSNR/SSIM & 25.71/0.680 &26.62/0.725 & 27.96/0.795 &27.53/0.782 &28.63/0.805 &27.74/0.789 &28.40/0.800 &\textbf{28.67}/\textbf{0.805} \\

			\includegraphics[width = 0.1\textwidth]{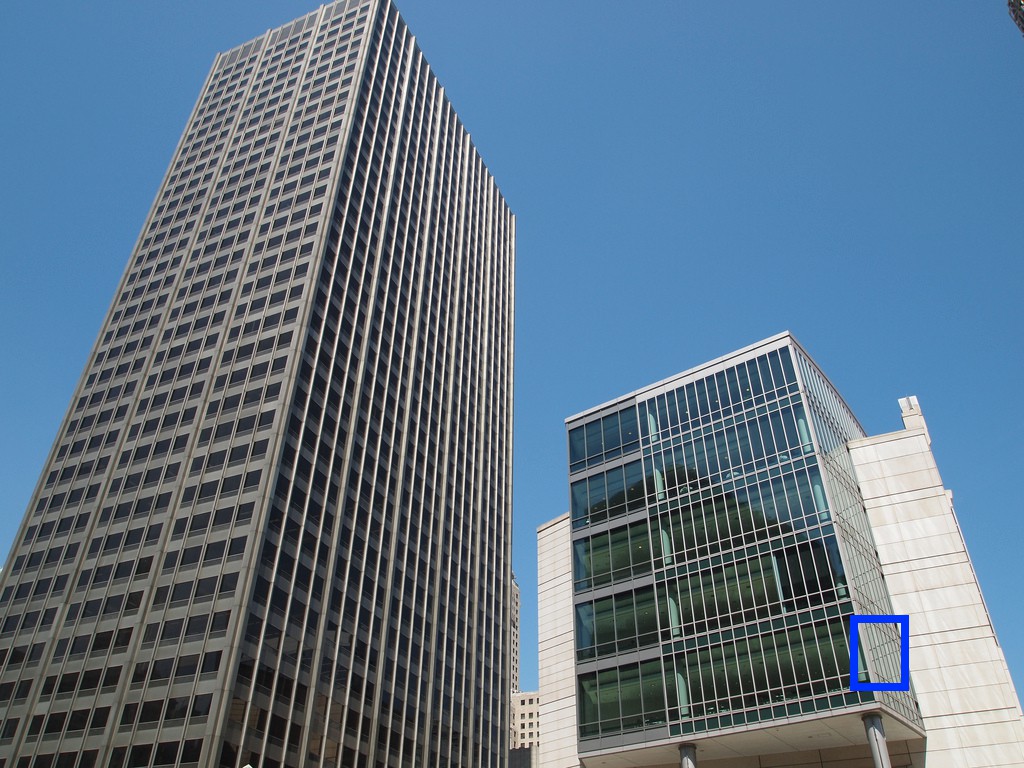}&
			\includegraphics[width = 0.1\textwidth]{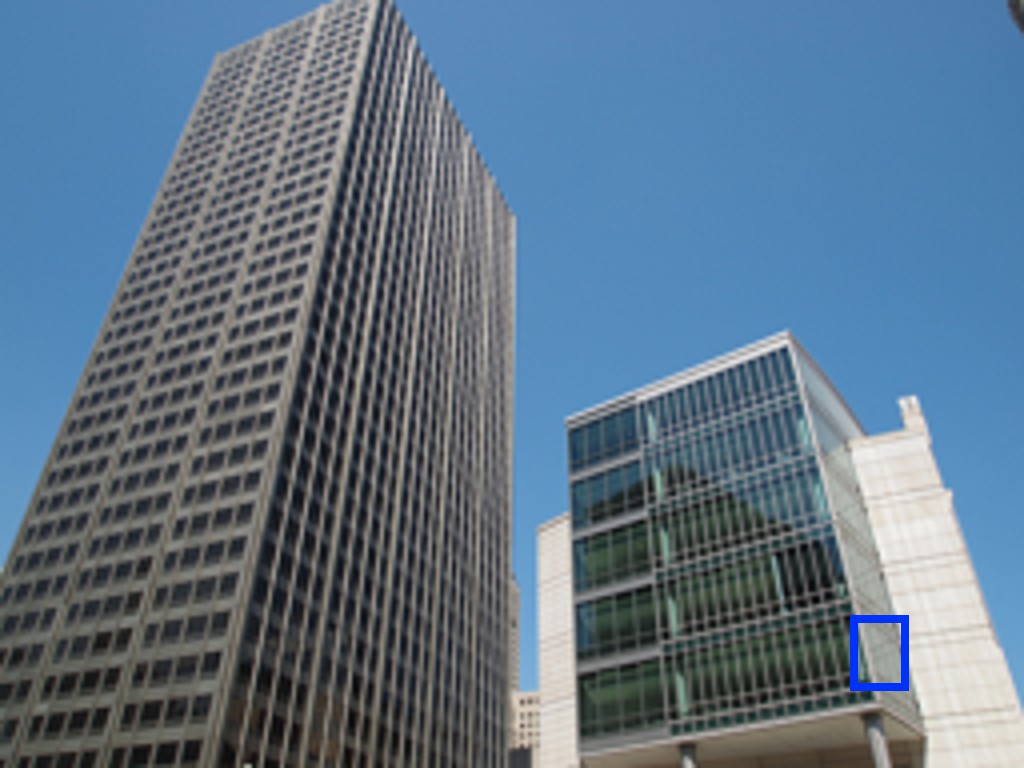} & 
			\includegraphics[width = 0.1\textwidth]{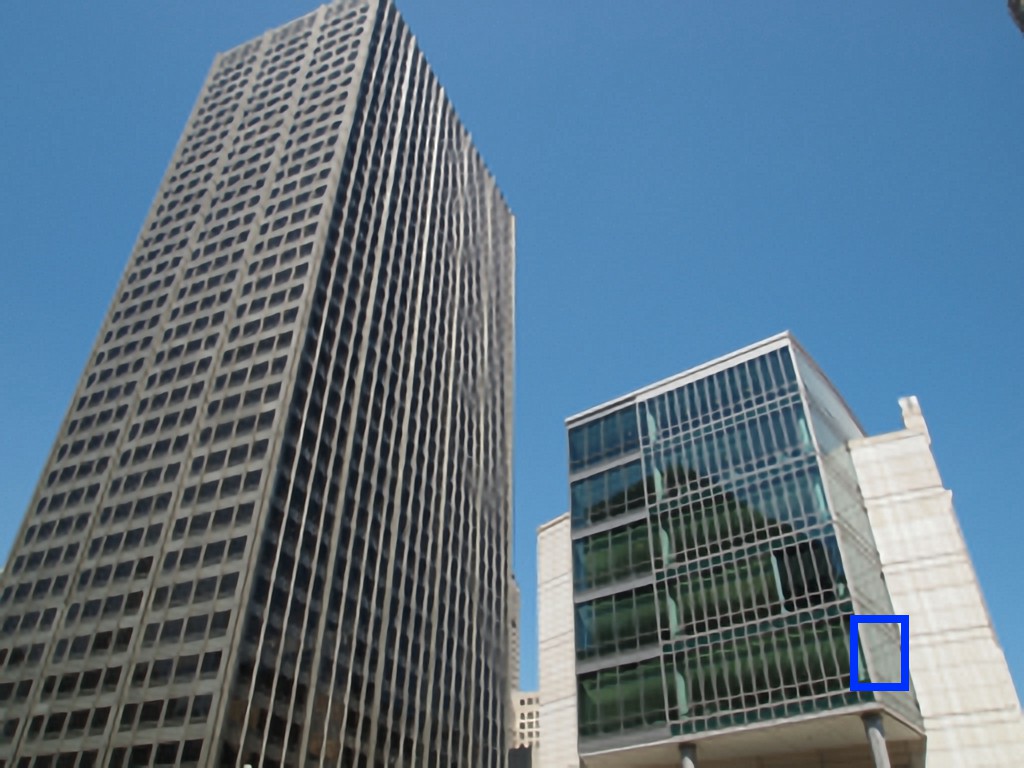} &
			\includegraphics[width = 0.1\textwidth]{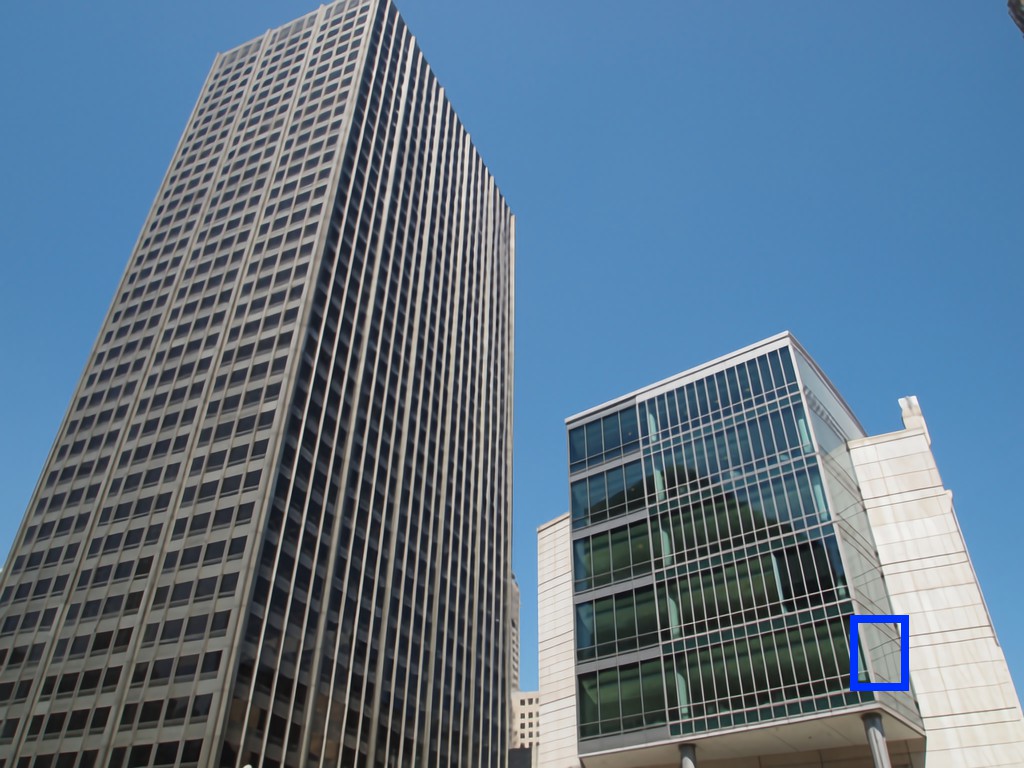} & 
			\includegraphics[width = 0.1\textwidth]{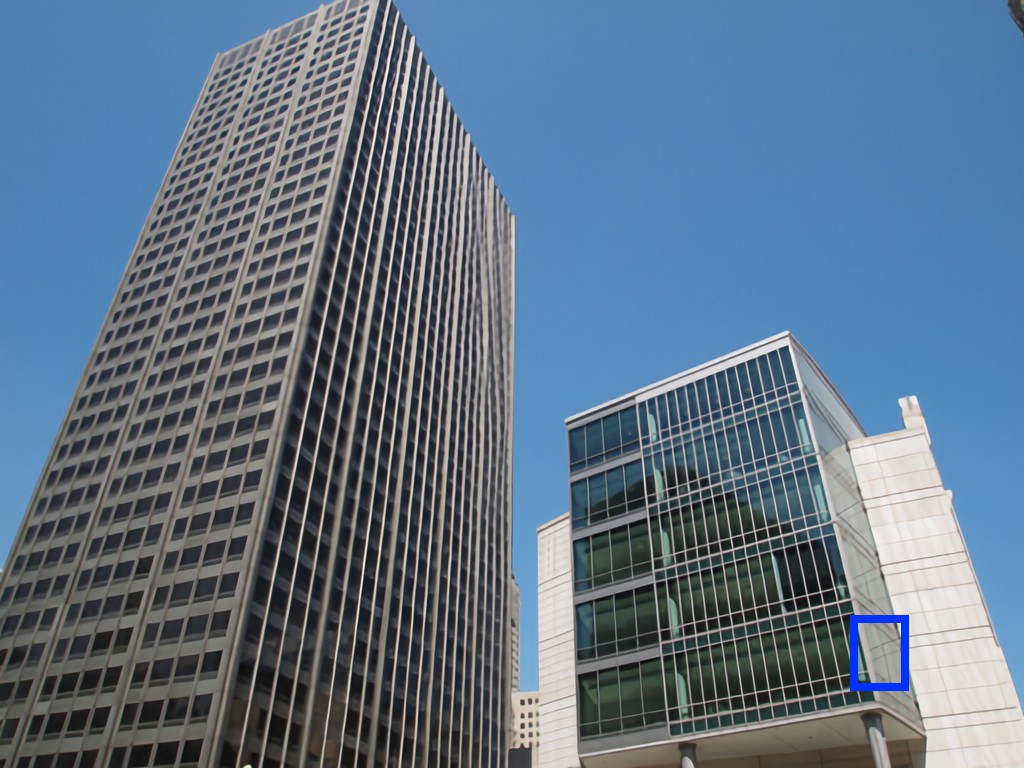} & 
			\includegraphics[width = 0.1\textwidth]{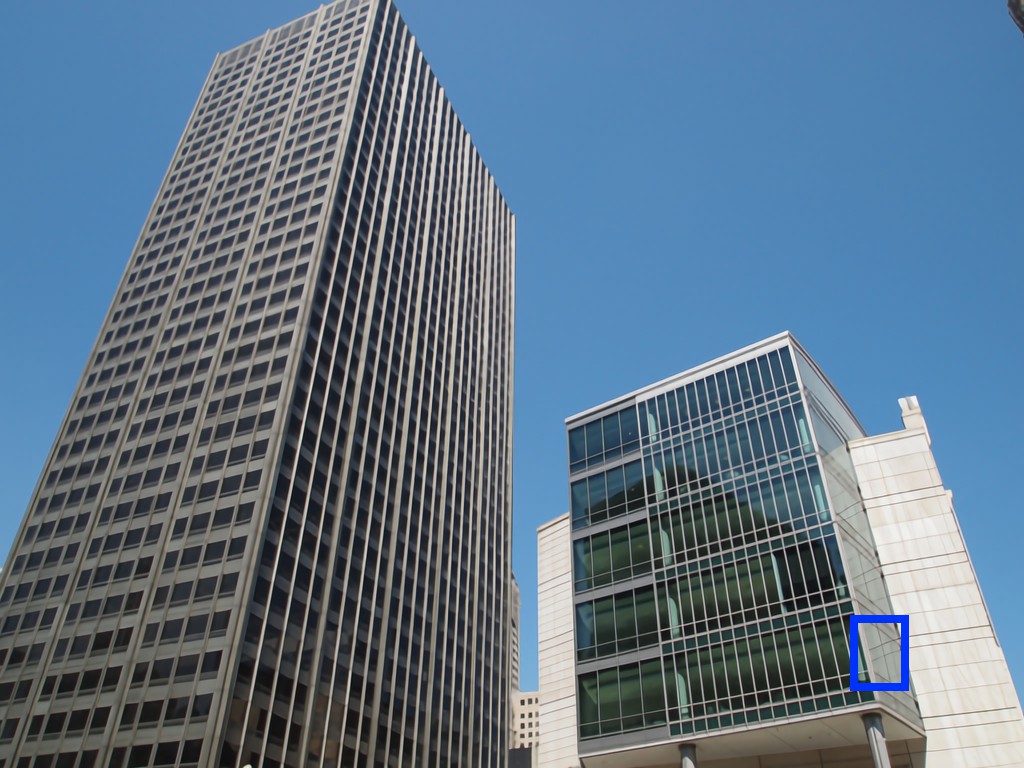} & 
			\includegraphics[width = 0.1\textwidth]{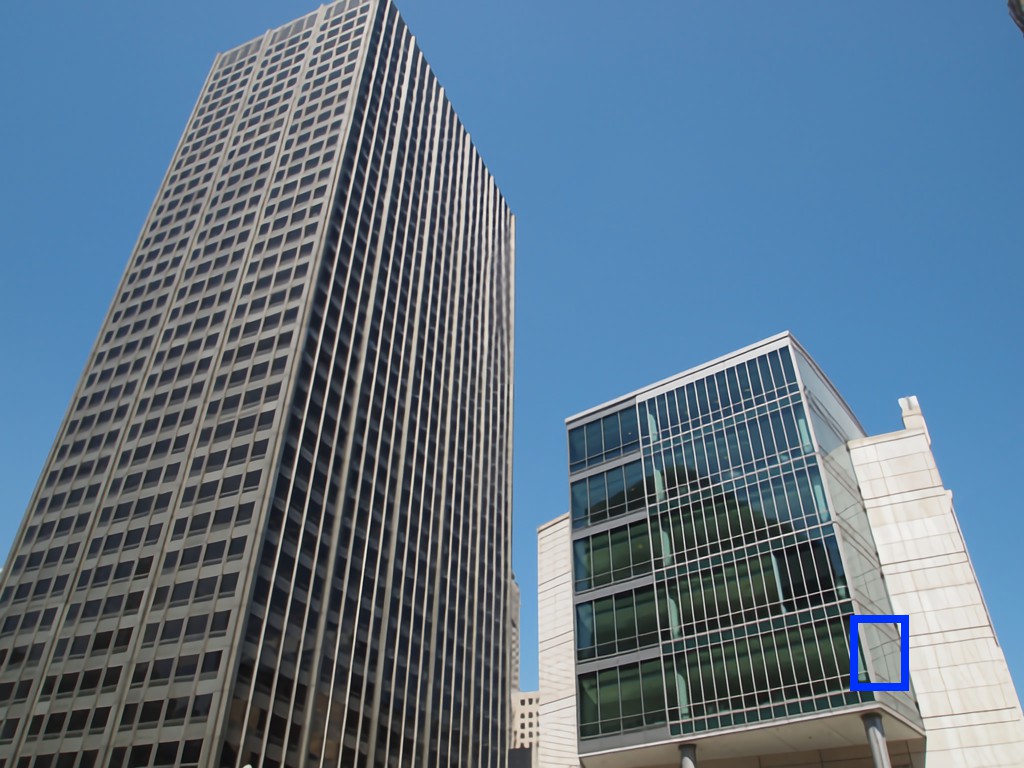}& 
			\includegraphics[width = 0.1\textwidth]{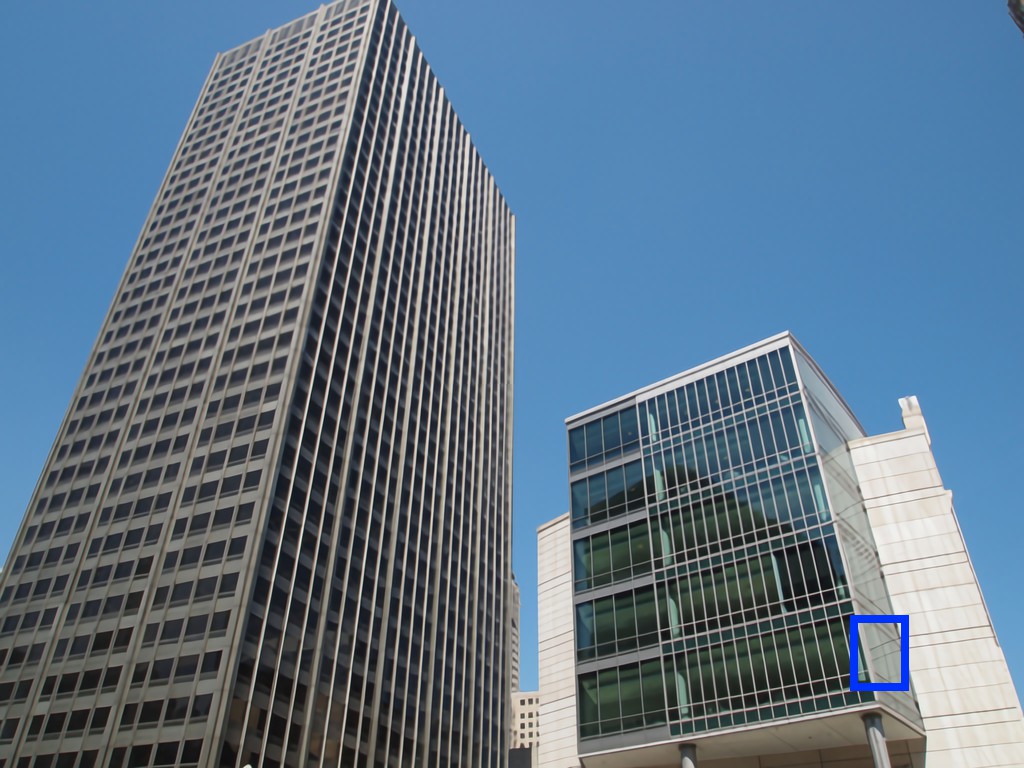} & 
			\includegraphics[width = 0.1\textwidth]{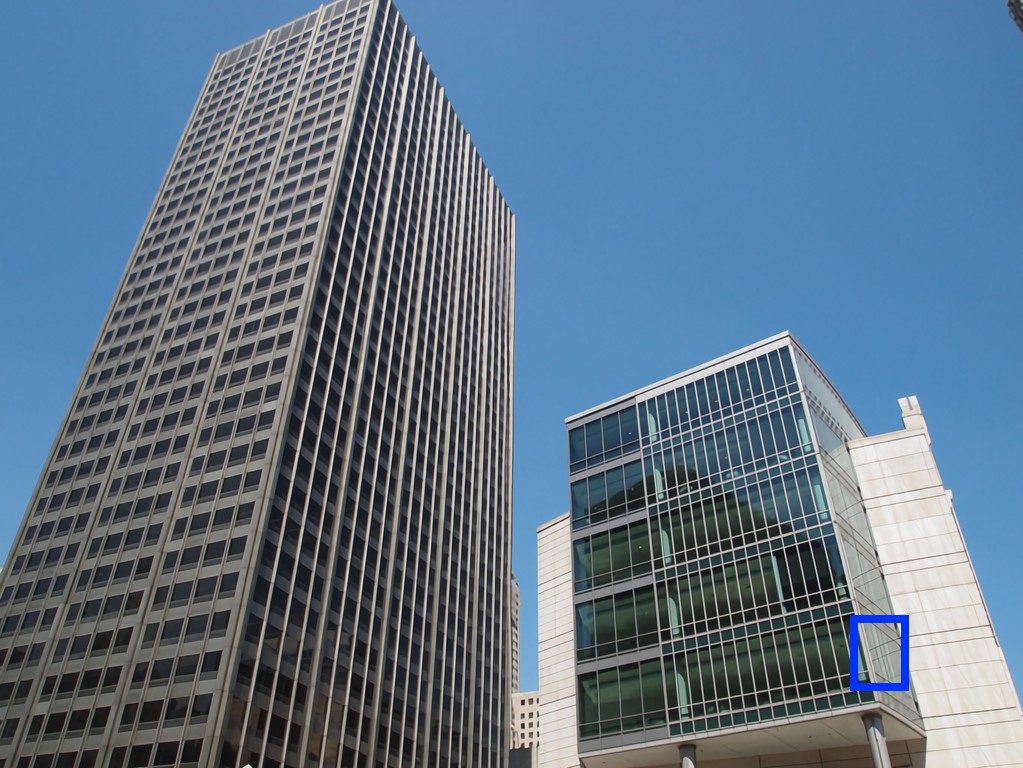} \\ 

			\includegraphics[width = 0.1\textwidth]{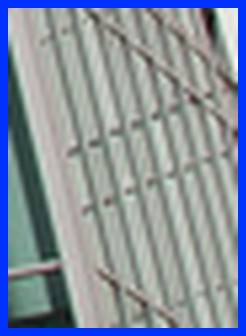}&
			\includegraphics[width = 0.1\textwidth]{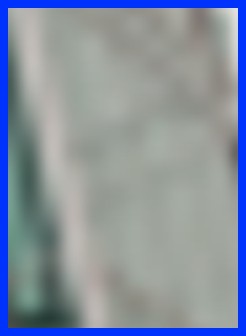} & 
			\includegraphics[width = 0.1\textwidth]{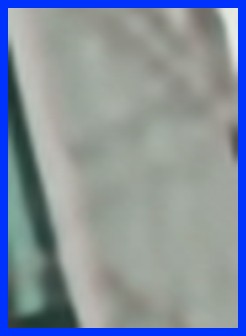} &
			\includegraphics[width = 0.1\textwidth]{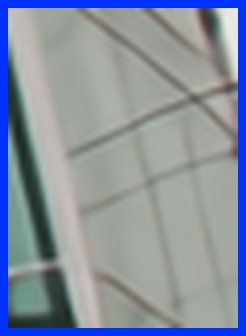} & 
			\includegraphics[width = 0.1\textwidth]{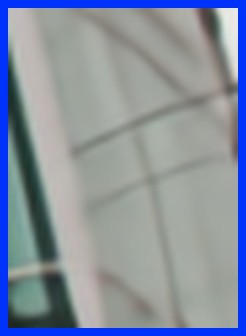} & 
			\includegraphics[width = 0.1\textwidth]{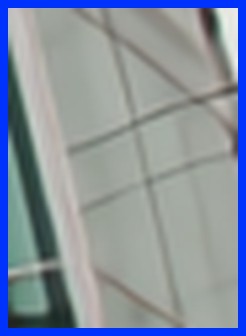} & 
			\includegraphics[width = 0.1\textwidth]{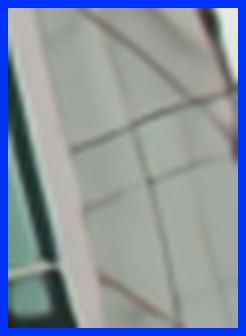}& 
			\includegraphics[width = 0.1\textwidth]{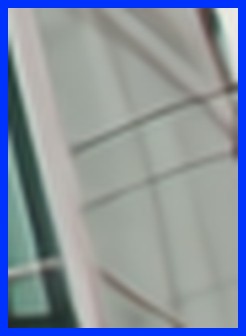} & 
			\includegraphics[width = 0.1\textwidth]{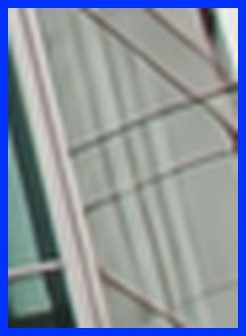} \\ 
			PSNR/SSIM & 21.32/0.686 &23.07/0.783 &26.33/0.895 &25.62/0.880 & 26.46/0.897  &26.57/0.897 &26.87/0.900 & \textbf{26.98}/\textbf{0.900} \\

		\end{tabular}
	\end{center}
	
	\caption{Visual comparison for 4$\times$ SR with BI degradation model on the Urban100 datasets. The best results are highlighted. Our method obtains better visual quality and recovers more image details compared with other state-of-the-art SR methods
	}

	\label{fig-BI}
\end{figure}

\begin{table}[t]\scriptsize
	\tabcolsep 6pt
	\caption{Effectiveness of the proposed LAM and CSAM for image super-resolution}
	
	\begin{center}
		\begin{tabular}{ccccc}
			\toprule
			& baseline & w/o CSAM & w/o LAM & Ours \\
			\midrule
			PSNR/SSIM  & 31.22/0.9173  & 31.38/0.9175  & 31.28/0.9174 & \textbf{31.42}/\textbf{0.9177} \\
			\bottomrule
		\end{tabular}
		\label{tab-psnr-ssim}
	\end{center}
	
\end{table}
\begin{table}[!t]\scriptsize
	
	\tabcolsep 10pt
	\caption{Ablation study about using different numbers of RGs}
	
	\begin{center}\scriptsize{
			\begin{tabular}{cccccc}
				\toprule
				&Set5 & Set14 & B100 & Urban100 & Manga100 \\
				\midrule
				RCAN       &32.63  & 28.87   & 27.77  & 26.82  &31.22   \\
				HAN 3RGs      &32.63  & 28.89   & 27.79  & 26.82  &31.40   \\
				HAN 6RGs      &\textbf{32.64}  & \textbf{28.90}    & 27.79  & 26.84  &\textbf{31.42}   \\
				HAN 10RGs      &\textbf{32.64}  & \textbf{28.90}   & \textbf{27.80}  & \textbf{26.85}  &\textbf{31.42}   \\
				\bottomrule
		\end{tabular}}
		\label{tab4}
	\end{center} 
\end{table}

\begin{table}[!ht]
	\scriptsize 
	\centering
	\caption{Quantitative results with BI degradation model. The best and second best results are highlighted in \textbf{bold} and \underline{underlined}}
	\begin{tabular}{|p{7em}|p{2.5em}|p{2.5em}|p{3em}|p{2.5em}|p{3em}|p{2.5em}|p{3em}|p{2.5em}|p{3em}|p{2.5em}|p{3em}|} 
		\hline
		\multicolumn{1}{|c|}{\multirow{2}{*}{{ Methods} }} & \multicolumn{1}{c|}{\multirow{2}{*}{Scale}} & \multicolumn{2}{c}{Set5} & \multicolumn{2}{c}{Set14} & \multicolumn{2}{c}{B100} & \multicolumn{2}{c}{Urban100} & \multicolumn{2}{c|}{ Manga109} \\
		\cline{3-12}   \multicolumn{1}{|c|}{} & \multicolumn{1}{c|}{} &  PSNR  & SSIM  & PSNR   & SSIM  & PSNR  & SSIM  & PSNR  & SSIM  & PSNR  & SSIM \\
		\hline
		
		Bicubic \newline{}SRCNN~\cite{dong2014learning} \newline{} FSRCNN~\cite{dong2016accelerating} \newline{} VDSR~\cite{kim2016accurate} \newline{} LapSRN~\cite{lai2017deep} \newline{} MemNet~\cite{tai2017memnet} \newline{} EDSR~\cite{lim2017enhanced} \newline{} SRMDNF~\cite{zhang2018learning}  \newline{} D-DBPN~\cite{haris2018deep} \newline{} RDN~\cite{zhang2018residual} \newline{} RCAN~\cite{zhang2018image} \newline{} SRFBN~\cite{li2019feedback} \newline{} SAN~\cite{dai2019second} \newline{} HAN(ours) \newline{} HAN+(ours) &
		$\times2$ \newline{}$\times2$ \newline{}$\times2$ \newline{}$\times2$ \newline{}$\times2$ \newline{}$\times2$ \newline{}$\times2$ \newline{}$\times2$ \newline{}$\times2$ \newline{}$\times2$ \newline{}$\times2$ $\times2$ \newline{} $\times2$ \newline{} $\times2$ \newline{} $\times2$ & 33.66 \newline{}36.66 \newline{}37.05 \newline{}37.53 \newline{}37.52 \newline{}37.78 \newline{}38.11 \newline{}37.79 \newline{}38.09 \newline{}38.24 \newline{}38.27 \newline{}38.11 \newline{}\underline{38.31} \newline{}38.27 \newline{}\bfseries{38.33} & 0.9299 \newline{}0.9542 \newline{}0.9560 \newline{}0.9590 \newline{}0.9591 \newline{}0.9597 \newline{}0.9602 \newline{}0.9601 \newline{}0.9600 \newline{}0.9614 \newline{}0.9614 \newline{}0.9609 \newline{}\textbf{0.9620} \newline{}0.9614 \newline{}\underline{0.9617}  & 30.24 \newline{}32.45 \newline{}32.66 \newline{}33.05 \newline{}33.08 \newline{}33.28 \newline{}33.92 \newline{}33.32 \newline{}33.85 \newline{}34.01 \newline{}34.12 \newline{}33.82 \newline{}34.07 \newline{}\underline{34.16} \newline{}\bfseries{34.24}  & 0.8688 \newline{}0.9067 \newline{}0.9090 \newline{}0.9130 \newline{}0.9130 \newline{}0.9142 \newline{}0.9195 \newline{}0.9159 \newline{}0.9190 \newline{}0.9212 \newline{}0.9216 \newline{}0.9196 \newline{}0.9213 \newline{}\underline{0.9217} \newline{}\bfseries{0.9224} & 29.56 \newline{}31.36 \newline{}31.53 \newline{}31.90 \newline{}31.08 \newline{}32.08 \newline{}32.32 \newline{}32.05 \newline{}32.27 \newline{}32.34 \newline{}32.41 \newline{}32.29 \newline{}\underline{32.42} \newline{}32.41 \newline{}\bfseries{32.45} & 0.8431 \newline{}0.8879 \newline{}0.8920 \newline{}0.8960 \newline{}0.8950 \newline{}0.8978 \newline{}0.9013 \newline{}0.8985 \newline{}0.9000 \newline{}0.9017 \newline{}0.9027 \newline{}0.9010 \newline{}\underline{0.9028} \newline{}0.9027 \newline{}\bfseries{0.9030} & 26.88 \newline{}29.50 \newline{}29.88 \newline{}30.77 \newline{}30.41 \newline{}31.31 \newline{}32.93 \newline{}31.33 \newline{}32.55 \newline{}32.89 \newline{}33.34 \newline{}32.62 \newline{}33.10 \newline{}\underline{33.35} \newline{}\bfseries{33.53} & 0.8403 \newline{}0.8946 \newline{}0.9020 \newline{}0.9140 \newline{}0.9101 \newline{}0.9195 \newline{}0.9351 \newline{}0.9204 \newline{}0.9324 \newline{}0.9353 \newline{}0.9384 \newline{}0.9328\newline{}0.9370 \newline{}\underline{0.9385} \newline{}\bfseries{0.9398}  & 30.80 \newline{}35.60 \newline{}36.67 \newline{}37.22 \newline{}37.27 \newline{}37.72 \newline{}39.10 \newline{}38.07 \newline{}38.89 \newline{}39.18 \newline{}39.44 \newline{}39.08 \newline{}39.32 \newline{}\underline{39.46} \newline{}\bfseries{39.62} & 0.9339 \newline{}0.9663 \newline{}0.9710 \newline{}0.9750 \newline{}0.9740 \newline{}0.9740 \newline{}0.9773 \newline{}0.9761 \newline{}0.9775 \newline{}0.9780 \newline{}\underline{0.9786} \newline{}0.9779 \newline{}0.9792 \newline{}0.9785 \newline{}\bfseries{0.9787} \\
		\hline
		\hline
		Bicubic \newline{}SRCNN~\cite{dong2014learning} \newline{} FSRCNN~\cite{dong2016accelerating} \newline{} VDSR~\cite{kim2016accurate} \newline{} LapSRN~\cite{lai2017deep} \newline{} MemNet~\cite{tai2017memnet} \newline{} EDSR~\cite{lim2017enhanced} \newline{} SRMDNF~\cite{zhang2018learning}  \newline{} RDN~\cite{zhang2018residual} \newline{} RCAN~\cite{zhang2018image} \newline{} SRFBN~\cite{li2019feedback} \newline{} SAN~\cite{dai2019second} \newline{} HAN(ours) \newline{} HAN+(ours)&
		$\times3$ \newline{}$\times3$ \newline{}$\times3$ \newline{}$\times3$ \newline{}$\times3$ \newline{}$\times3$ \newline{}$\times3$ \newline{}$\times3$ \newline{}$\times3$ \newline{}$\times3$ \newline{}$\times3$ \newline{}$\times3$ \newline{}$\times3$ \newline{}$\times3$ & 30.39 \newline{}32.75 \newline{}33.18 \newline{}33.67 \newline{}33.82 \newline{}34.09 \newline{}34.65 \newline{}34.12 \newline{}34.71 \newline{}34.74 \newline{}34.70 \newline{}34.75 \newline{}\underline{34.75} \newline{}\bfseries{34.85} & 0.8682 \newline{}0.9090 \newline{}0.9140 \newline{}0.9210 \newline{}0.9227 \newline{}0.9248 \newline{}0.9280 \newline{}0.9254 \newline{}0.9296 \newline{}0.9299 \newline{}0.9292 \newline{}\underline{0.9300} \newline{}0.9299 \newline{}\bfseries{0.9305} & 27.55 \newline{}29.30 \newline{}29.37 \newline{}29.78 \newline{}29.87 \newline{}30.00 \newline{}30.52 \newline{}30.04 \newline{}30.57 \newline{}30.65 \newline{}30.51 \newline{}30.59 \newline{}\underline{30.67} \newline{}\bfseries{30.77} & 0.7742 \newline{}0.8215 \newline{}0.8240 \newline{}0.8320 \newline{}0.8320 \newline{}0.8350 \newline{}0.8462 \newline{}0.8382 \newline{}0.8468 \newline{}0.8482 \newline{}0.8461 \newline{}0.8476 \newline{}\underline{0.8483} \newline{}\bfseries{0.8495} & 27.21 \newline{}28.41 \newline{}28.53 \newline{}28.83 \newline{}28.82 \newline{}28.96 \newline{}29.25 \newline{}28.97 \newline{}29.26 \newline{}29.32 \newline{}29.24 \newline{}\underline{29.33} \newline{}29.32 \newline{}\bfseries{29.39}&  0.7385 \newline{}0.7863 \newline{}0.7910 \newline{}0.7990 \newline{}0.7980 \newline{}0.8001 \newline{}0.8093 \newline{}0.8025 \newline{}0.8093 \newline{}0.8111 \newline{}0.8084 \newline{}\underline{0.8112} \newline{}0.8110 \newline{}\bfseries{0.8120} & 24.46 \newline{}26.24 \newline{}26.43 \newline{}27.14 \newline{}27.07 \newline{}27.56 \newline{}28.80 \newline{}27.57 \newline{}28.80 \newline{}29.09 \newline{}28.73 \newline{}28.93 \newline{}\underline{29.10} \newline{}\bfseries{29.30} & 0.7349 \newline{}0.7989 \newline{}0.8080 \newline{}0.8290 \newline{}0.8280 \newline{}0.8376 \newline{}0.8653 \newline{}0.8398 \newline{}0.8653 \newline{}0.8702 \newline{}0.8641 \newline{}0.8671 \newline{}\underline{0.8705} \newline{}\bfseries{0.8735} & 26.95 \newline{}30.48 \newline{}31.10 \newline{}32.01 \newline{}32.21 \newline{}32.51 \newline{}34.17 \newline{}33.00 \newline{}34.13 \newline{}34.44 \newline{}34.18 \newline{}34.30 \newline{}\underline{34.48} \newline{}\bfseries{34.80} & 0.8556 \newline{}0.9117 \newline{}0.9210 \newline{}0.9340 \newline{}0.9350 \newline{}0.9369 \newline{}0.9476 \newline{}0.9403 \newline{}0.9484 \newline{}0.9499 \newline{}0.9481 \newline{}0.9494 \newline{}\underline{0.9500} \newline{}\bfseries{0.9514}\\
		\hline
		\hline
		Bicubic \newline{}SRCNN~\cite{dong2014learning} \newline{} FSRCNN~\cite{dong2016accelerating} \newline{} VDSR~\cite{kim2016accurate} \newline{} LapSRN~\cite{lai2017deep} \newline{} MemNet~\cite{tai2017memnet} \newline{} EDSR~\cite{lim2017enhanced} \newline{} SRMDNF~\cite{zhang2018learning}  \newline{} D-DBPN~\cite{haris2018deep} \newline{} RDN~\cite{zhang2018residual} \newline{} RCAN~\cite{zhang2018image} \newline{} SRFBN~\cite{li2019feedback} \newline{} SAN~\cite{dai2019second} \newline{} HAN(ours) \newline{} HAN+(ours) &
		$\times4$ \newline{}$\times4$ \newline{}$\times4$ \newline{}$\times4$ \newline{}$\times4$ \newline{}$\times4$ \newline{}$\times4$ \newline{}$\times4$ \newline{}$\times4$ \newline{}$\times4$ \newline{}$\times4$ \newline{}$\times4$ \newline{}$\times4$ \newline{}$\times4$ \newline{}$\times4$  & 28.42 \newline{}30.48 \newline{}30.72 \newline{}31.35 \newline{}31.54 \newline{}31.74 \newline{}32.46 \newline{}31.96 \newline{}32.47 \newline{}32.47 \newline{}32.63 \newline{}32.47 \newline{}32.64 \newline{}\underline{32.64} \newline{}\bfseries{32.75} & 0.8104 \newline{}0.8628 \newline{}0.8660 \newline{}0.8830 \newline{}0.8850 \newline{}0.8893 \newline{}0.8968 \newline{}0.8925 \newline{}0.8980 \newline{}0.8990 \newline{}0.9002 \newline{}0.8983 \newline{}\underline{0.9003} \newline{}0.9002 \newline{}\bfseries{0.9016} & 26.00 \newline{}27.50 \newline{}27.61 \newline{}28.02 \newline{}28.19 \newline{}28.26 \newline{}28.80 \newline{}28.35 \newline{}28.82 \newline{}28.81 \newline{}28.87 \newline{}28.81 \newline{}\underline{28.92} \newline{}28.90 \newline{}\bfseries{28.99} & 0.7027 \newline{}0.7513 \newline{}0.7550 \newline{}0.7680 \newline{}0.7720 \newline{}0.7723 \newline{}0.7876 \newline{}0.7787 \newline{}0.7860 \newline{}0.7871 \newline{}0.7889 \newline{}0.7868 \newline{}0.7888 \newline{}\underline{0.7890} \newline{}\bfseries{0.7907} & 25.96 \newline{}26.90 \newline{}26.98 \newline{}27.29 \newline{}27.32 \newline{}27.40 \newline{}27.71 \newline{}27.49 \newline{}27.72 \newline{}27.72 \newline{}27.77 \newline{}27.72 \newline{}27.78 \newline{}\underline{27.80} \newline{}\bfseries{27.85} & 0.6675 \newline{}0.7101 \newline{}0.7150 \newline{}0.0726 \newline{}0.7270 \newline{}0.7281 \newline{}0.7420 \newline{}0.7337 \newline{}0.7400 \newline{}0.7419 \newline{}0.7436 \newline{}0.7409 \newline{}0.7436 \newline{}\underline{0.7442} \newline{}\bfseries{0.7454} & 23.14 \newline{}24.52 \newline{}24.62 \newline{}25.18 \newline{}25.21 \newline{}25.50 \newline{}26.64 \newline{}25.68 \newline{}26.38 \newline{}26.61 \newline{}26.82 \newline{}26.60 \newline{}26.79 \newline{}\underline{26.85} \newline{}\bfseries{27.02} & 0.6577 \newline{}0.7221 \newline{}0.7280 \newline{}0.7540 \newline{}0.7560 \newline{}0.7630 \newline{}0.8033 \newline{}0.7731 \newline{}0.7946 \newline{}0.8028 \newline{}0.8087 \newline{}0.8015 \newline{}0.8068 \newline{}\underline{0.8094} \newline{}\bfseries{0.8131} & 24.89 \newline{}27.58 \newline{}27.90 \newline{}28.83 \newline{}29.09 \newline{}29.42 \newline{}31.02 \newline{}30.09 \newline{}30.91 \newline{}31.00 \newline{}31.22 \newline{}31.15 \newline{}31.18 \newline{}\underline{31.42} \newline{}\bfseries{31.73} & 0.7866\newline{} 0.8555\newline{} 0.8610\newline{} 0.8870\newline{} 0.8900\newline{} 0.8942\newline{} 0.9148\newline{} 0.9024\newline{} 0.9137\newline{} 0.9151\newline{} 0.9173 \newline{}0.9160 \newline{}0.9169 \newline{}\underline{0.9177} \newline{}\bfseries{0.9207} \\
		\hline
		\hline
		Bicubic \newline{}SRCNN~\cite{dong2014learning} \newline{} FSRCNN~\cite{dong2016accelerating} \newline{} SCN~\cite{wang2015deep} \newline{} VDSR~\cite{kim2016accurate} \newline{} LapSRN~\cite{lai2017deep} \newline{} MemNet~\cite{tai2017memnet} \newline{} MSLapSRN\cite{lai2017deep} \newline{} EDSR~\cite{lim2017enhanced} \newline{}  D-DBPN~\cite{haris2018deep} \newline{}  RCAN~\cite{zhang2018image} \newline{}  SAN~\cite{dai2019second} \newline{} HAN(ours) \newline{} HAN+(ours) &
		$\times8$ \newline{}$\times8$ \newline{}$\times8$ \newline{}$\times8$ \newline{}$\times8$ \newline{}$\times8$ \newline{}$\times8$ \newline{}$\times8$ \newline{}$\times8$ \newline{}$\times8$ \newline{}$\times8$ \newline{}$\times8$ \newline{}$\times8$ \newline{}$\times8$  & 24.40 \newline{}25.33 \newline{}20.13 \newline{}25.59 \newline{}25.93 \newline{}26.15 \newline{}26.16 \newline{}26.34 \newline{}26.96 \newline{}27.21 \newline{}27.31 \newline{}27.22 \newline{}\underline{27.33} \newline{}\bfseries{27.47}& 0.6580 \newline{}0.6900 \newline{}0.5520 \newline{}0.7071 \newline{}0.7240 \newline{}0.7380 \newline{}0.7414 \newline{}0.7558 \newline{}0.7762 \newline{}0.7840 \newline{}0.7878 \newline{}0.7829 \newline{}\underline{0.7884} \newline{}\bfseries{0.7920} & 23.10 \newline{}23.76 \newline{}19.75 \newline{}24.02 \newline{}24.26 \newline{}24.35 \newline{}24.38 \newline{}24.57 \newline{}24.91 \newline{}25.13 \newline{}25.23 \newline{}25.14 \newline{}\underline{25.24} \newline{}\bfseries{25.39} & 0.5660 \newline{}0.5910 \newline{}0.4820 \newline{}0.6028 \newline{}0.6140 \newline{}0.6200 \newline{}0.6199 \newline{}0.6273 \newline{}0.6420 \newline{}0.6480 \newline{}\underline{0.6511} \newline{}0.6476 \newline{}0.6510 \newline{}\bfseries{0.6552} & 23.67 \newline{}24.13 \newline{}24.21 \newline{}24.30 \newline{}24.49 \newline{}24.54 \newline{}24.58 \newline{}24.65 \newline{}24.81 \newline{}24.88 \newline{}24.98 \newline{}24.88 \newline{}\underline{24.98} \newline{}\bfseries{25.04} & 0.5480 \newline{}0.5660 \newline{}0.5680 \newline{}0.5698 \newline{}0.5830 \newline{}0.5860 \newline{}0.5842 \newline{}0.5895 \newline{}0.5985 \newline{}0.6010 \newline{}0.6058 \newline{}0.6011 \newline{}\underline{0.6059} \newline{}\bfseries{0.6075} & 20.74 \newline{}21.29 \newline{}21.32 \newline{}21.52 \newline{}21.70 \newline{}21.81 \newline{}21.89 \newline{}22.06 \newline{}22.51 \newline{}22.73 \newline{}\underline{23.00} \newline{}22.70 \newline{}22.98 \newline{}\bfseries{23.20} & 0.5160 \newline{}0.5440 \newline{}0.5380 \newline{}0.5571 \newline{}0.5710 \newline{}0.5810 \newline{}0.5825 \newline{}0.5963 \newline{}0.6221 \newline{}0.6312 \newline{}\underline{0.6452} \newline{}0.6314 \newline{}0.6437 \newline{}\bfseries{0.6518} & 21.47 \newline{}22.46 \newline{}22.39 \newline{}22.68 \newline{}23.16 \newline{}23.39 \newline{}23.56 \newline{}23.90 \newline{}24.69 \newline{}25.14 \newline{}\underline{25.24} \newline{}24.85 \newline{}25.20 \newline{}\bfseries{25.54} & 0.6500\newline{} 0.6950\newline{} 0.6730\newline{} 0.6963\newline{} 0.7250\newline{} 0.7350\newline{} 0.7387\newline{} 0.7564\newline{} 0.7841\newline{} 0.7987\newline{} \underline{0.8029} \newline{}0.7906 \newline{}0.8011 \newline{}\bfseries{0.8080}  \\
		\hline
	\end{tabular}%
	\label{table-BI}%
\end{table}

\section{Experiments}
\label{sec-exp}
In this section, we first analyze the contributions of the proposed two attention modules. We then compare our HAN with state-of-the-art algorithms on five benchmark datasets. 
The implementation code will be made available to the public. Results on more images can be found in the supplementary material.

\subsection{Settings}
\noindent\textbf{Datasets.} We selecte DIV2K~\cite{timofte2017ntire} as the training set as like in \cite{zhang2018image,dai2019second,zhang2018residual,lim2017enhanced}.
%which included 800 images for training and 100 images for testing.
%
For the testing set, we choose five standard datasets: Set5~\cite{bevilacqua2012low}, Set14~\cite{zeyde2010single}, B100~\cite{martin2001database}, Urban100~\cite{huang2015single}, and Manga109~\cite{matsui2017sketch}. Degraded data was obtained by bilinear interpolation and Blur-downscale Degradation model. 
Following \cite{zhang2018image}, the reconstruct RGB results by the proposed HAN are first converted to YCbCr space, and then we only consider the luminance channel to calculate PSNR and SSIM in our experiments.

%{\flushleft \textbf{Evaluation metrics.}} The RGB space results of network reconstruction are first converted to YCbCr space, and then the PSNR and SSIM of Y channel are calculated as the Evaluation methods.

%

\begin{figure}[t]\tiny
	\begin{center}
		\tabcolsep 1pt
		\begin{tabular}{@{}ccccccccc@{}}
			HR&
			Bicubic & 
			VDSR~\cite{kim2016accurate} & 
			DBPN~\cite{haris2018deep}   &
			EDSR~\cite{lim2017enhanced} & 
			RCAN~\cite{zhang2018image}  &
			SRFBN~\cite{li2019feedback} &
			DBPNLL~\cite{haris2018deep}   &
			HAN(our) \\
			%HAN+(our) \\
			\includegraphics[width = 0.1\textwidth]{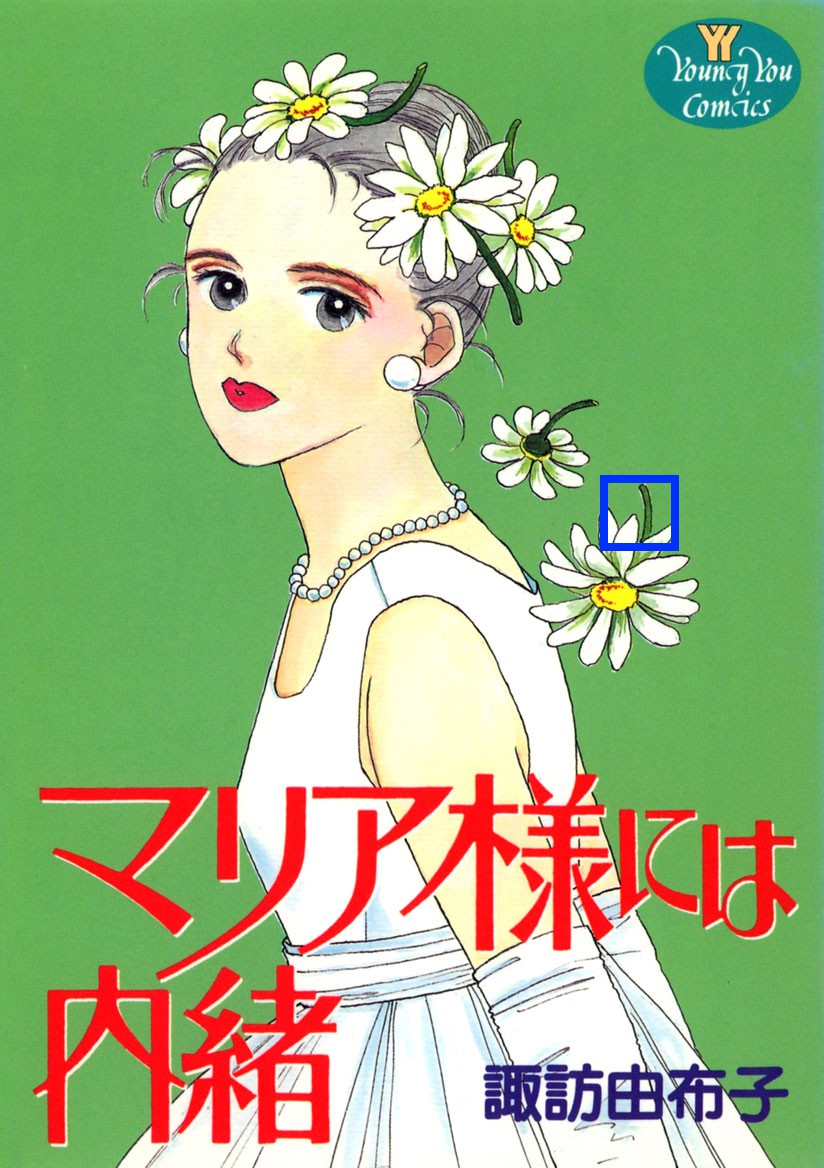}&
			\includegraphics[width = 0.1\textwidth]{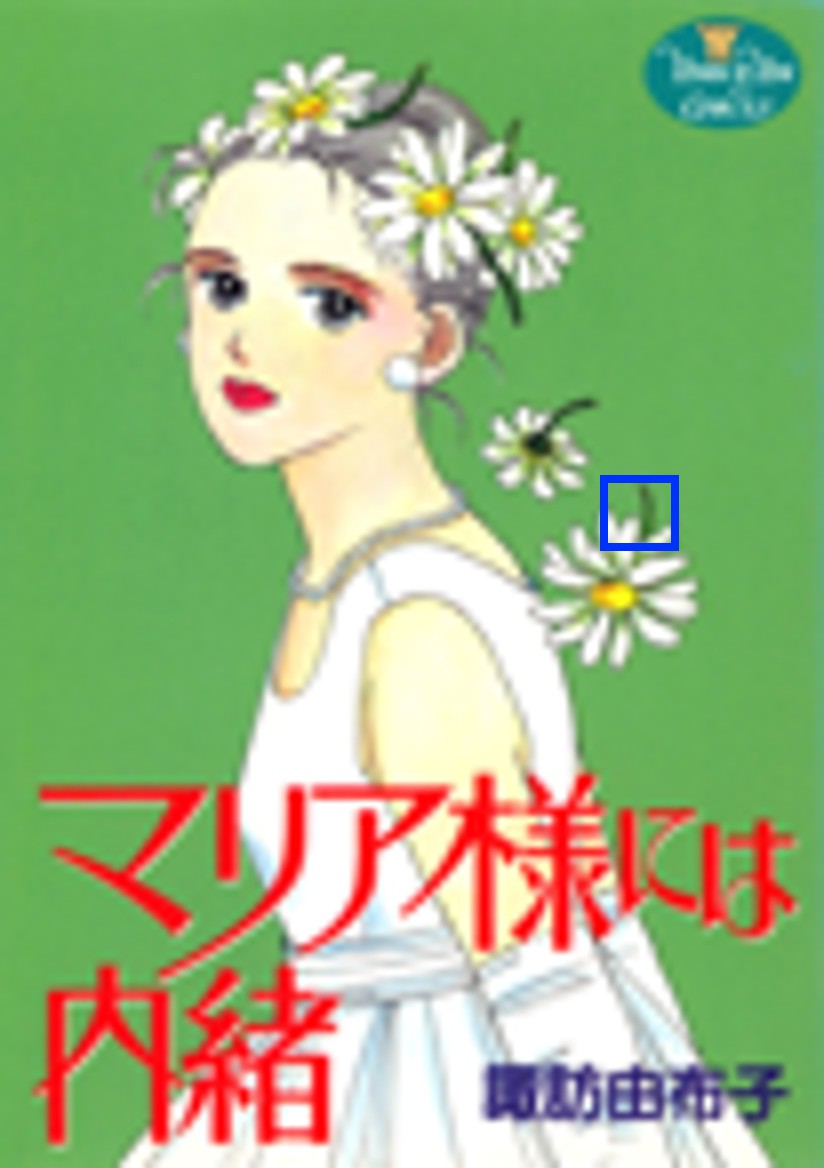} & 
			\includegraphics[width = 0.1\textwidth]{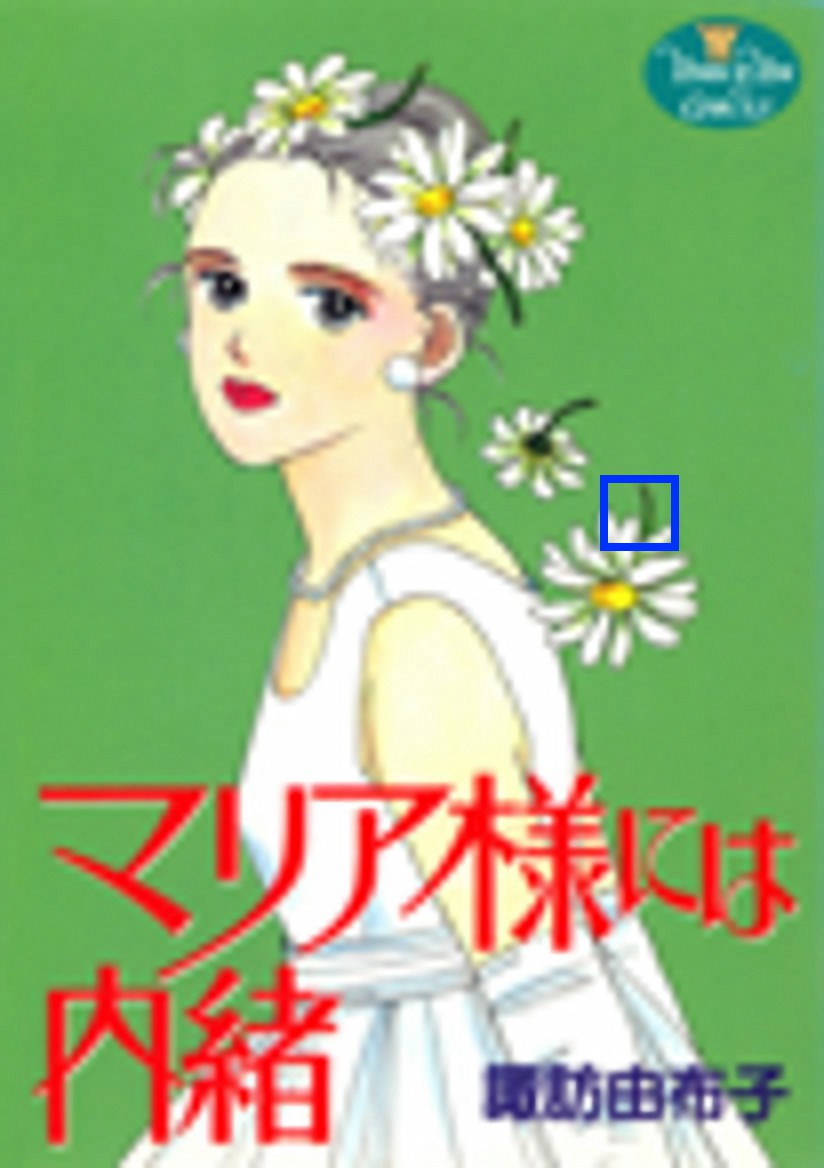} &
			\includegraphics[width = 0.1\textwidth]{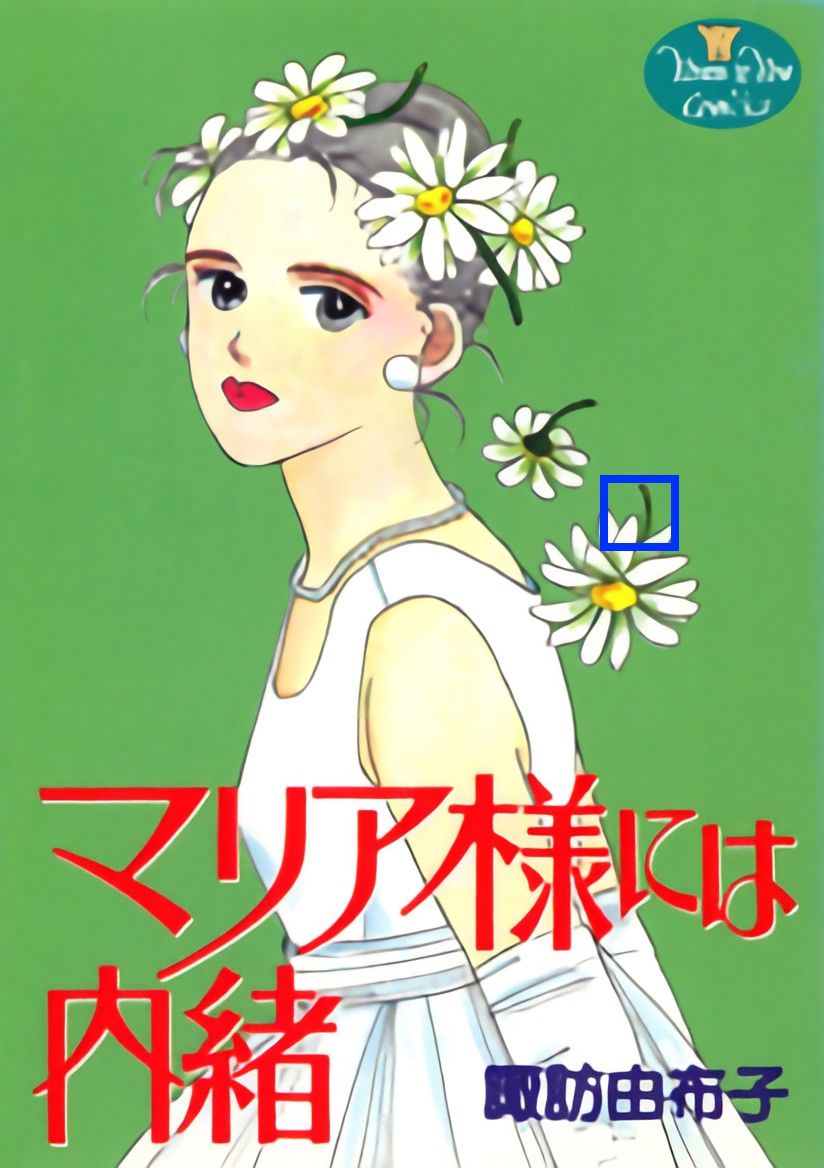} & 
			\includegraphics[width = 0.1\textwidth]{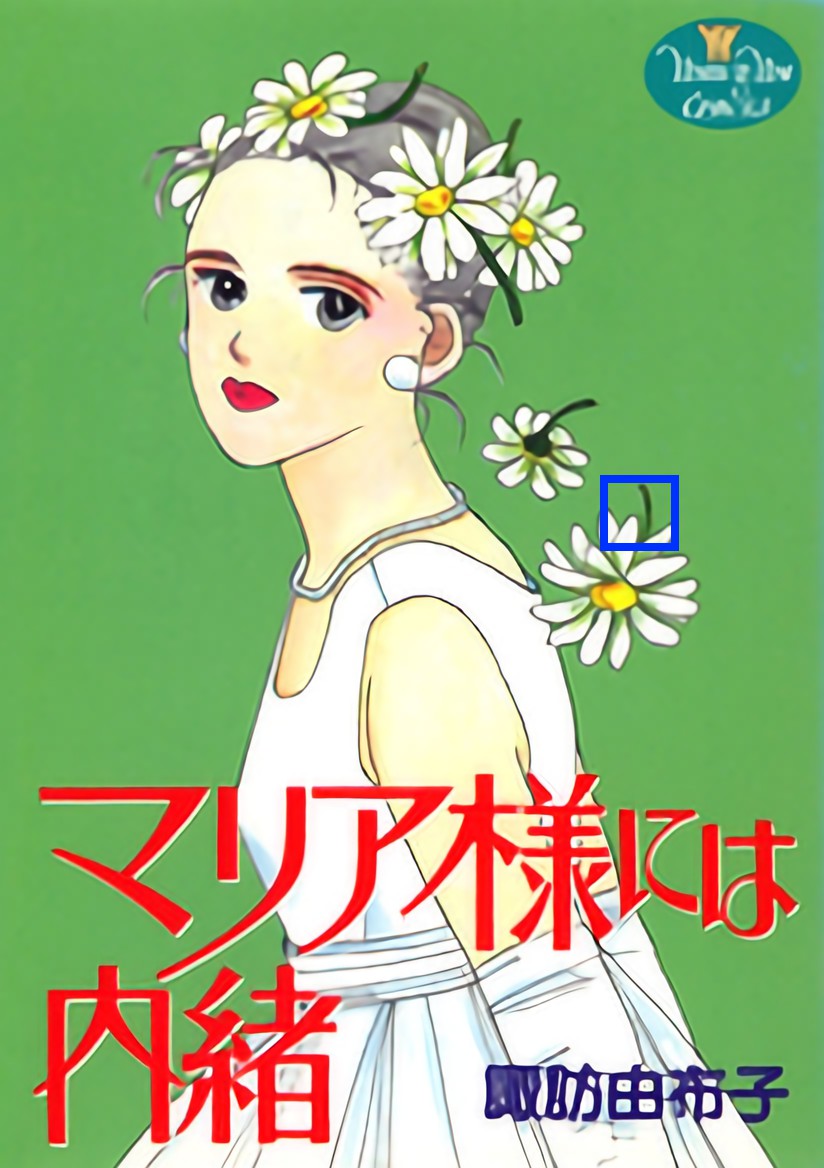}&
			\includegraphics[width = 0.1\textwidth]{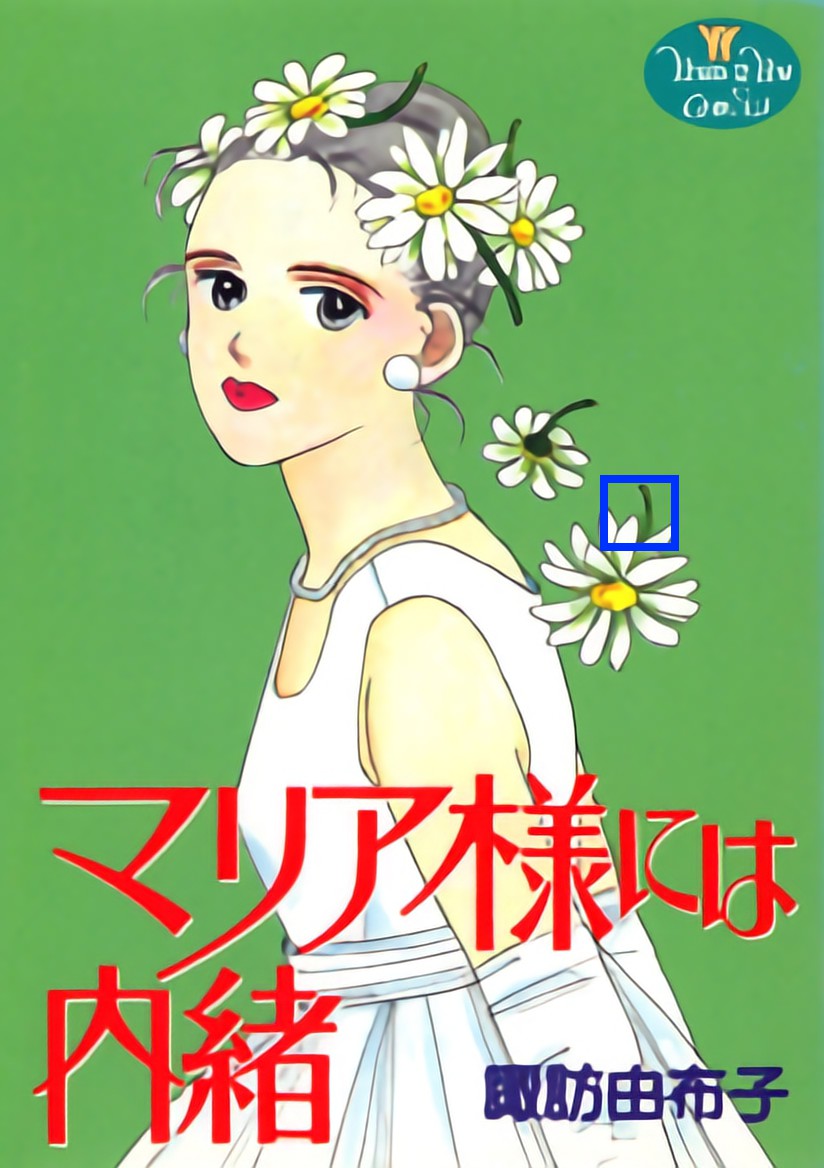} & 
			\includegraphics[width = 0.1\textwidth]{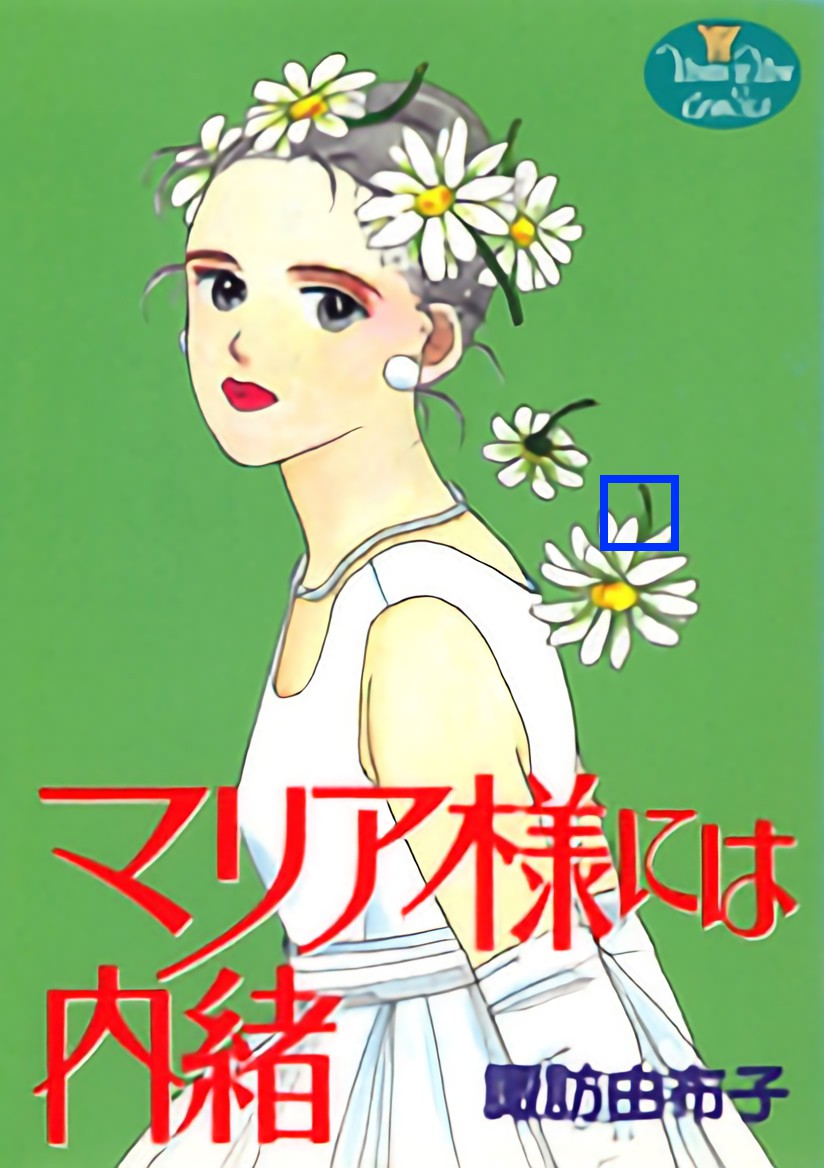}&
			\includegraphics[width = 0.1\textwidth]{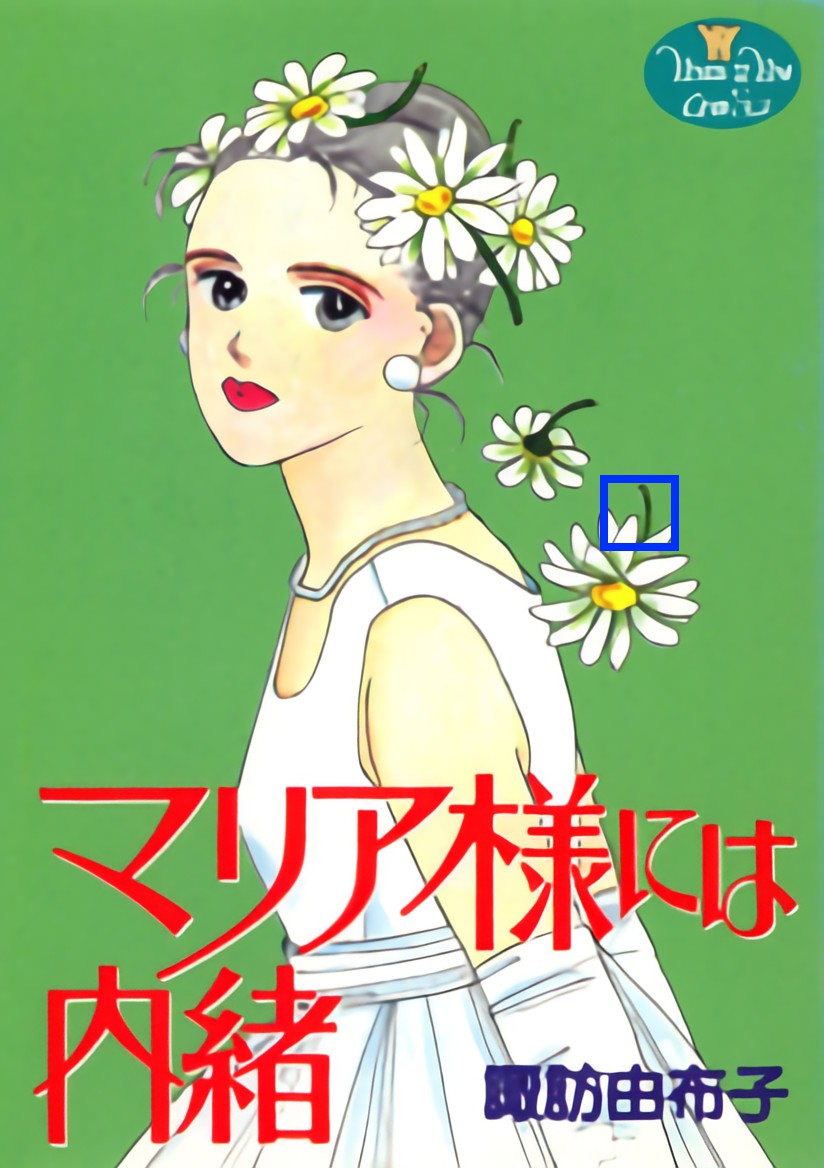}& 
			\includegraphics[width = 0.1\textwidth]{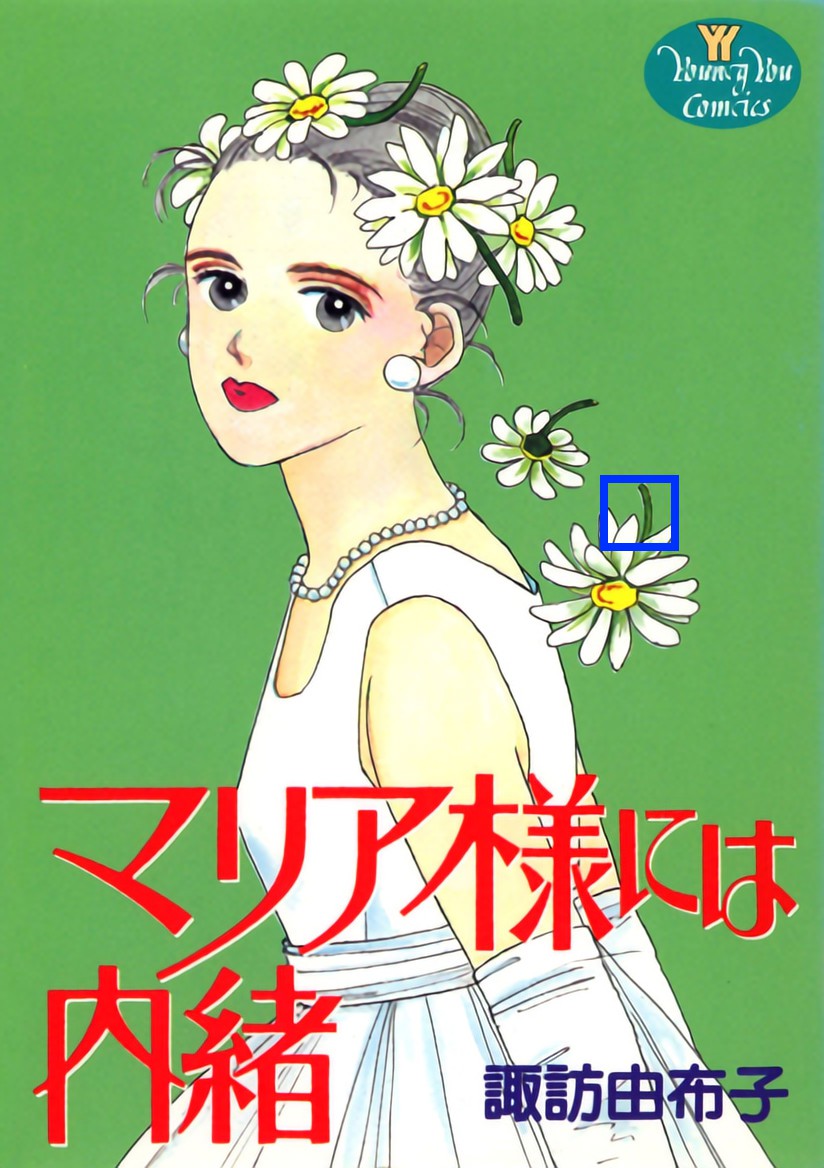} \\ 

			\includegraphics[width = 0.1\textwidth]{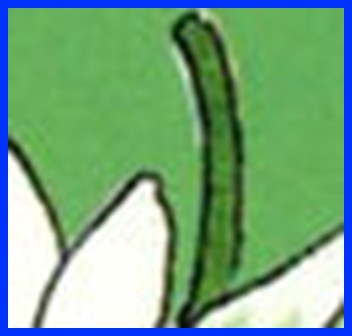}&
			\includegraphics[width = 0.1\textwidth]{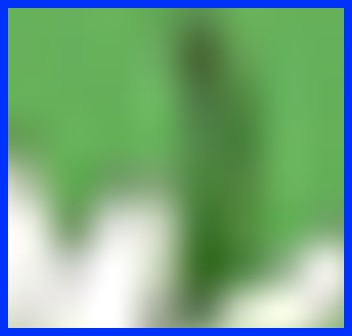} & 
			\includegraphics[width = 0.1\textwidth]{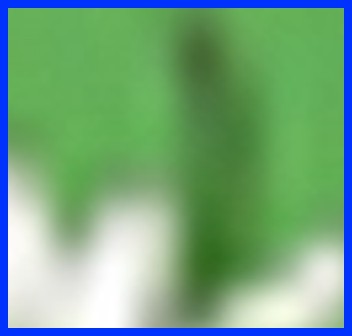} &
			\includegraphics[width = 0.1\textwidth]{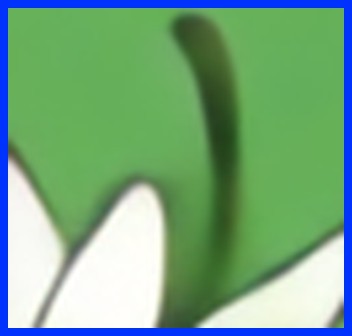} &
			\includegraphics[width = 0.1\textwidth]{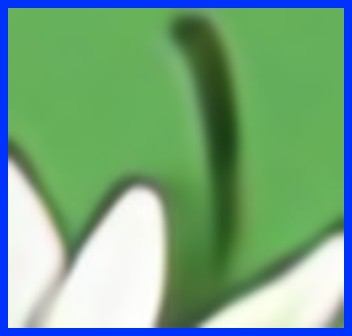} & 
			\includegraphics[width = 0.1\textwidth]{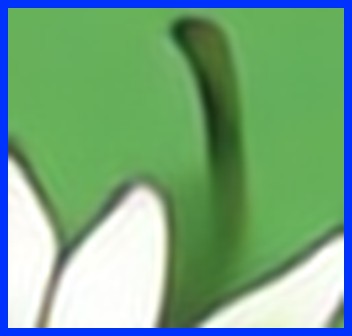} & 
			\includegraphics[width = 0.1\textwidth]{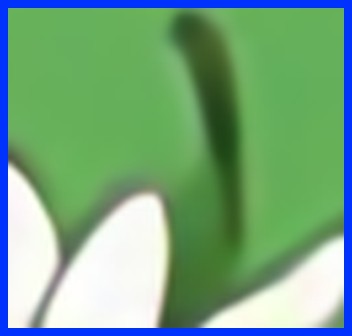} &
			\includegraphics[width = 0.1\textwidth]{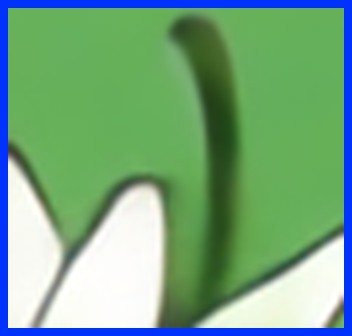}& 
			\includegraphics[width = 0.1\textwidth]{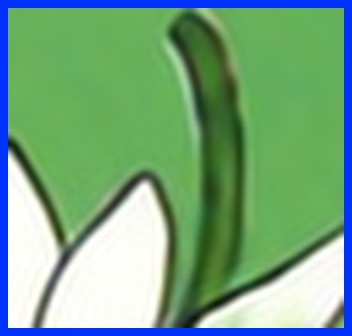} \\ 
			PSNR/SSIM & 21.22/ 0.737 &21.20/0.733 &24.92/ 0.881 & 24.54/0.873 &25.08/0.886 &24.26/0.866 &25.25 0.889 &\textbf{25.78}/\textbf{0.902} \\
			
			\includegraphics[width = 0.1\textwidth]{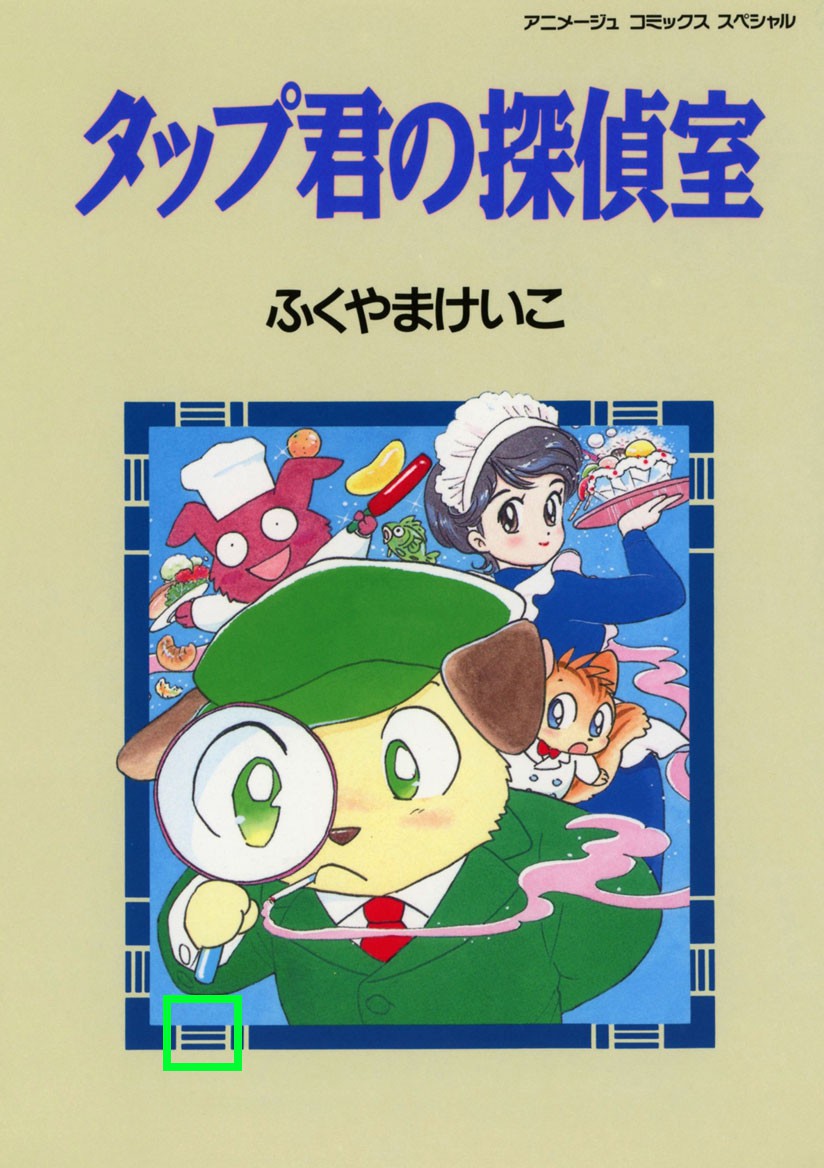}&
			\includegraphics[width = 0.1\textwidth]{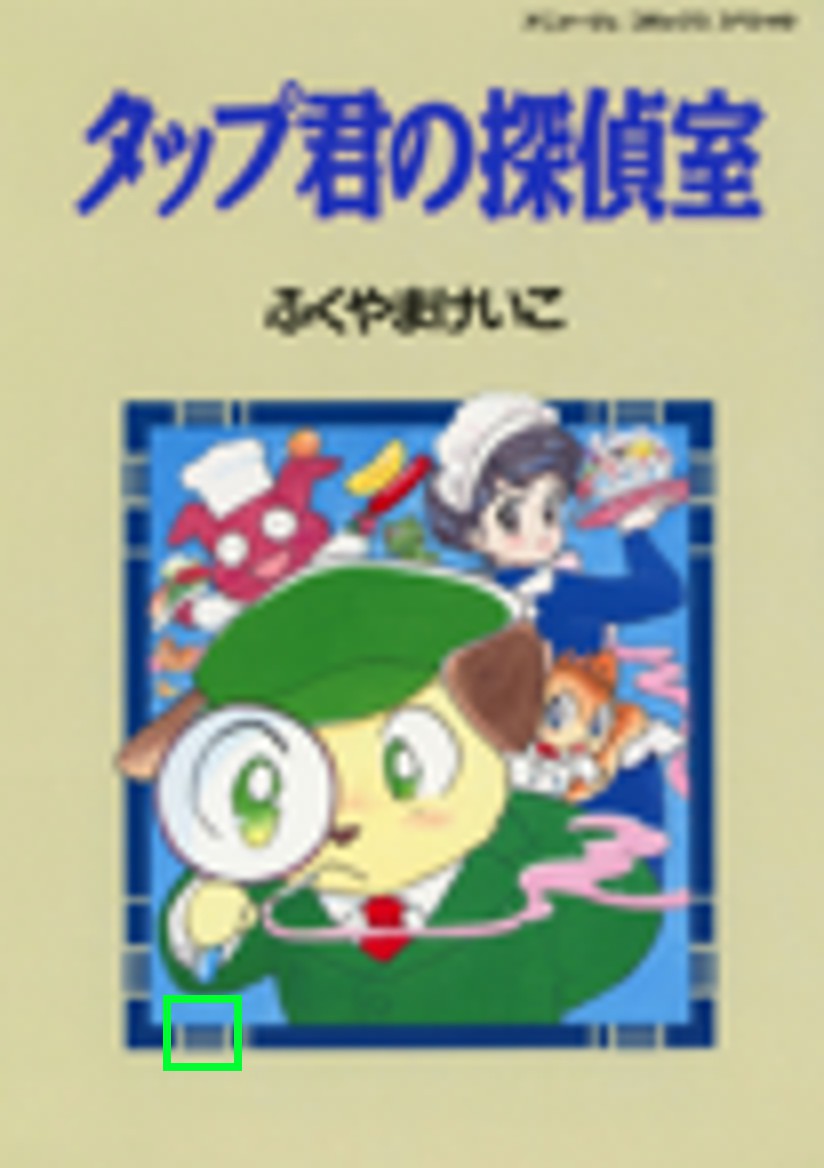} & 
			\includegraphics[width = 0.1\textwidth]{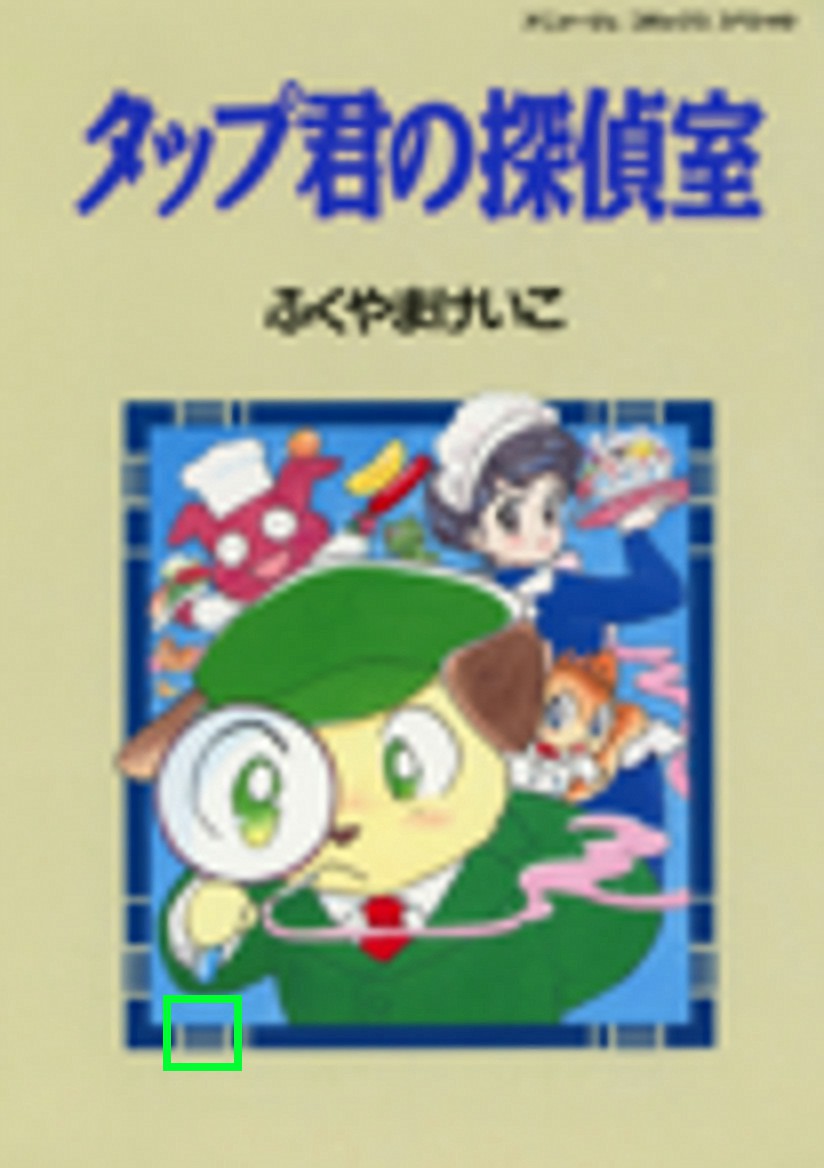} &
			\includegraphics[width = 0.1\textwidth]{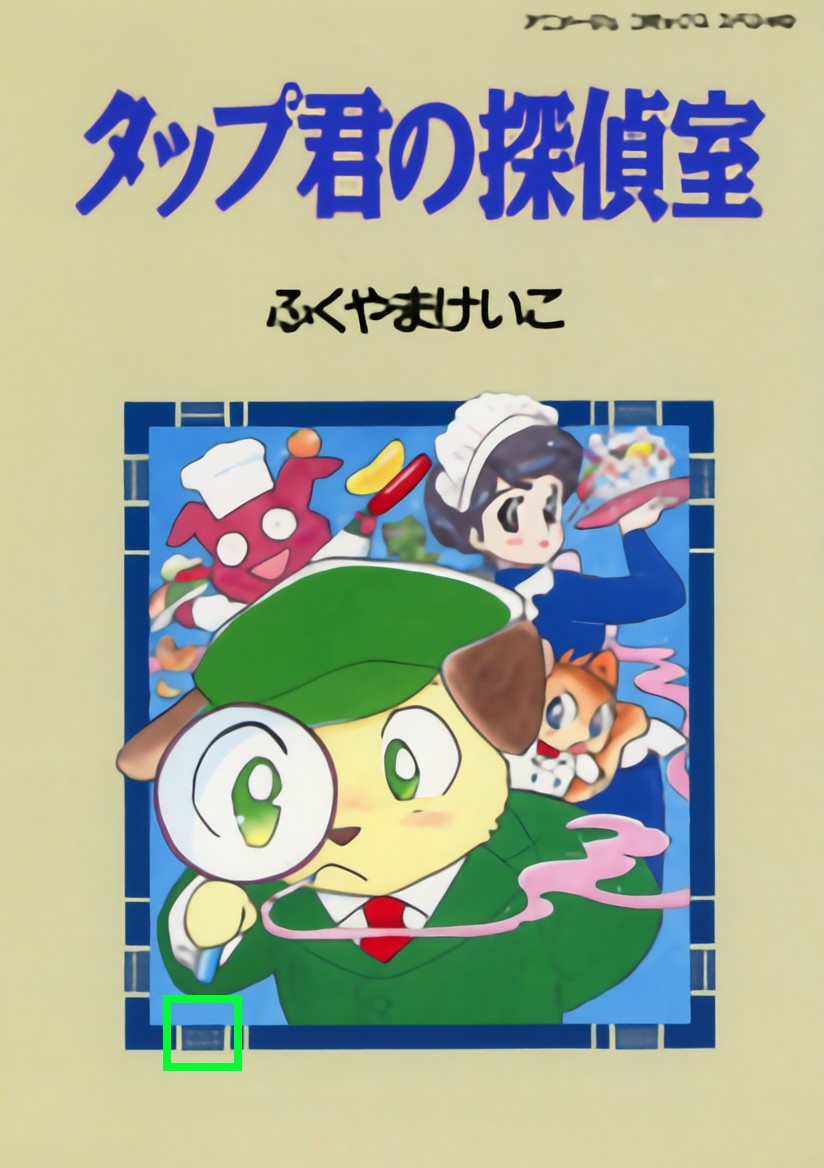} & 
			\includegraphics[width = 0.1\textwidth]{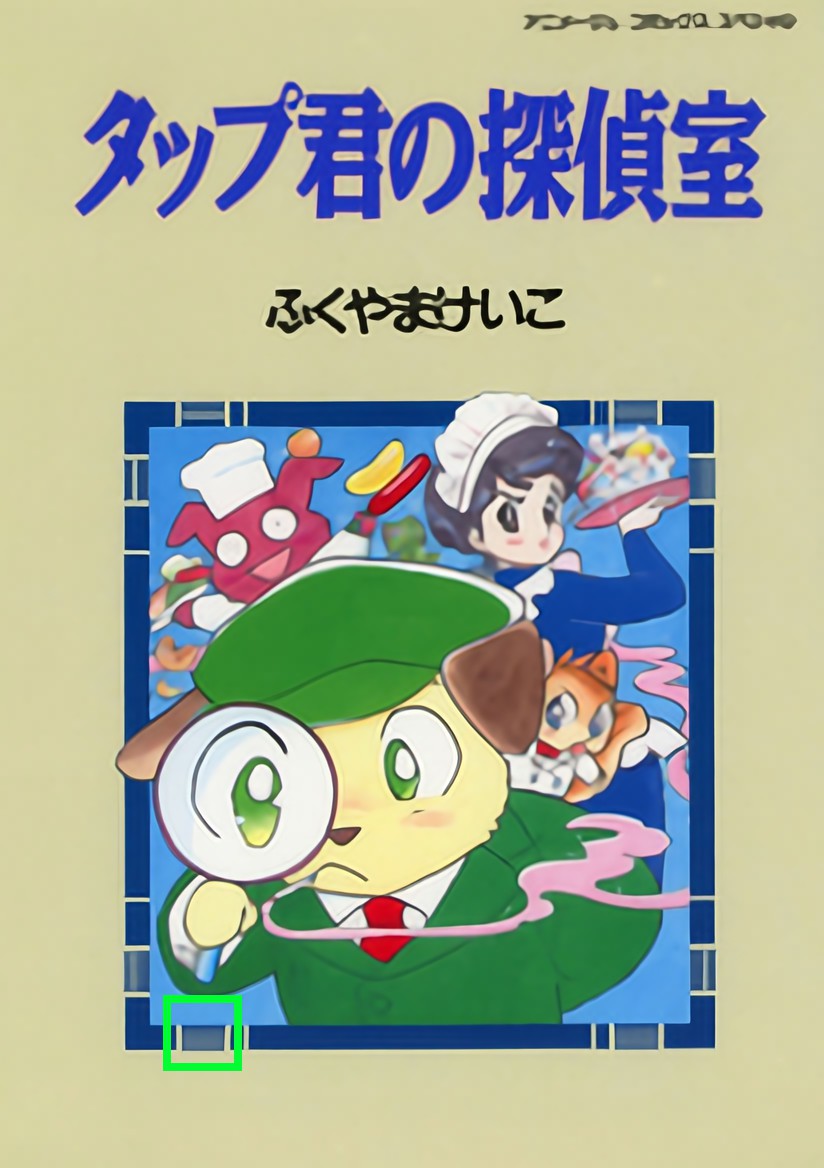} &
			\includegraphics[width = 0.1\textwidth]{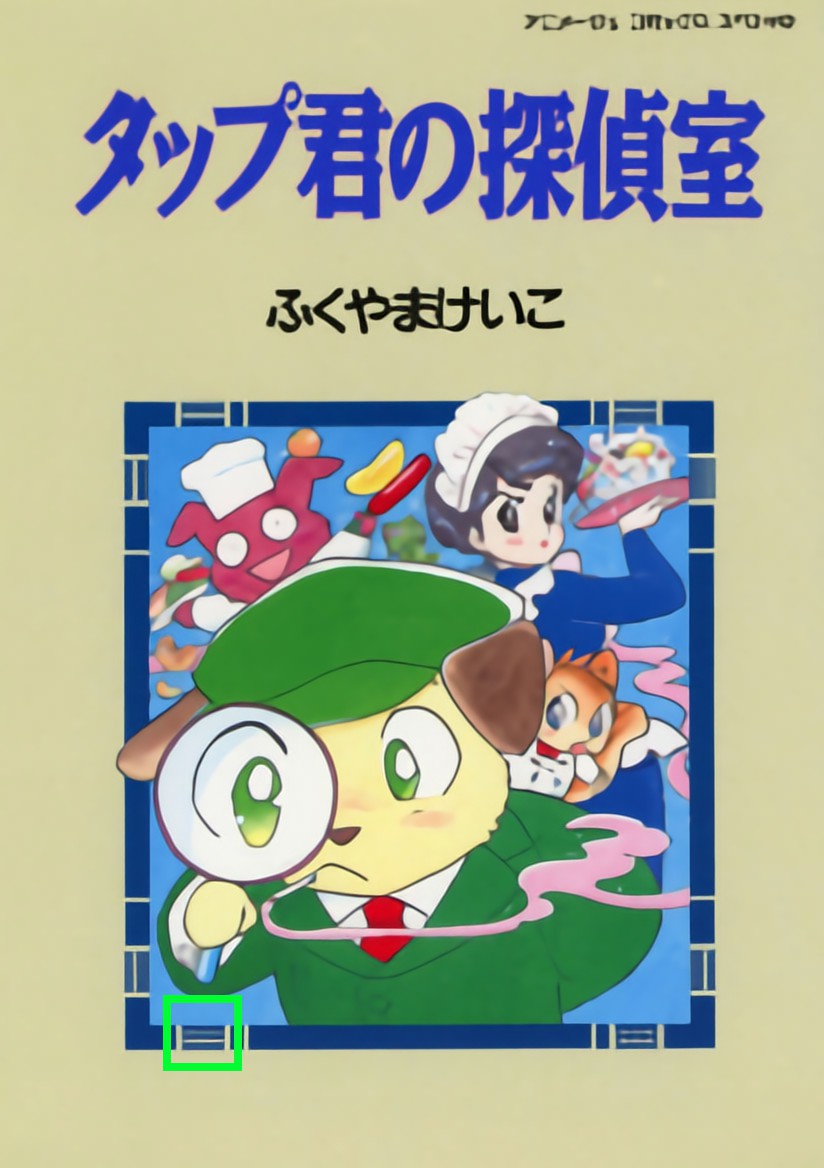} & 
			\includegraphics[width = 0.1\textwidth]{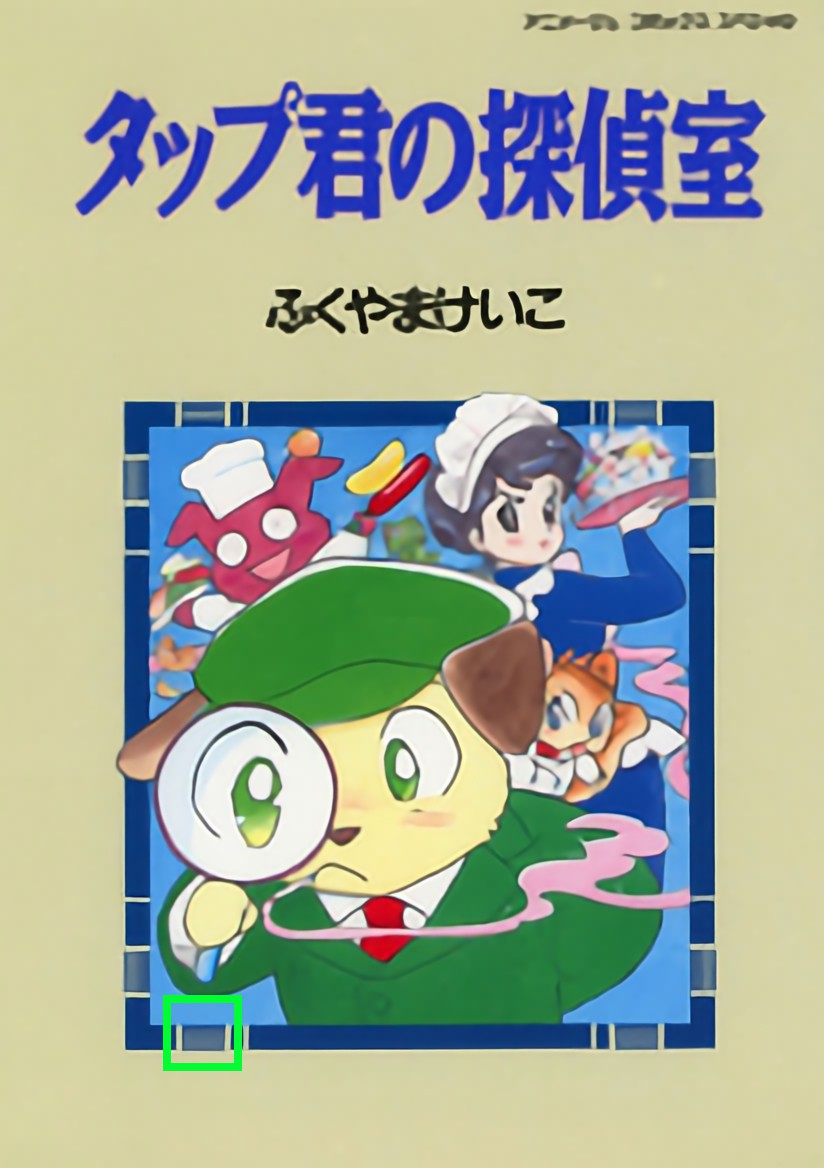} &
			\includegraphics[width = 0.1\textwidth]{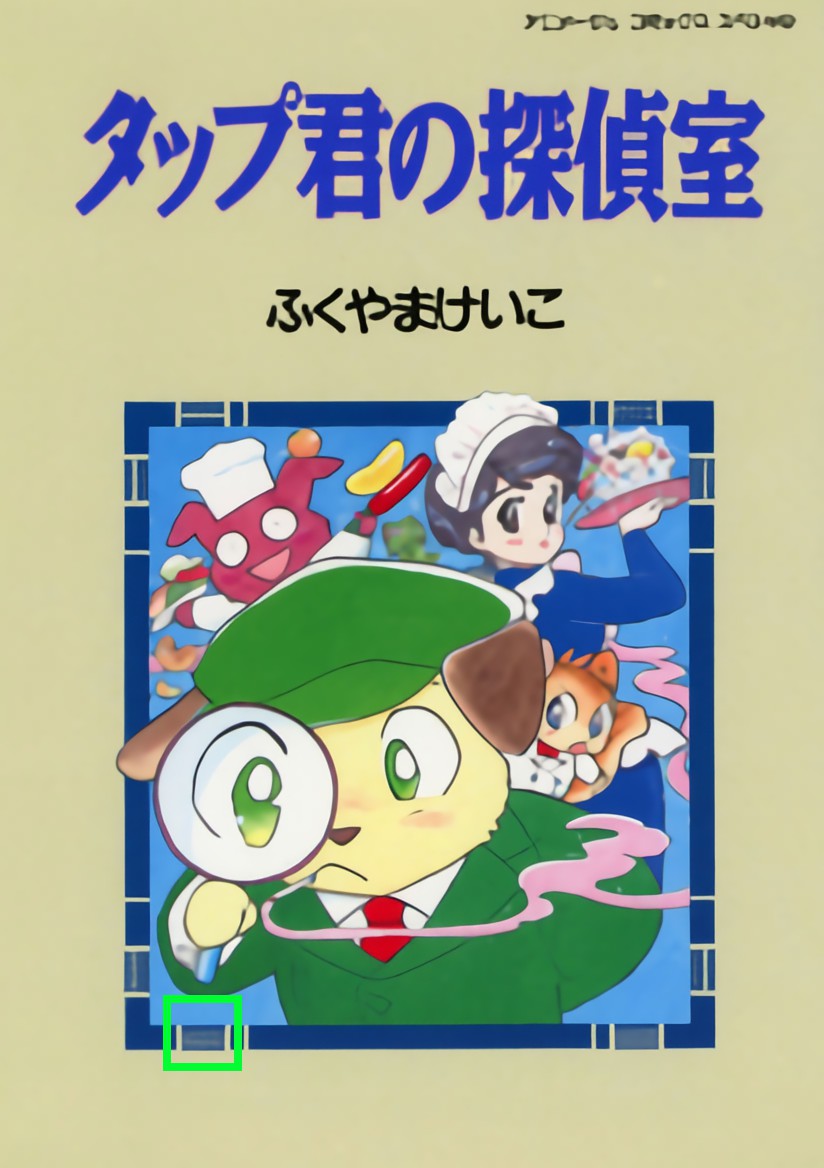}& 
			
			\includegraphics[width = 0.1\textwidth]{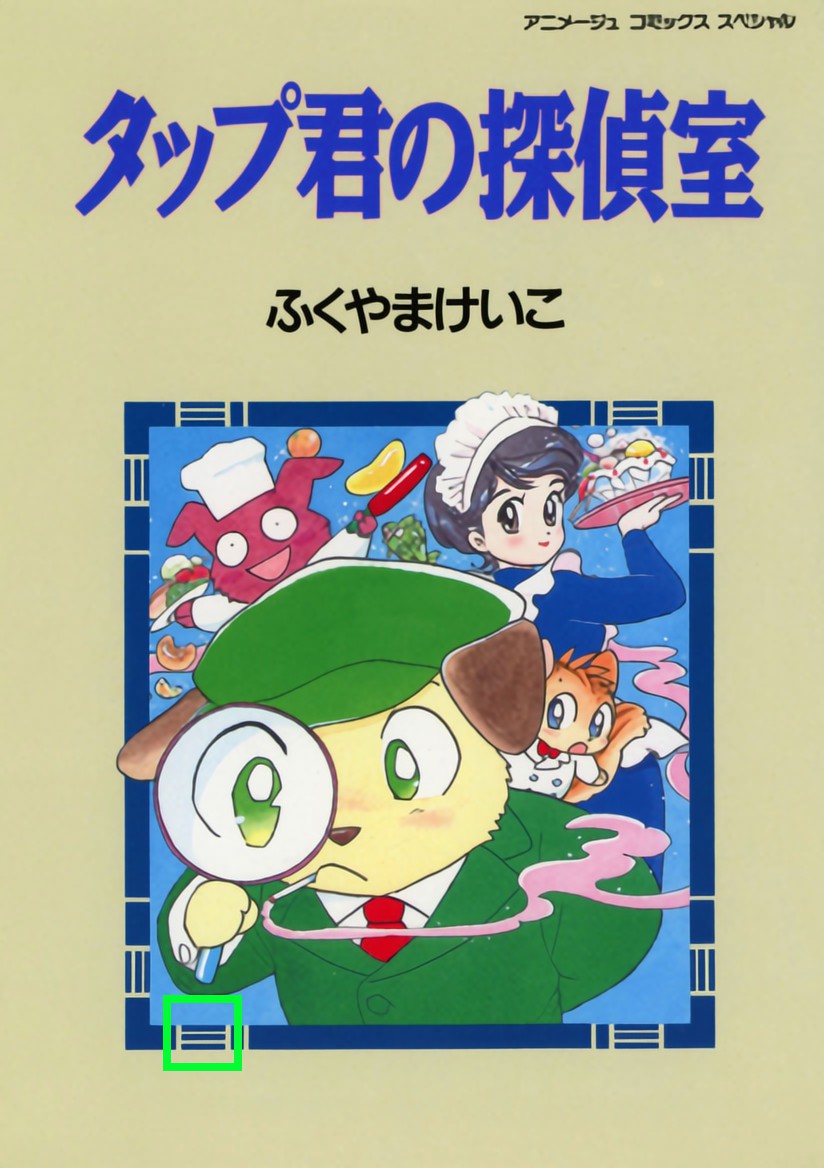} \\ 

			\includegraphics[width = 0.1\textwidth]{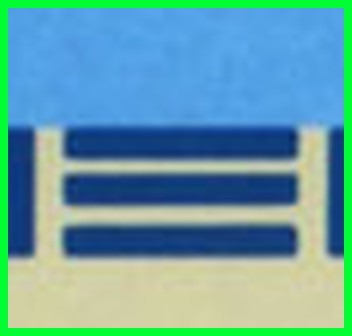}&
			\includegraphics[width = 0.1\textwidth]{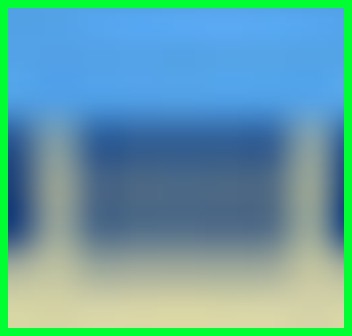} & 
			\includegraphics[width = 0.1\textwidth]{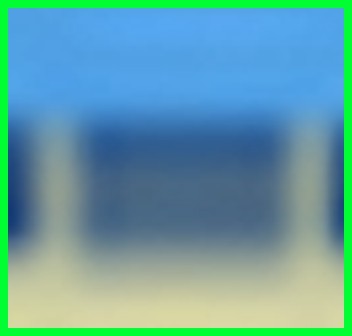} &
			\includegraphics[width = 0.1\textwidth]{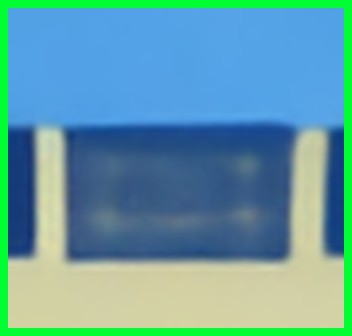} & 
			\includegraphics[width = 0.1\textwidth]{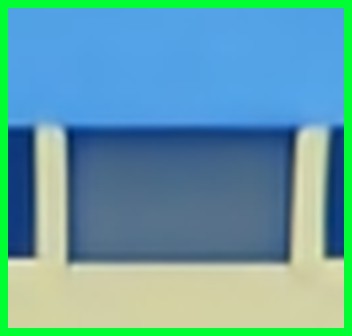} &
			\includegraphics[width = 0.1\textwidth]{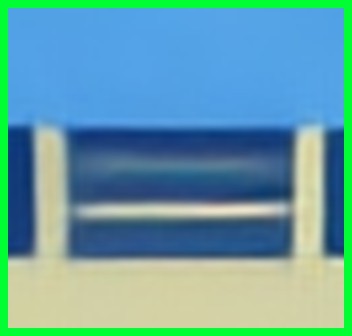} &
			\includegraphics[width = 0.1\textwidth]{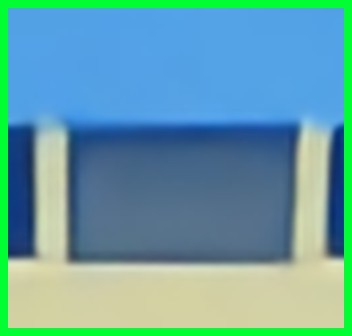} & 
			\includegraphics[width = 0.1\textwidth]{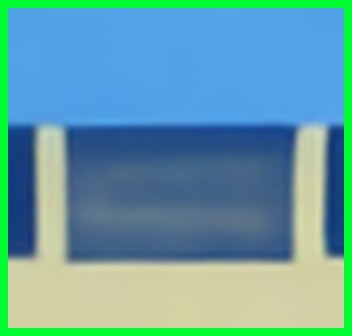}& 
			\includegraphics[width = 0.1\textwidth]{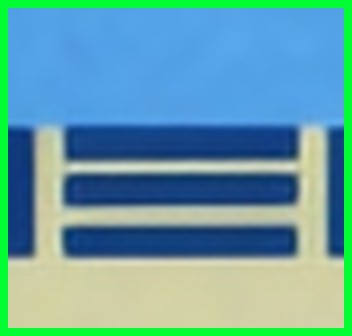} \\

			PSNR/SSIM & 22.88/0.768 & 24.86/0.845 &27.52/0.913 & 27.01/0.900 &27.56/0.914 &26.69/0.893 &27.75/0.918 &\textbf{27.77}/\textbf{0.935} \\
			
			\includegraphics[width = 0.1\textwidth]{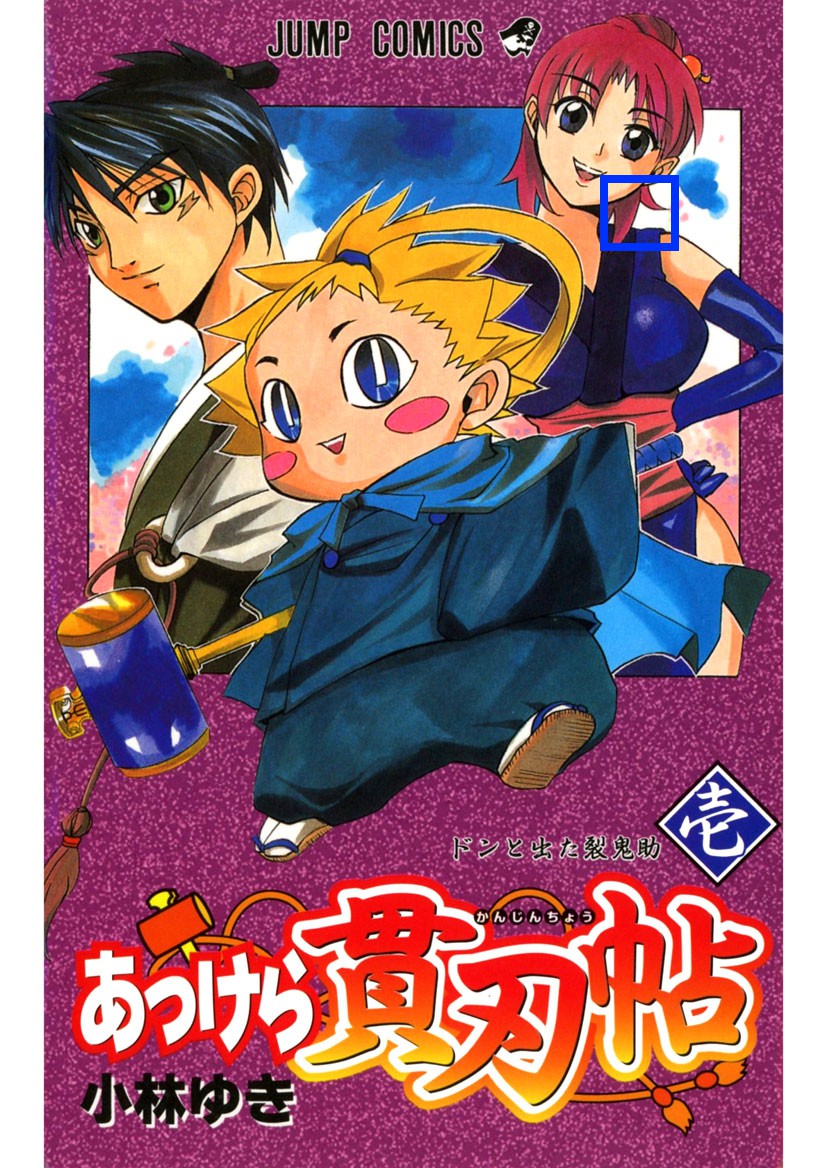}&
			\includegraphics[width = 0.1\textwidth]{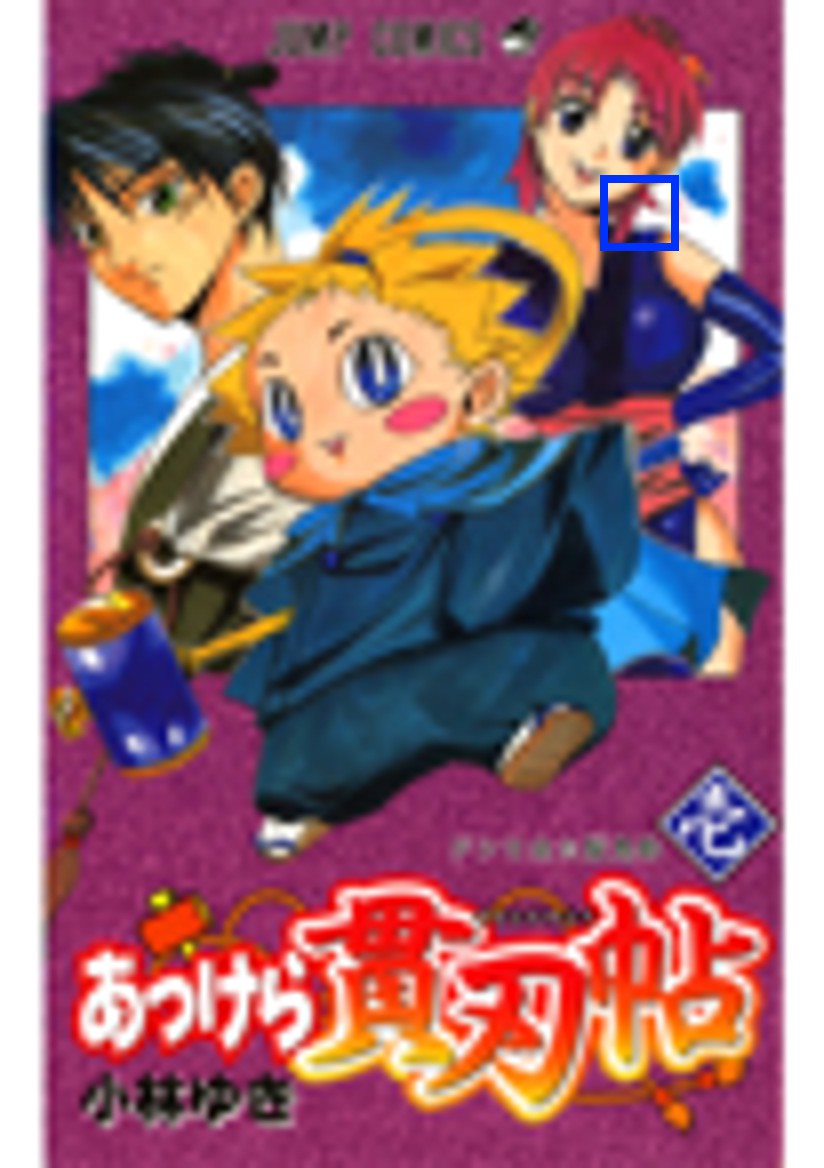} & 
			\includegraphics[width = 0.1\textwidth]{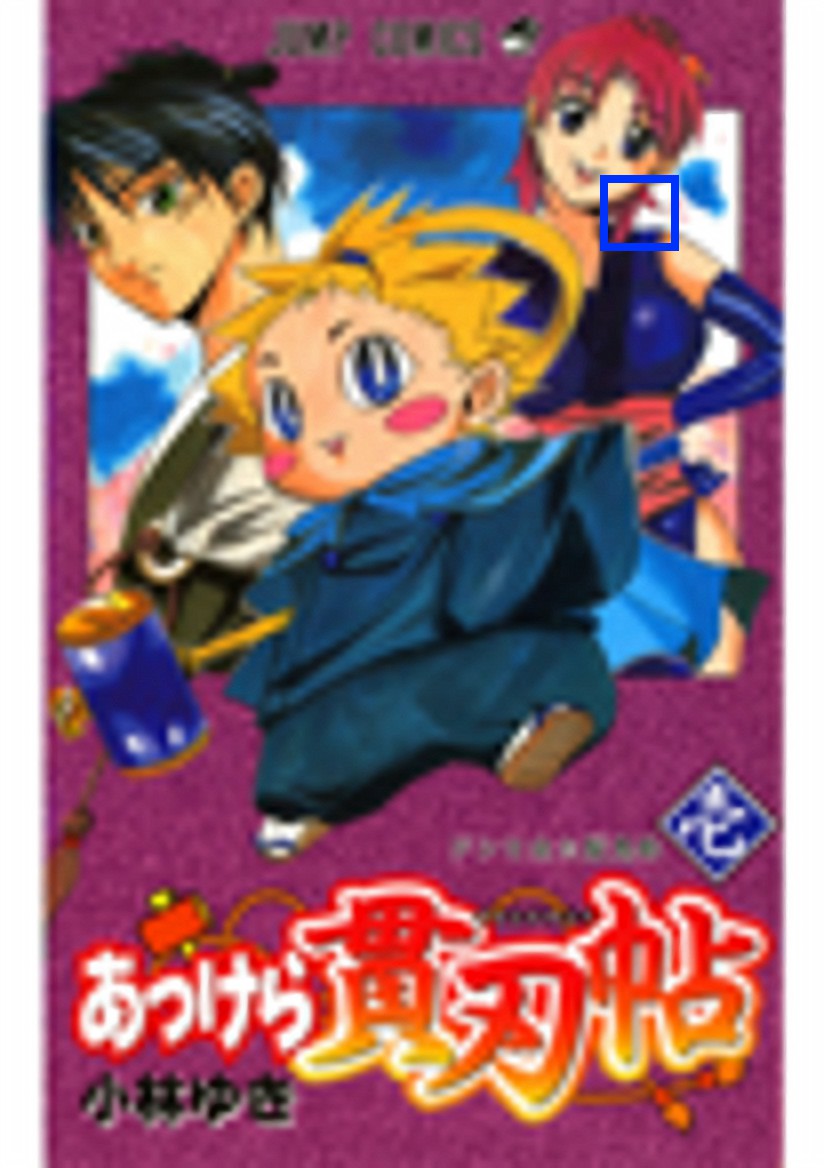} &
			\includegraphics[width = 0.1\textwidth]{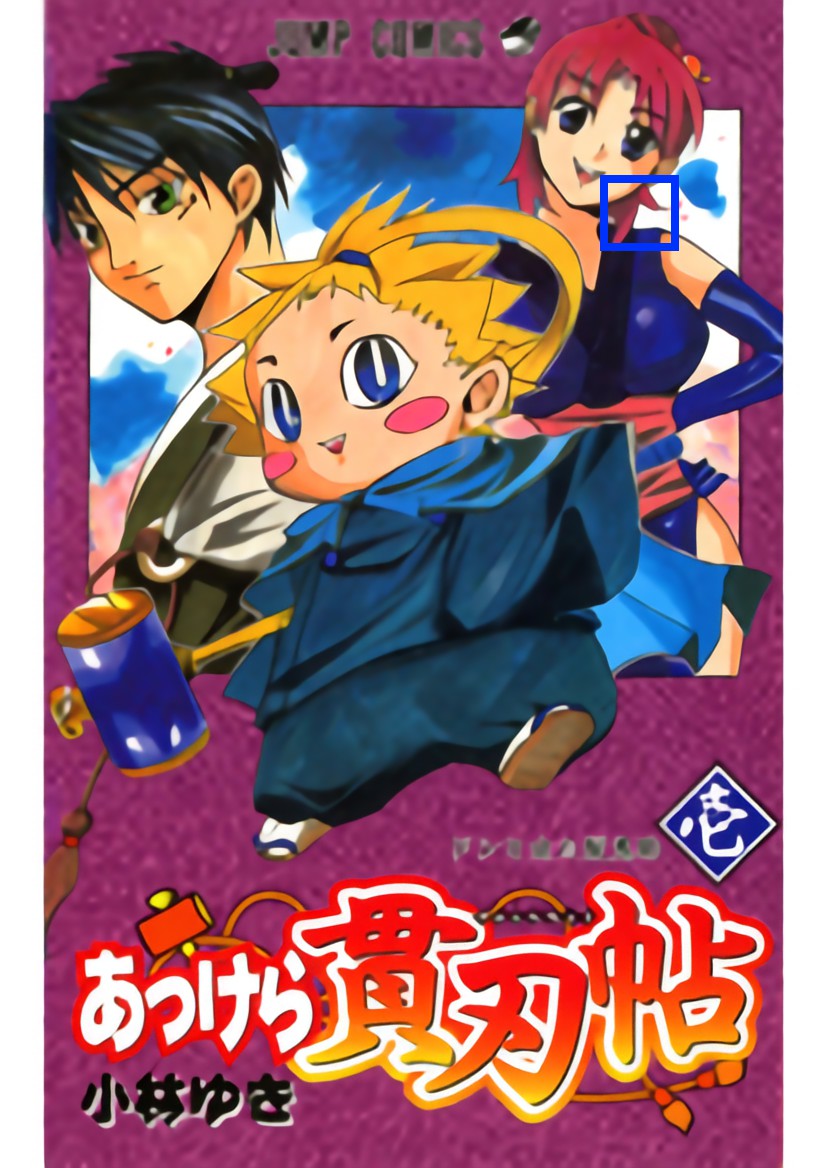} & 
			\includegraphics[width = 0.1\textwidth]{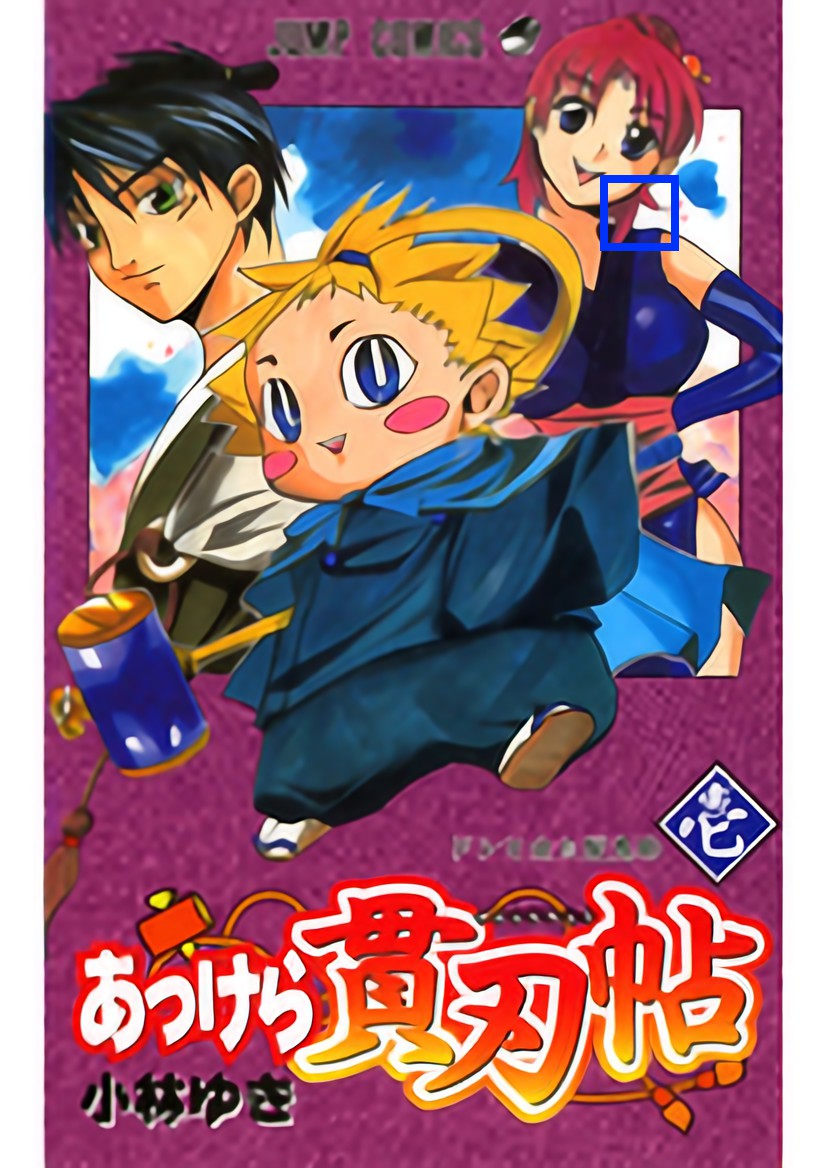} &
			\includegraphics[width = 0.1\textwidth]{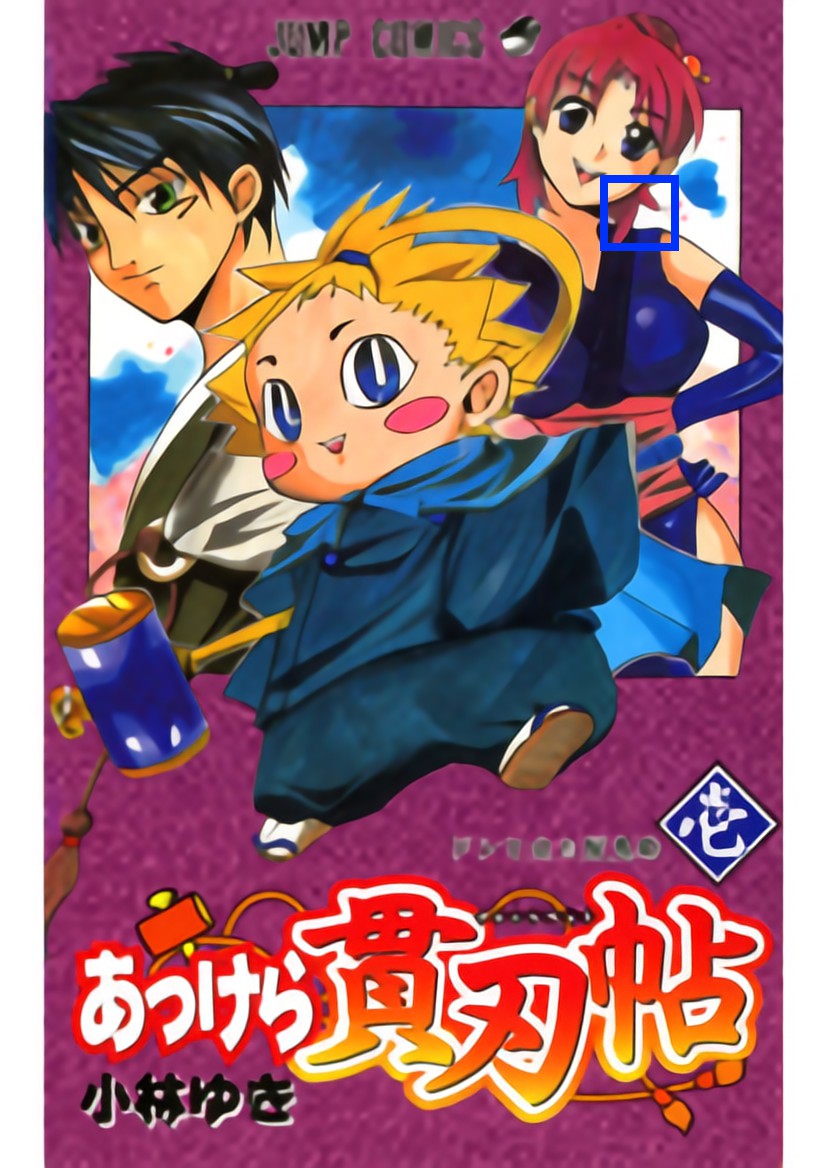} & 
			\includegraphics[width = 0.1\textwidth]{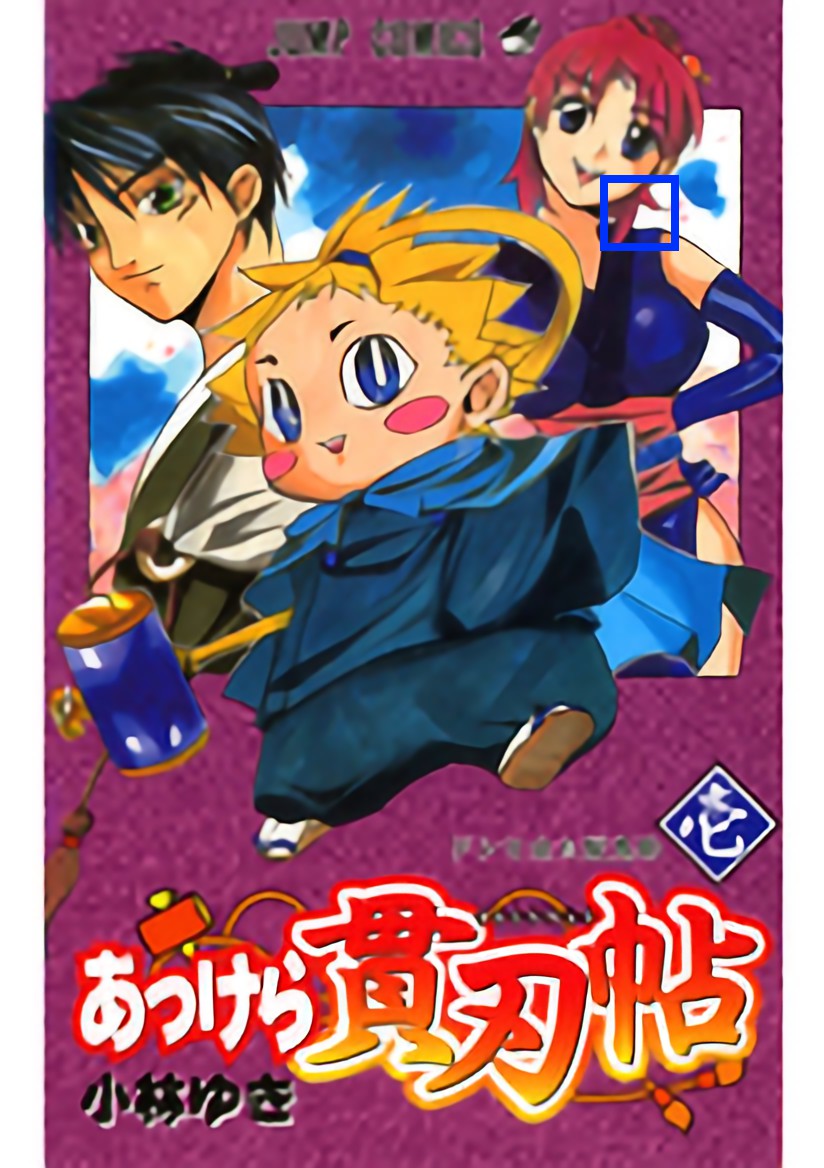} &
			\includegraphics[width = 0.1\textwidth]{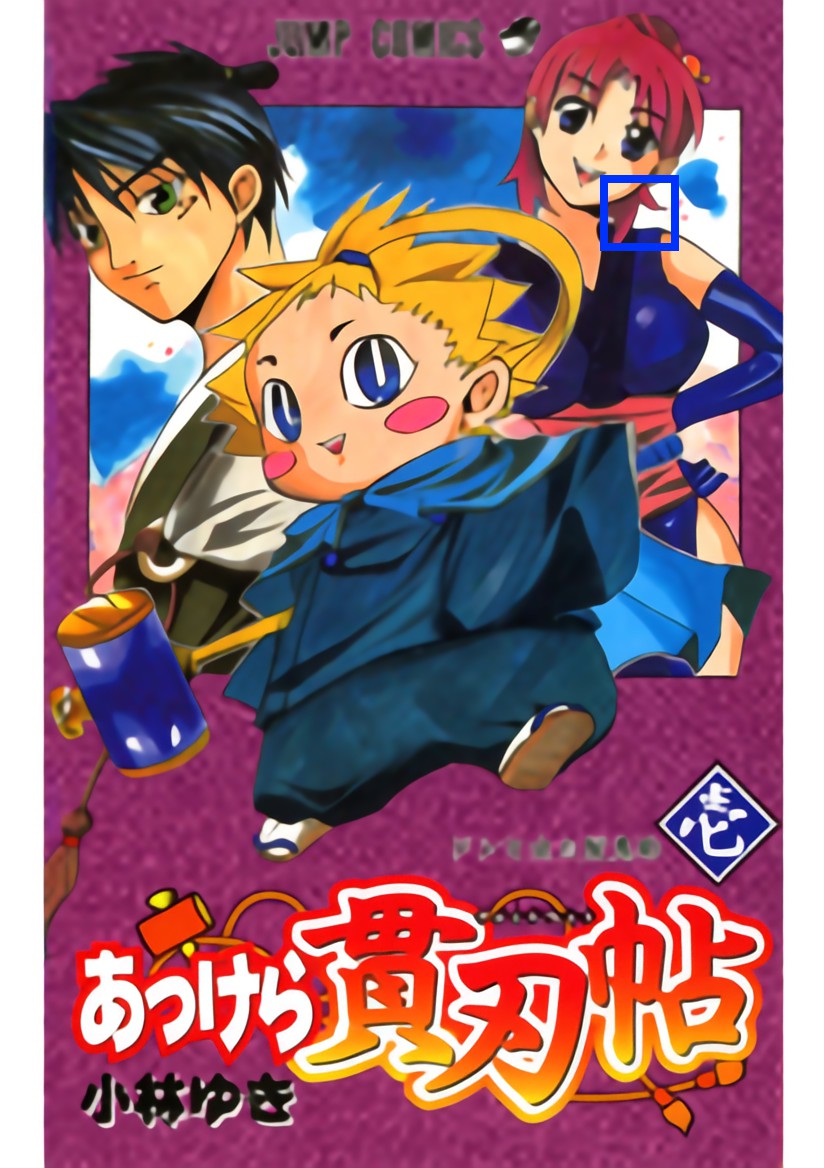}& 
			\includegraphics[width = 0.1\textwidth]{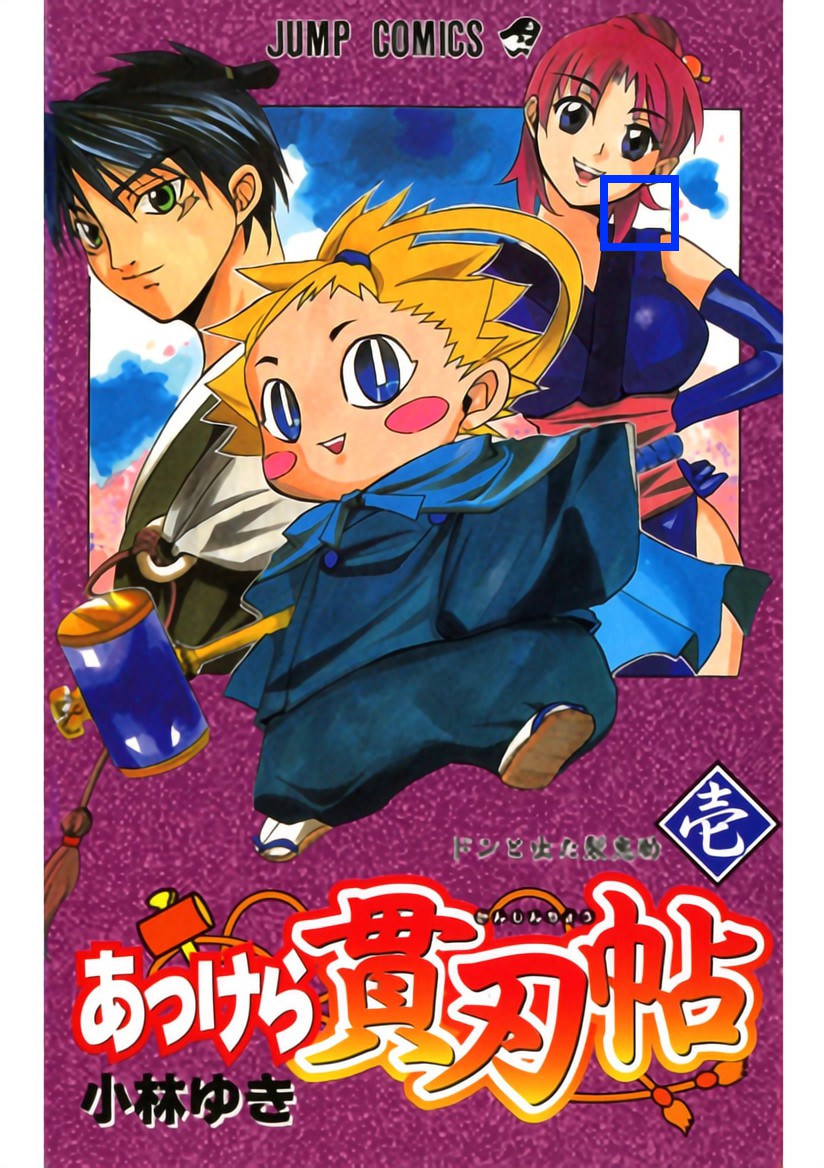} \\ 

			\includegraphics[width = 0.1\textwidth]{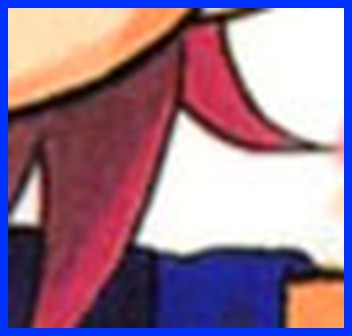}&
			\includegraphics[width = 0.1\textwidth]{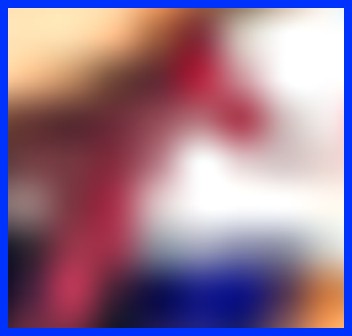} & 
			\includegraphics[width = 0.1\textwidth]{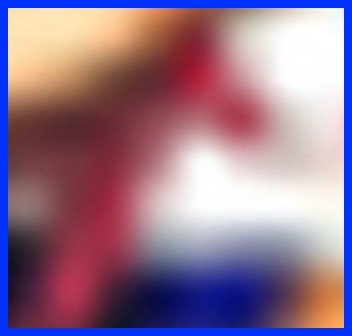} &
			\includegraphics[width = 0.1\textwidth]{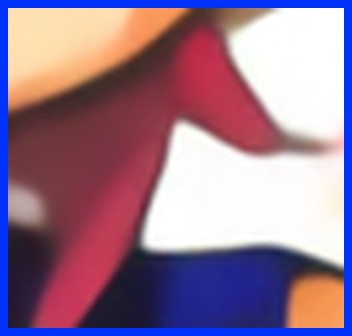} & 
			\includegraphics[width = 0.1\textwidth]{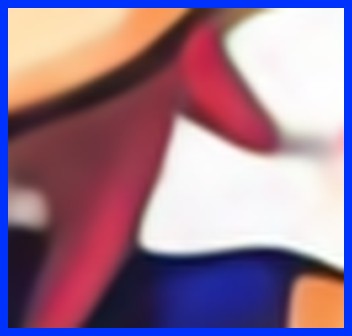} &
			\includegraphics[width = 0.1\textwidth]{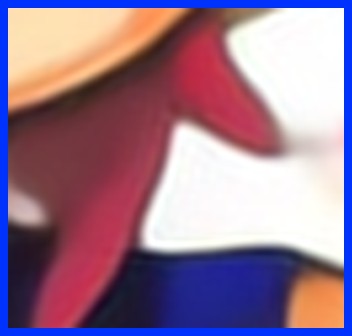} &
			\includegraphics[width = 0.1\textwidth]{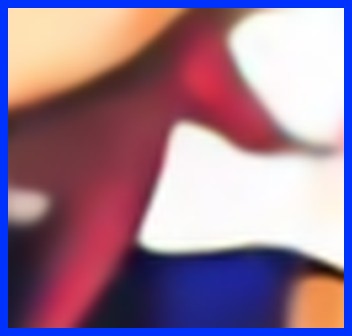} & 
			\includegraphics[width = 0.1\textwidth]{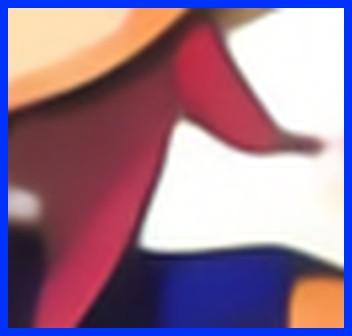}& 
			
			\includegraphics[width = 0.1\textwidth]{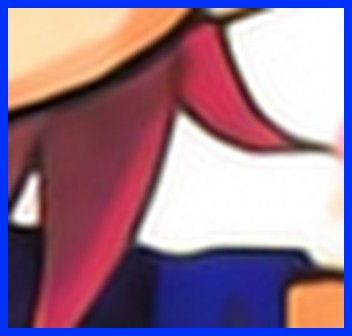} \\ 

			PSNR/SSIM & 20.09/0.525 &21.07/0.523 &23.79/0.700 &23.47/0.688 &23.87/0.703 &23.12/ 0.673 &24.00/0.708 &\textbf{24.24}/\textbf{0.746} \\

		\end{tabular}
	\end{center}

	\caption{Visual comparison for 8$\times$ SR with BI model on the Manga109 dataset.The best results are highlighted 
	}

	\label{fig-BD8}
\end{figure}

\begin{figure}[t]\tiny
	\begin{center}
		\tabcolsep 1pt
		\begin{tabular}{@{}ccccccccc@{}}
			
			HR&
			Bicubic & 
			VDSR~\cite{kim2016accurate} & 
			EDSR~\cite{lim2017enhanced} & 
			RCAN~\cite{zhang2018image}  &
			SRFBN~\cite{li2019feedback} &
			SAN~\cite{dai2019second} &
			HAN(our) &
			HAN+(our) \\
			\includegraphics[width = 0.1\textwidth]{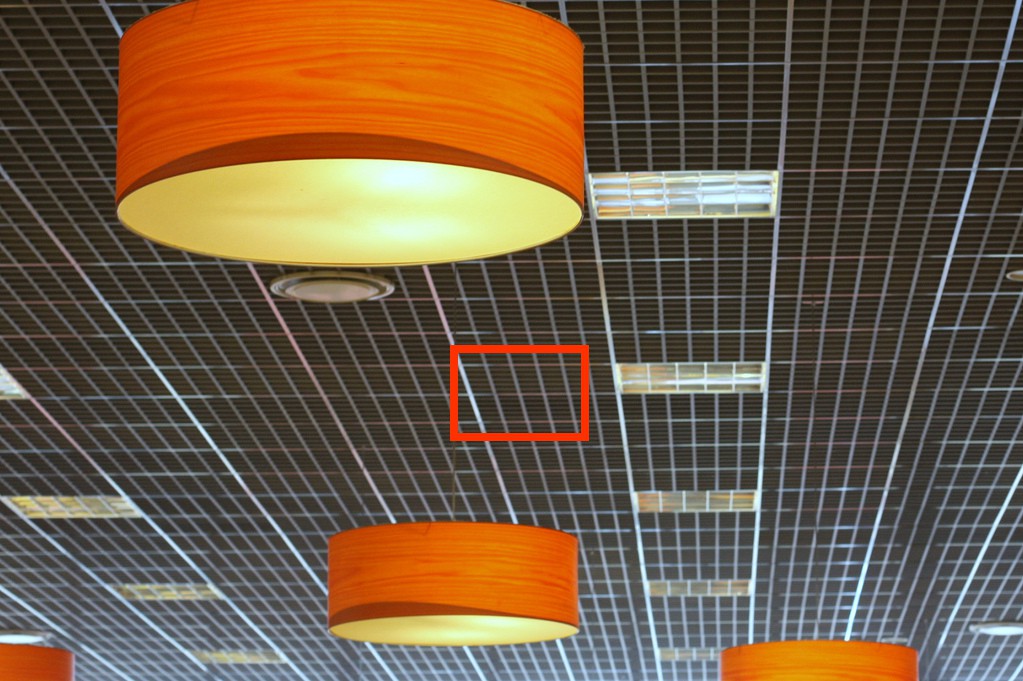}&
			\includegraphics[width = 0.1\textwidth]{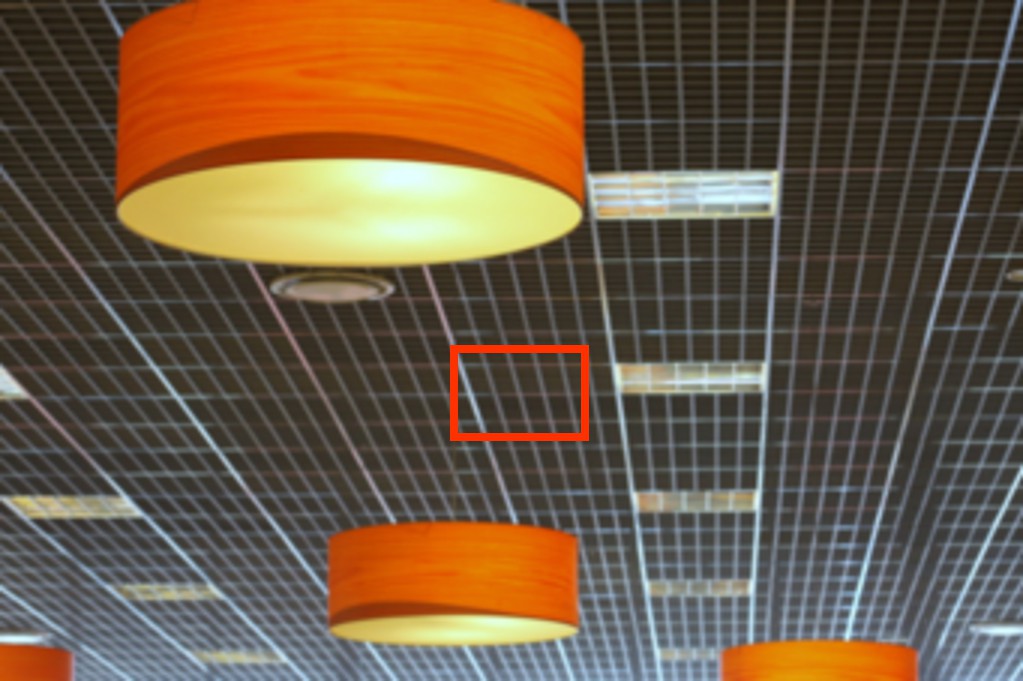} & 
			\includegraphics[width = 0.1\textwidth]{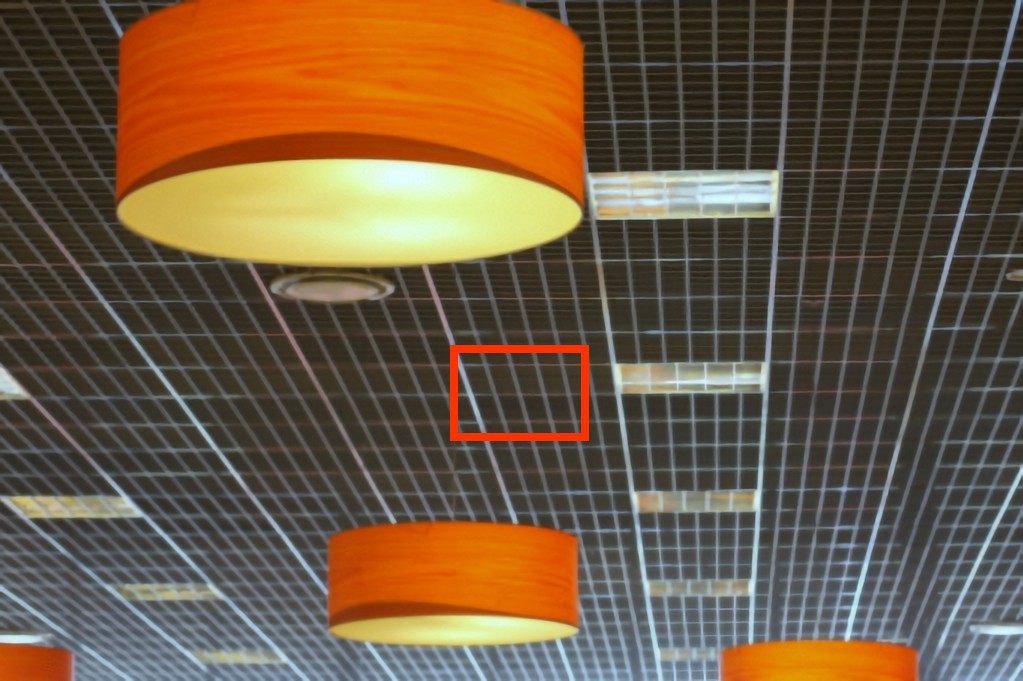} &
			\includegraphics[width = 0.1\textwidth]{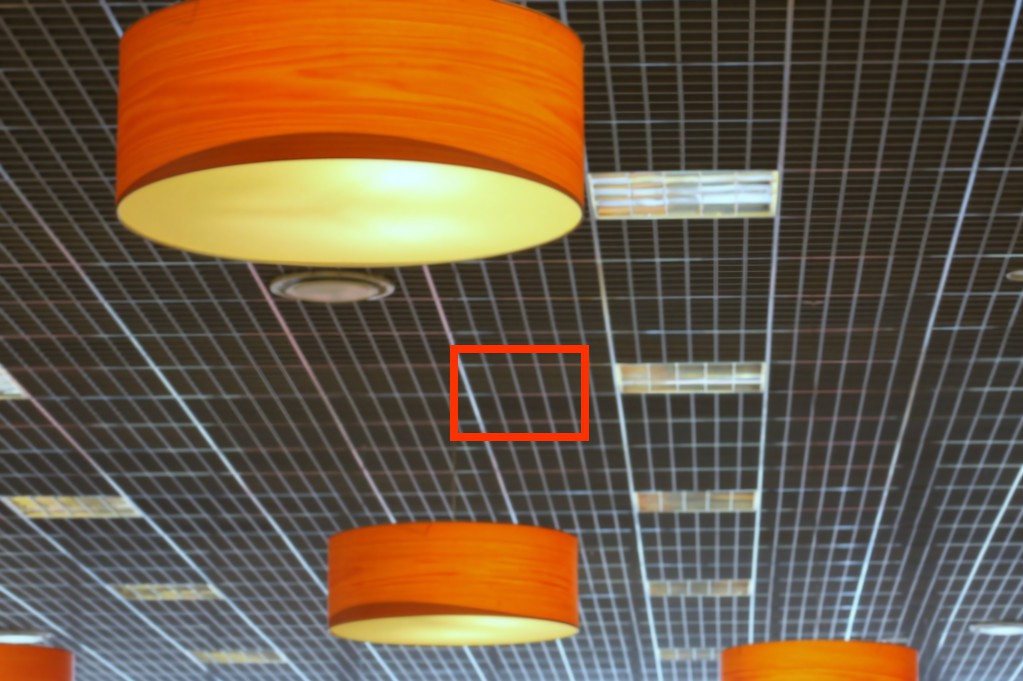} & 
			\includegraphics[width = 0.1\textwidth]{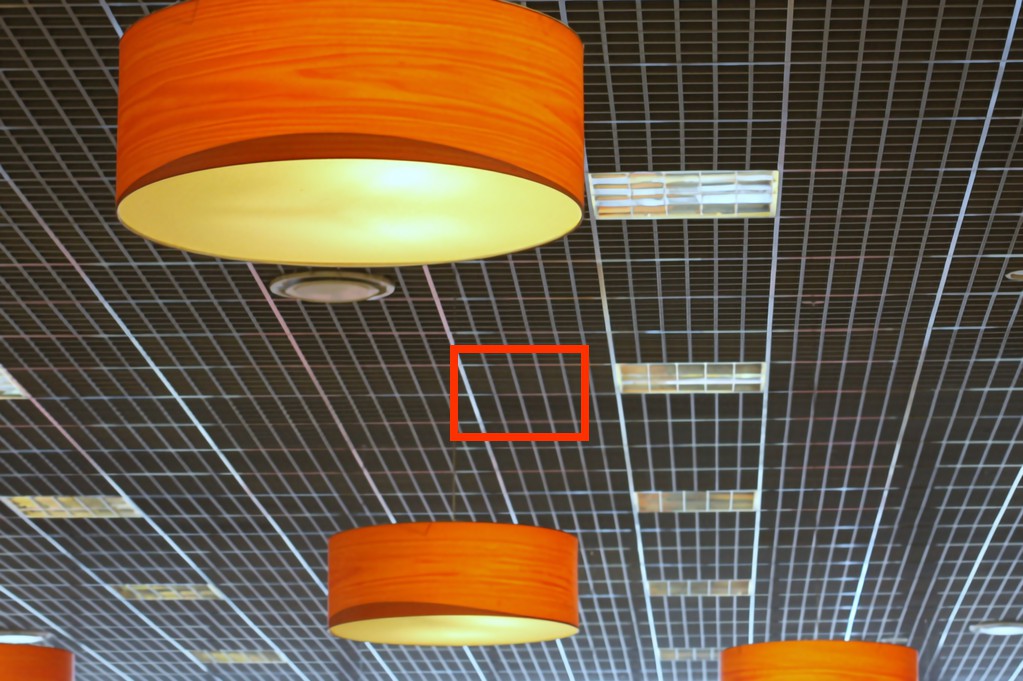} & 
			\includegraphics[width = 0.1\textwidth]{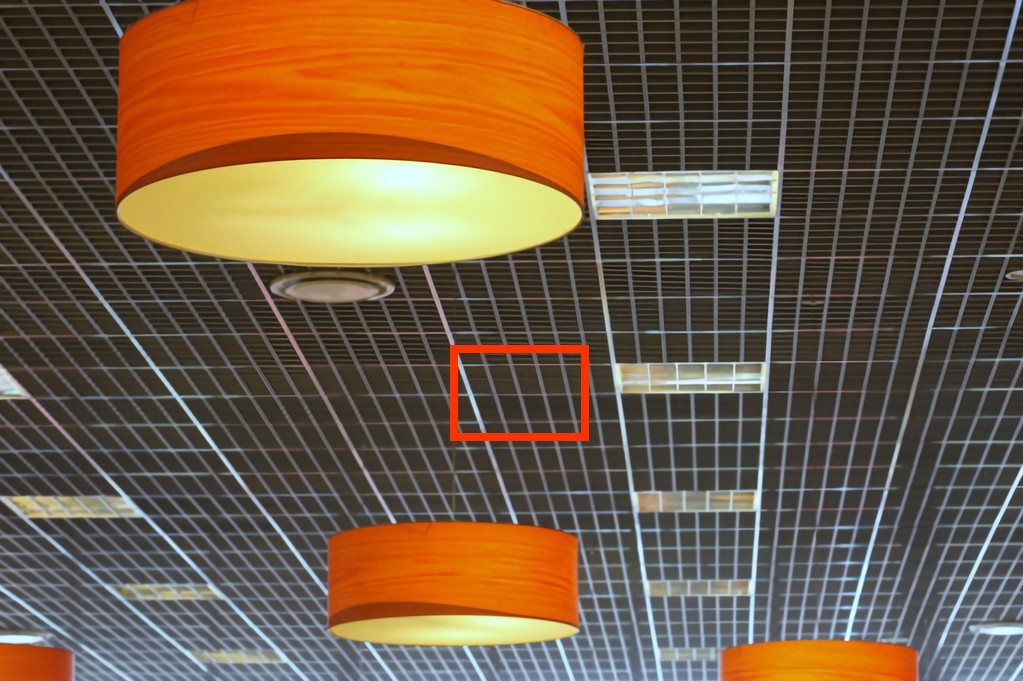}& 
			\includegraphics[width = 0.1\textwidth]{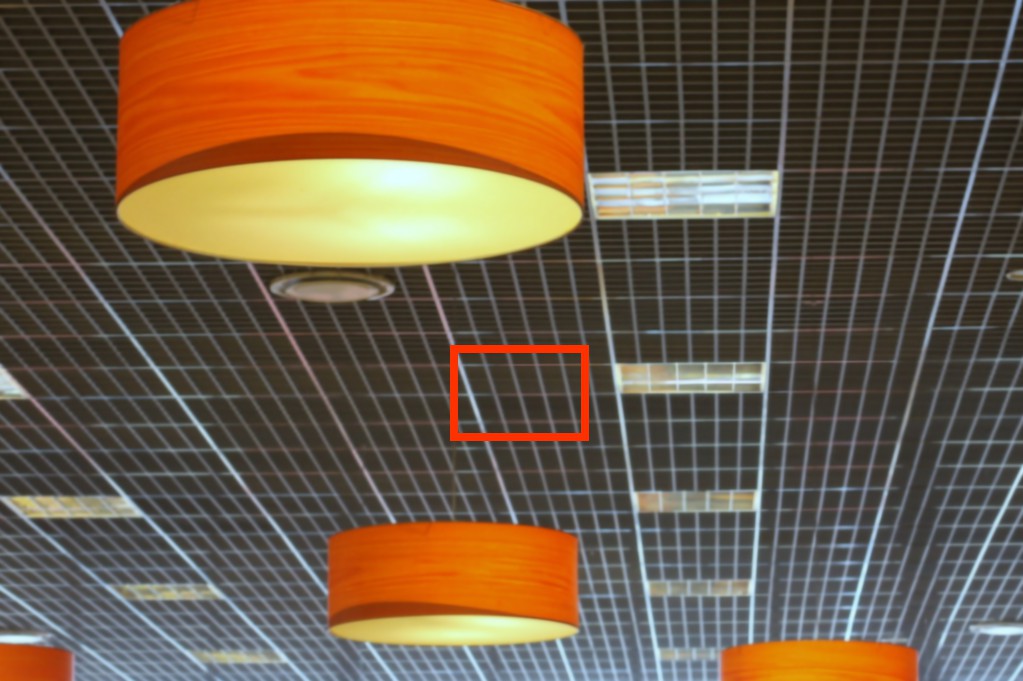} & 
			\includegraphics[width = 0.1\textwidth]{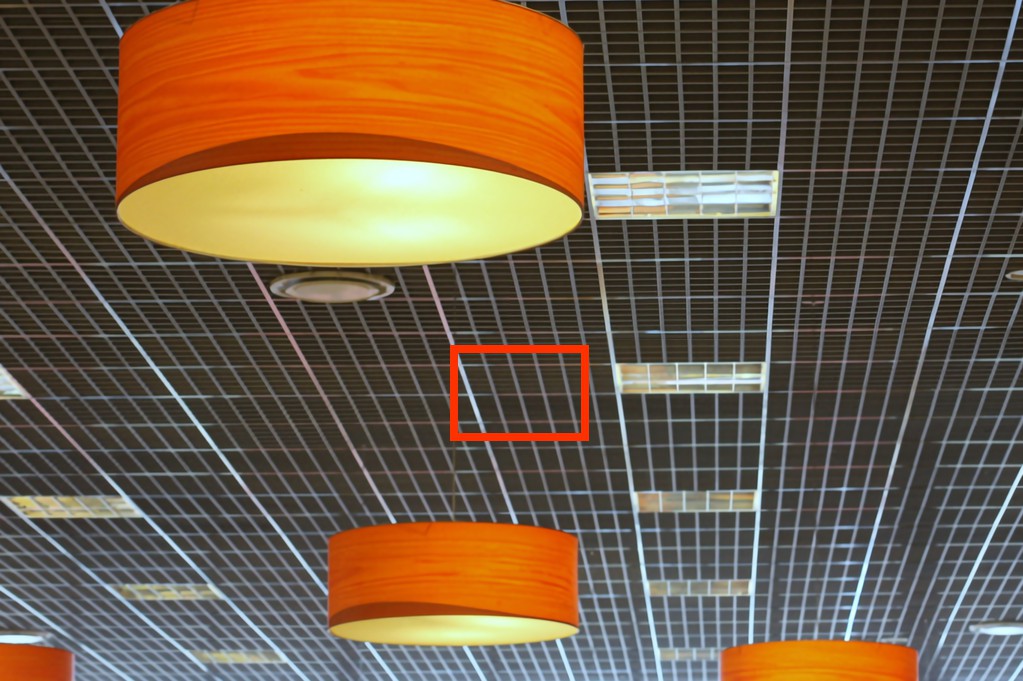} & 
			\includegraphics[width = 0.1\textwidth]{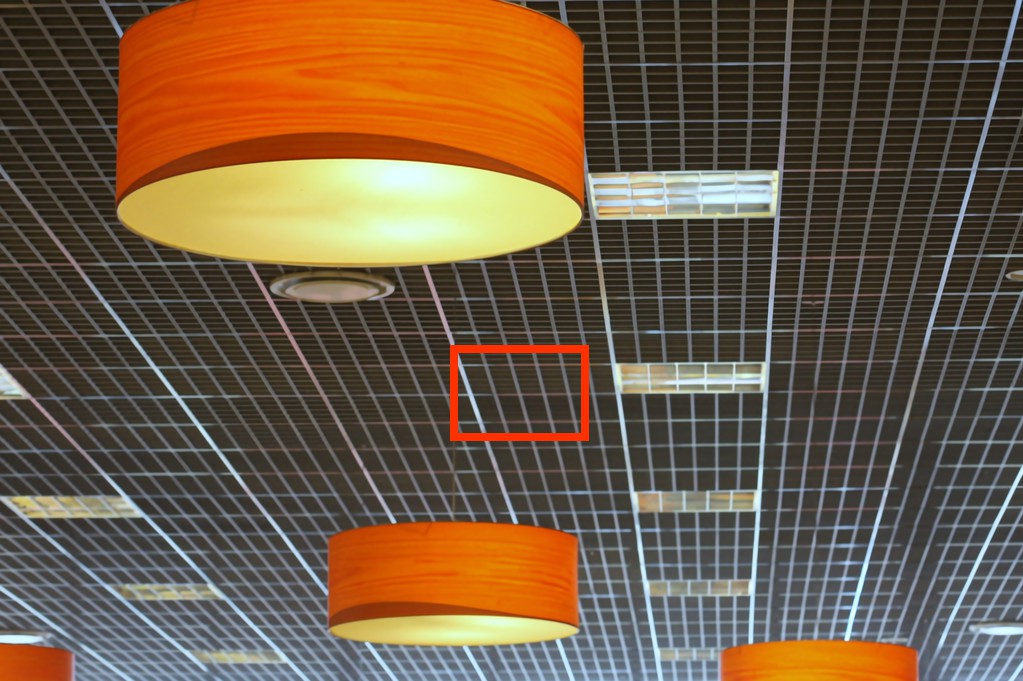} \\
			
			\includegraphics[width = 0.1\textwidth]{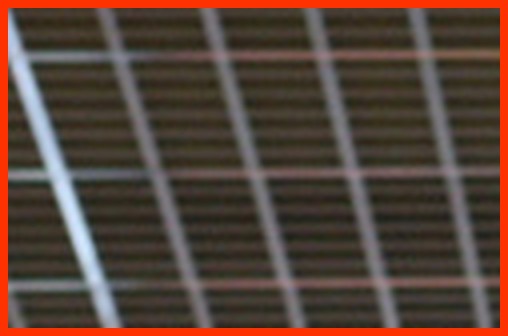}&
			\includegraphics[width = 0.1\textwidth]{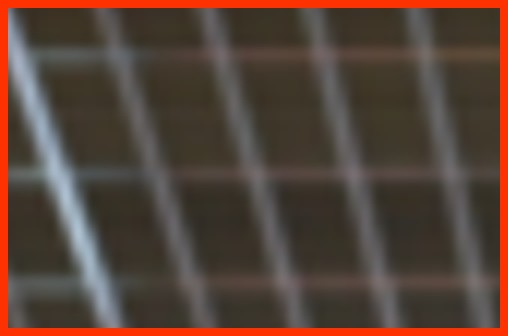} & 
			\includegraphics[width = 0.1\textwidth]{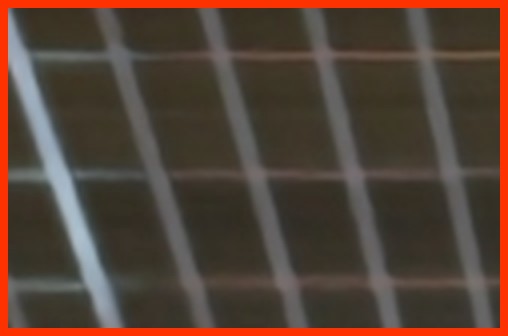} &
			\includegraphics[width = 0.1\textwidth]{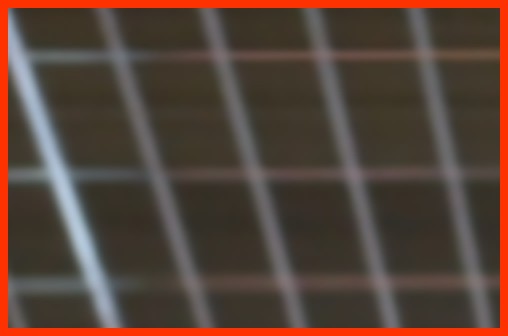} & 
			\includegraphics[width = 0.1\textwidth]{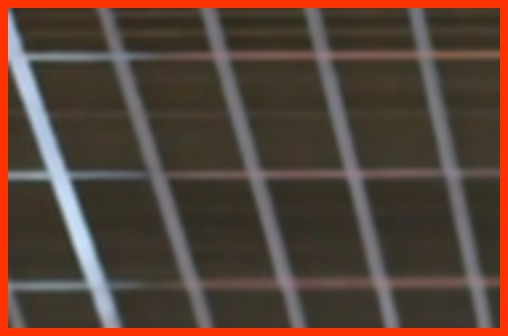} & 
			\includegraphics[width = 0.1\textwidth]{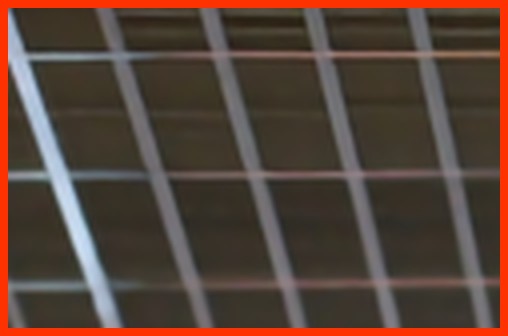}& 
			\includegraphics[width = 0.1\textwidth]{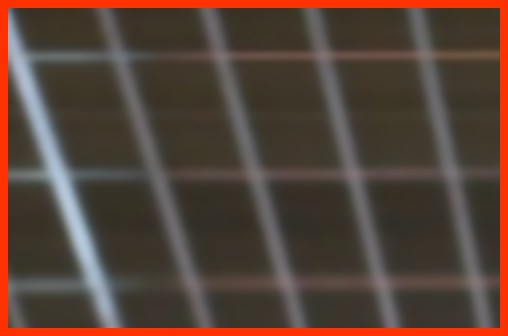} & 
			\includegraphics[width = 0.1\textwidth]{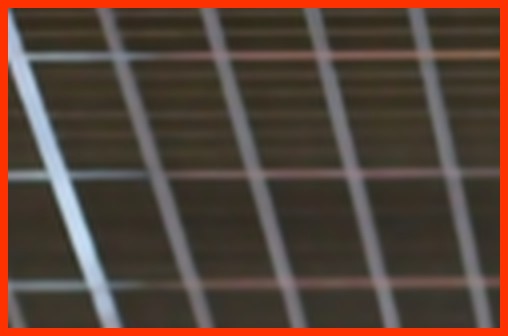} & 
			\includegraphics[width = 0.1\textwidth]{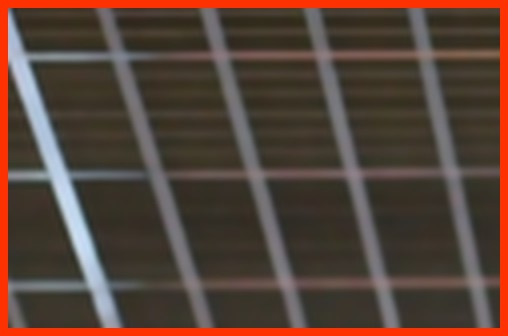} \\
			
			PSNR/SSIM & 27.70/ 0.774 &30.10/0.854 &30.64/0.878 & 36.39/0.951 &30.75/.879 &34.31/0.930 &36.44/ 0.955 &\textbf{36.62}/\textbf{0.956}\\
			
			\includegraphics[width = 0.1\textwidth]{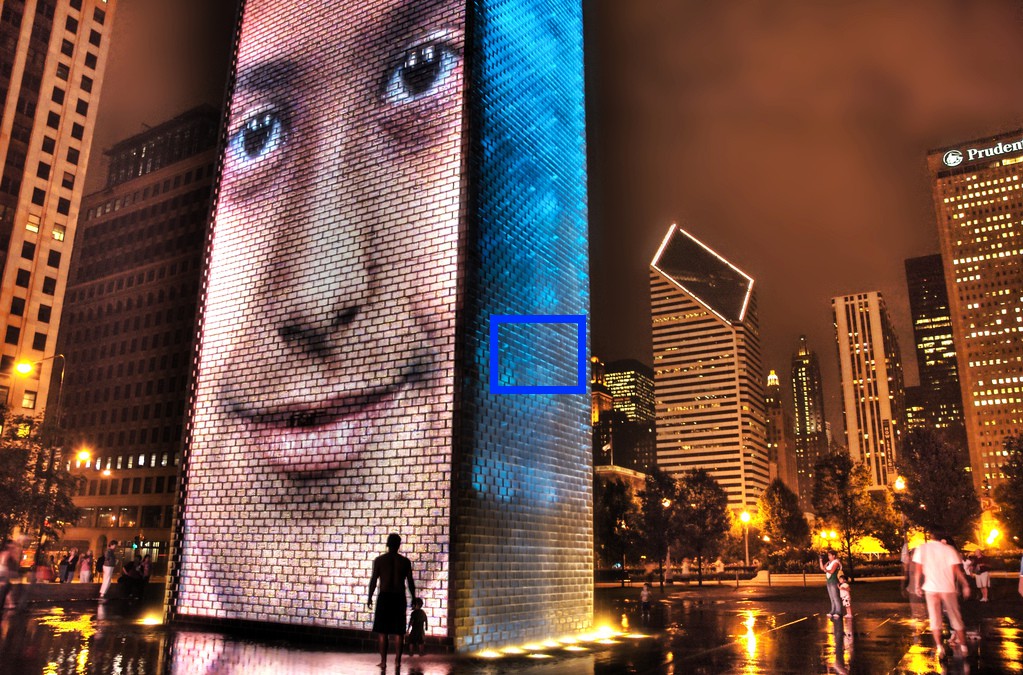}&
			\includegraphics[width = 0.1\textwidth]{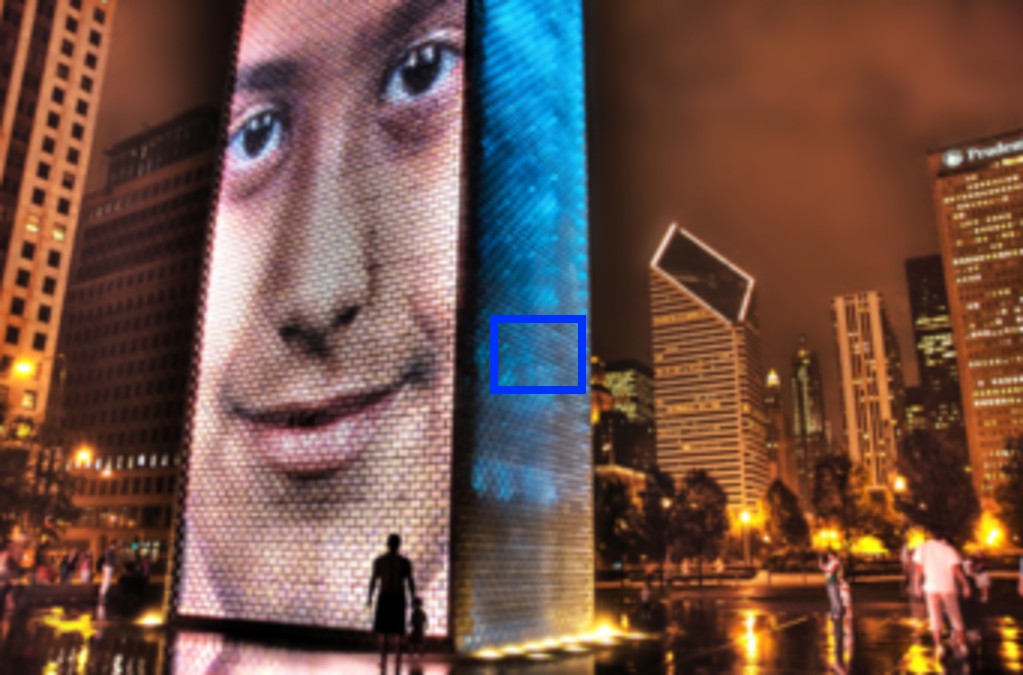} & 
			\includegraphics[width = 0.1\textwidth]{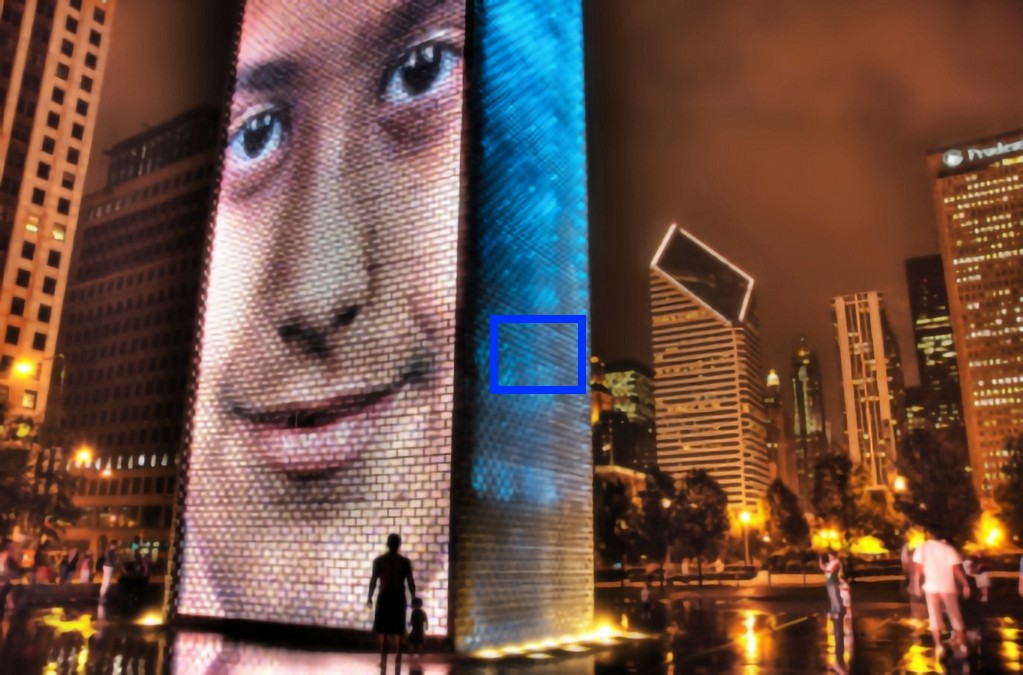} &
			\includegraphics[width = 0.1\textwidth]{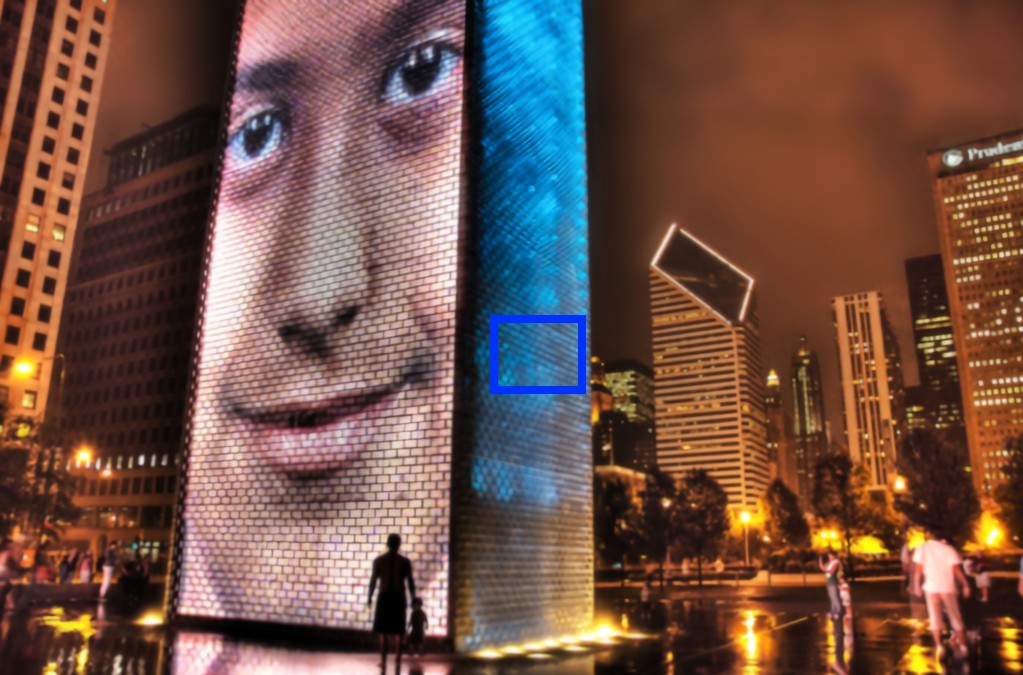} & 
			\includegraphics[width = 0.1\textwidth]{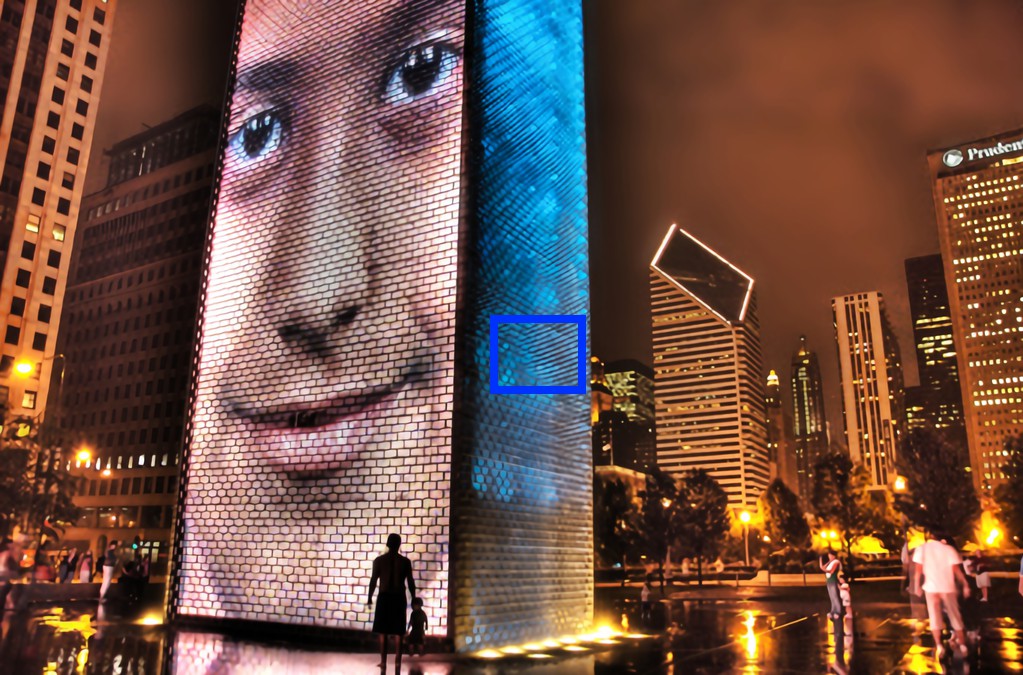} & 
			\includegraphics[width = 0.1\textwidth]{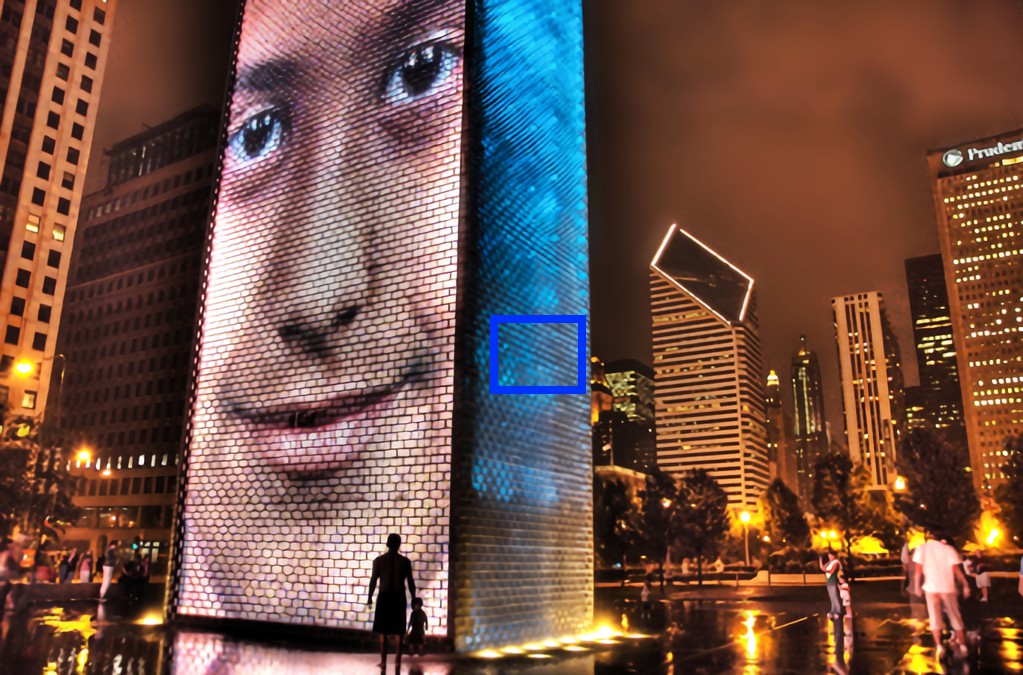}& 
			\includegraphics[width = 0.1\textwidth]{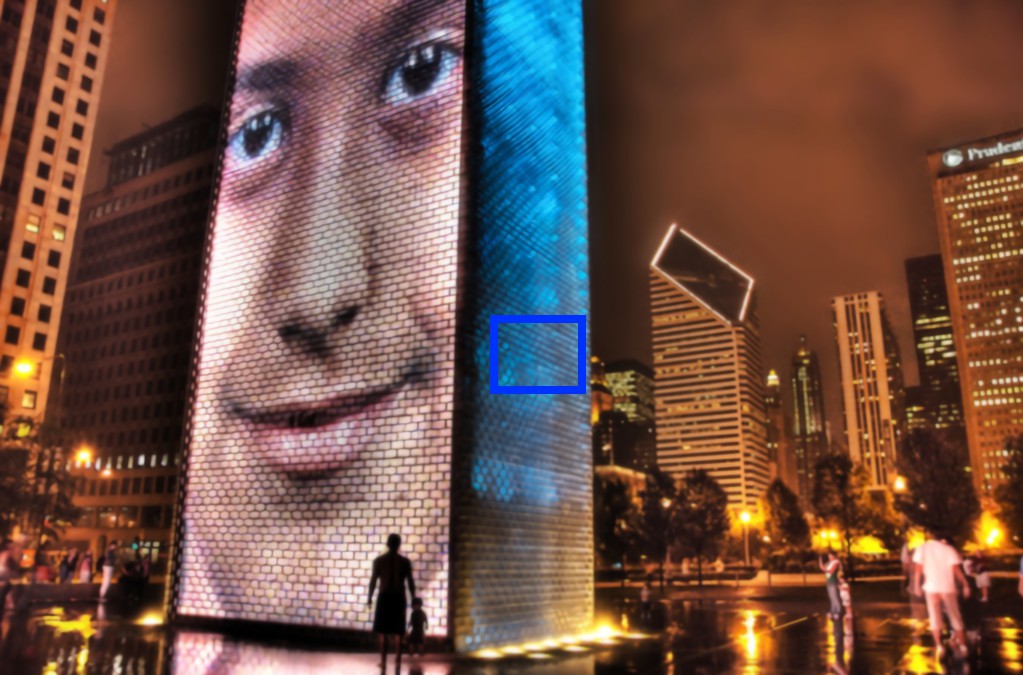} & 
			\includegraphics[width = 0.1\textwidth]{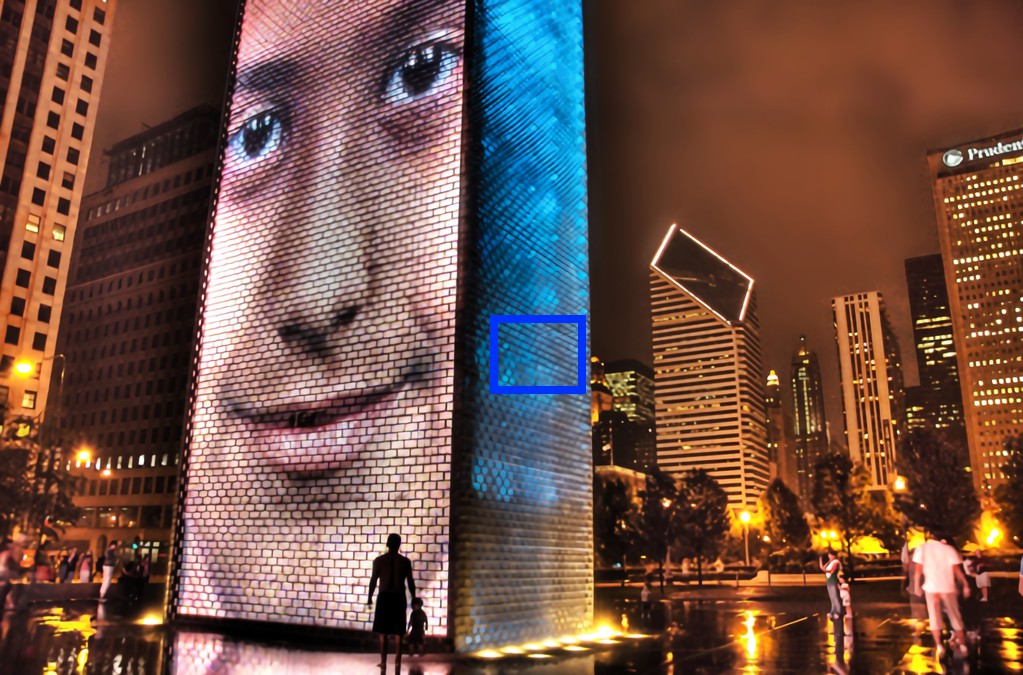} & 
			\includegraphics[width = 0.1\textwidth]{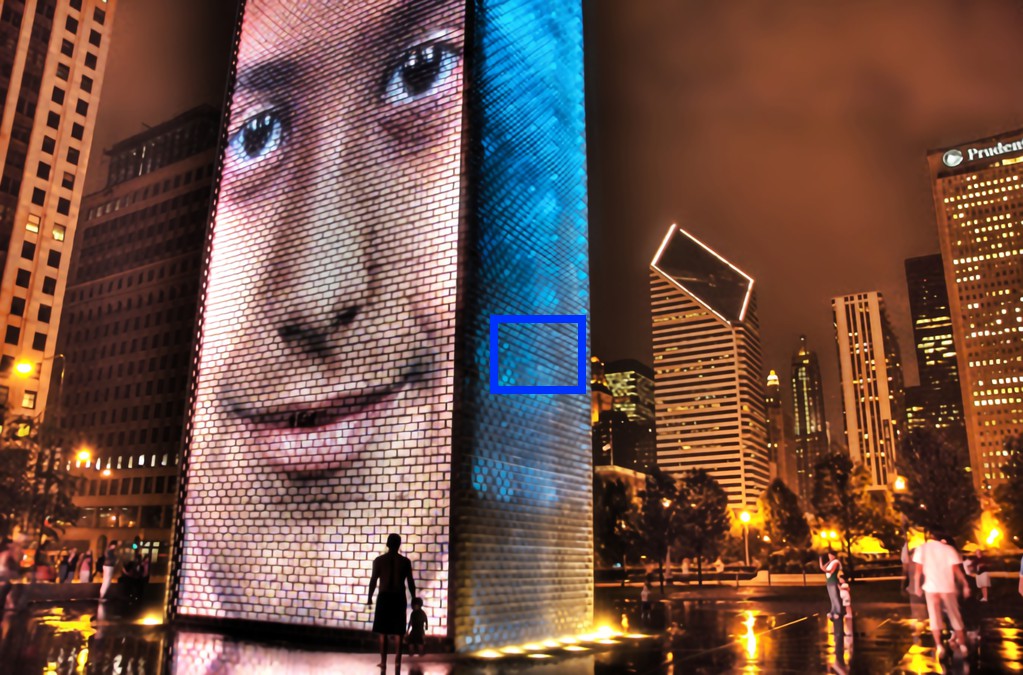} \\
			
			\includegraphics[width = 0.1\textwidth]{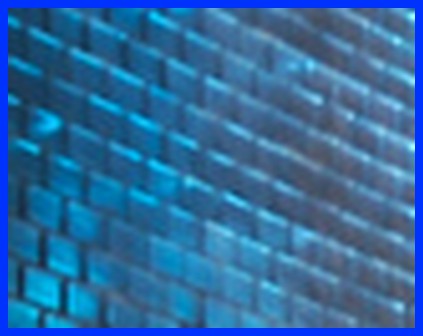}&
			\includegraphics[width = 0.1\textwidth]{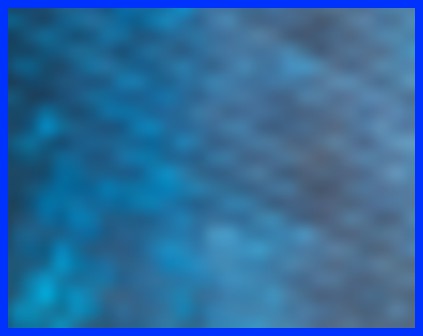} & 
			\includegraphics[width = 0.1\textwidth]{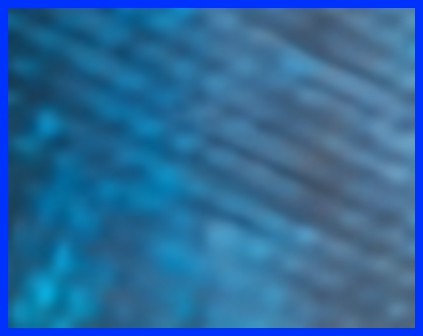} &
			\includegraphics[width = 0.1\textwidth]{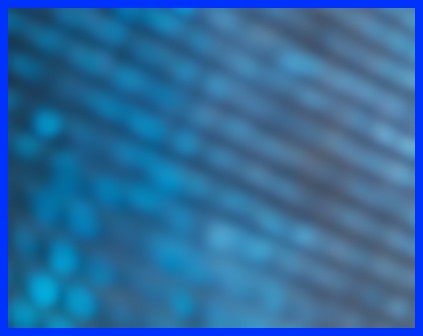} & 
			\includegraphics[width = 0.1\textwidth]{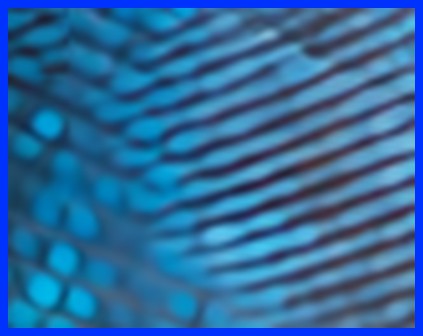} & 
			\includegraphics[width = 0.1\textwidth]{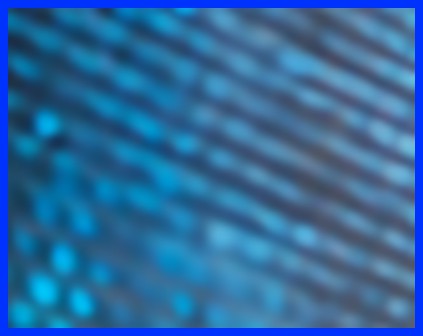}& 
			\includegraphics[width = 0.1\textwidth]{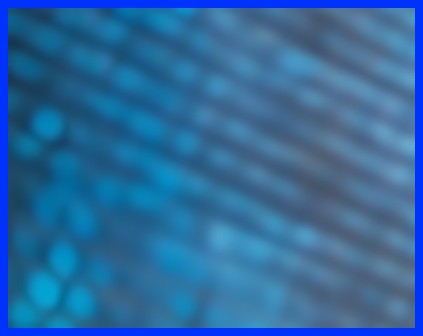} & 
			\includegraphics[width = 0.1\textwidth]{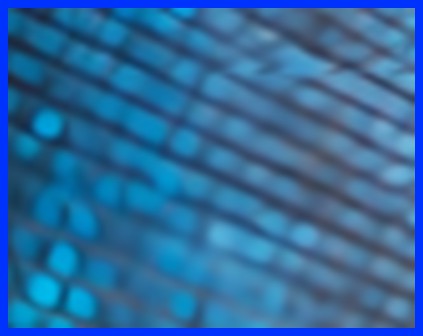} & 
			\includegraphics[width = 0.1\textwidth]{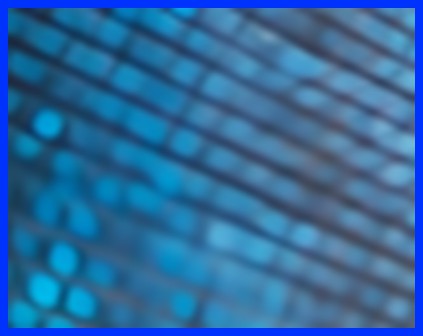} \\
			
			PSNR/SSIM & 22.17/0.674 &23.39/ 0.747 &24.19/0.785 & 27.18/0.882 &24.20/0.788 &26.56/0.873 &27.40/0.889 & \textbf{27.67}/\textbf{0.893}\\
			
			\includegraphics[width = 0.1\textwidth]{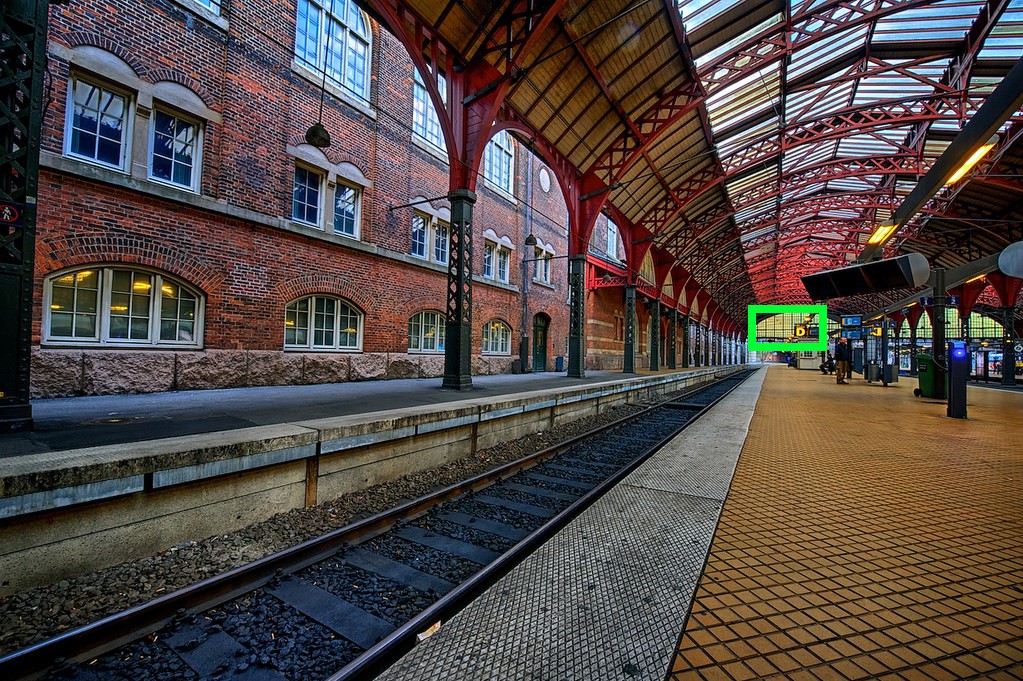}&
			\includegraphics[width = 0.1\textwidth]{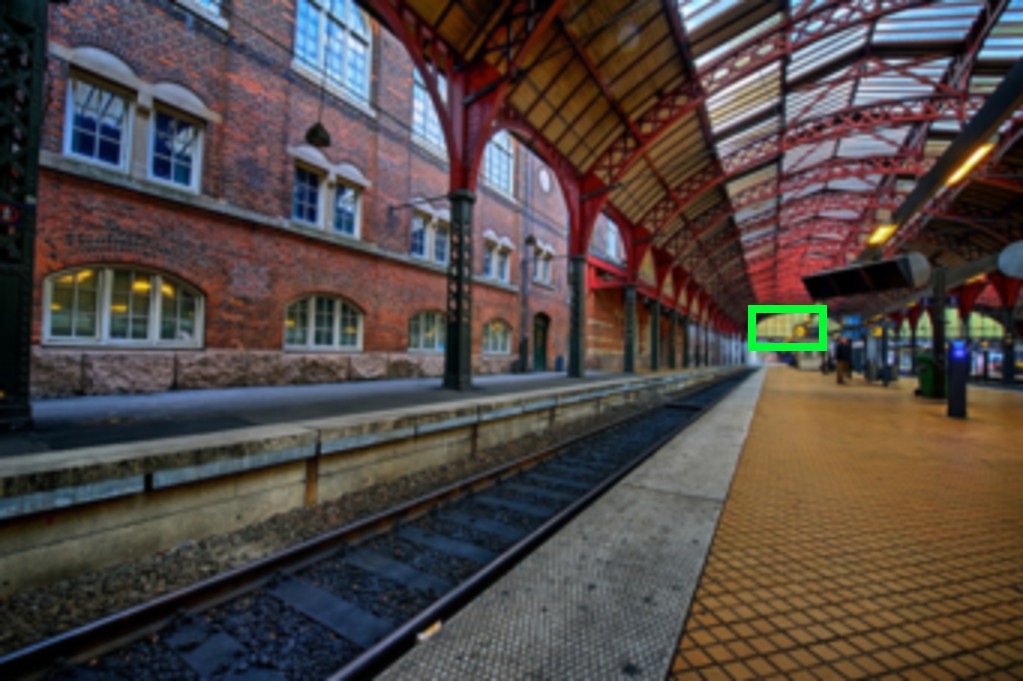} & 
			\includegraphics[width = 0.1\textwidth]{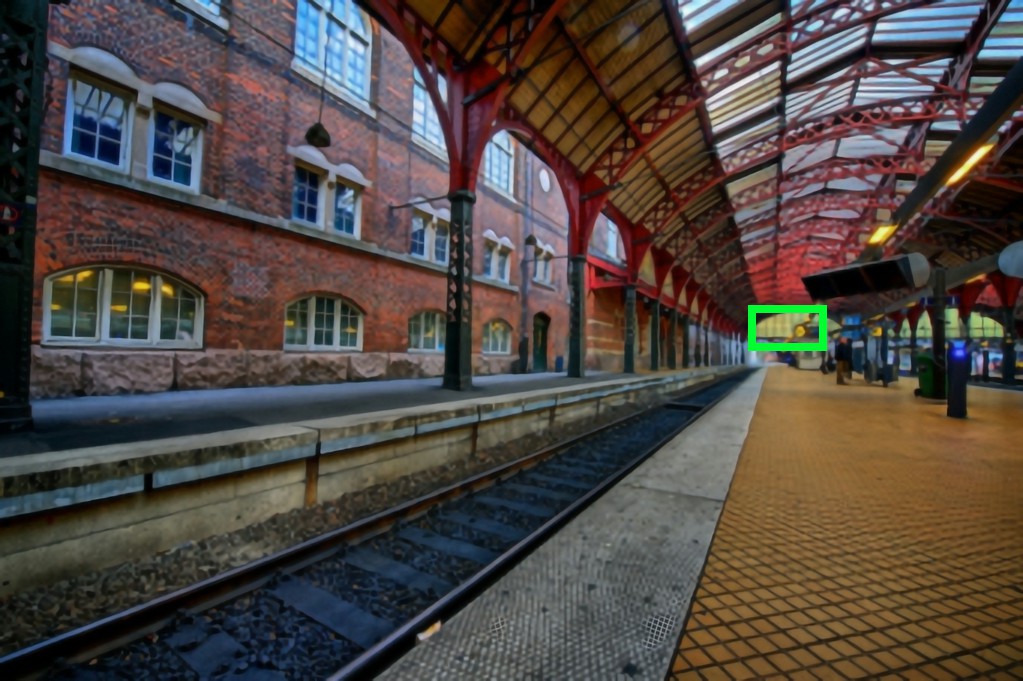} &
			\includegraphics[width = 0.1\textwidth]{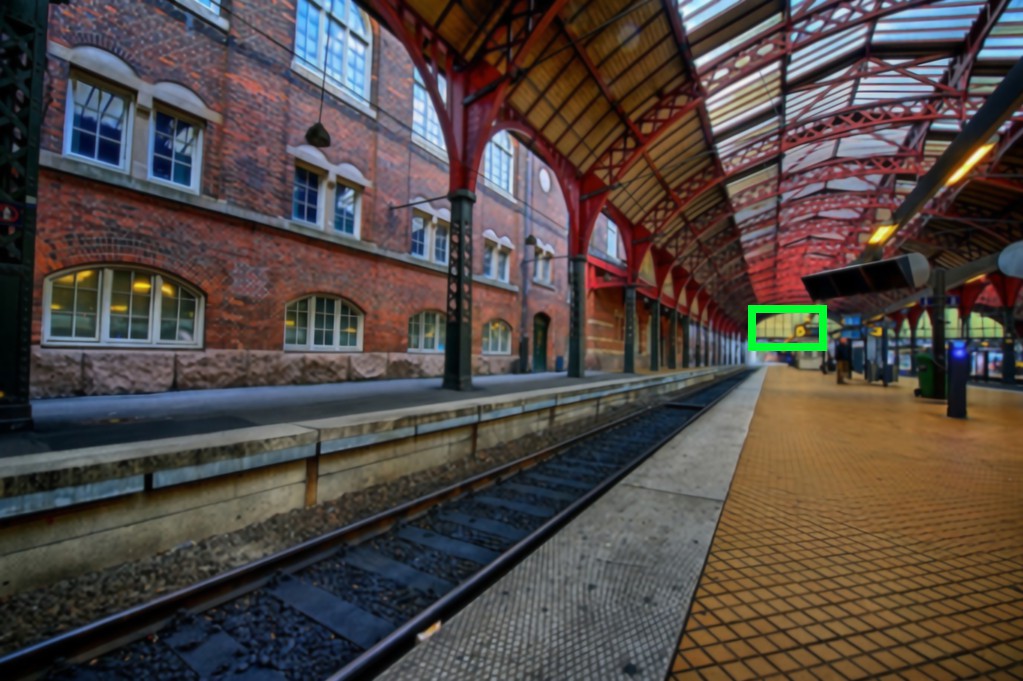} & 
			\includegraphics[width = 0.1\textwidth]{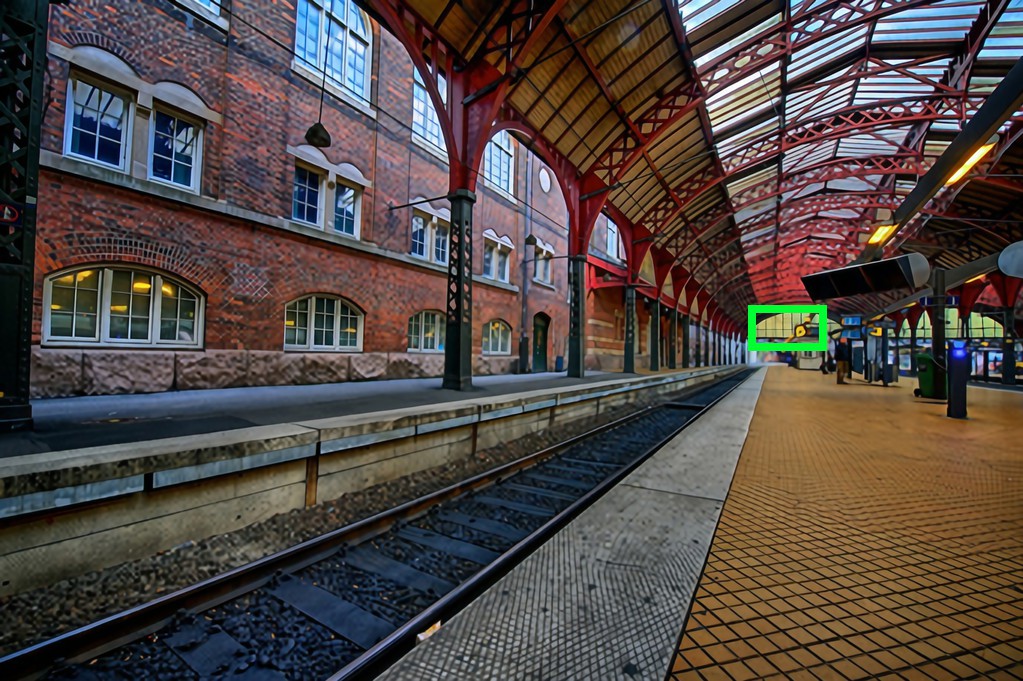} & 
			\includegraphics[width = 0.1\textwidth]{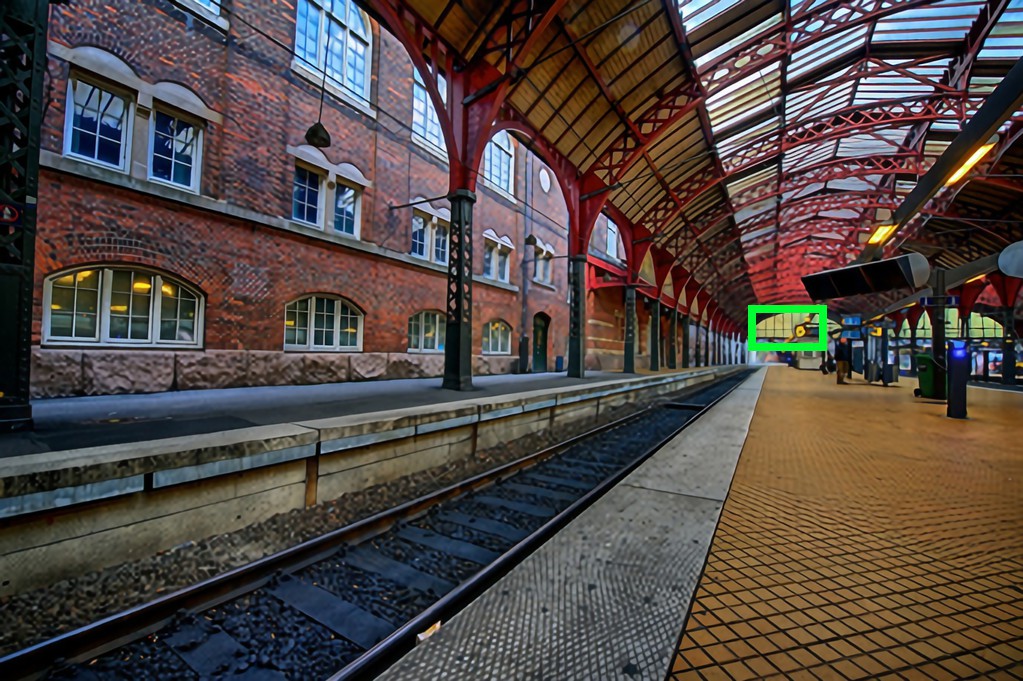}& 
			\includegraphics[width = 0.1\textwidth]{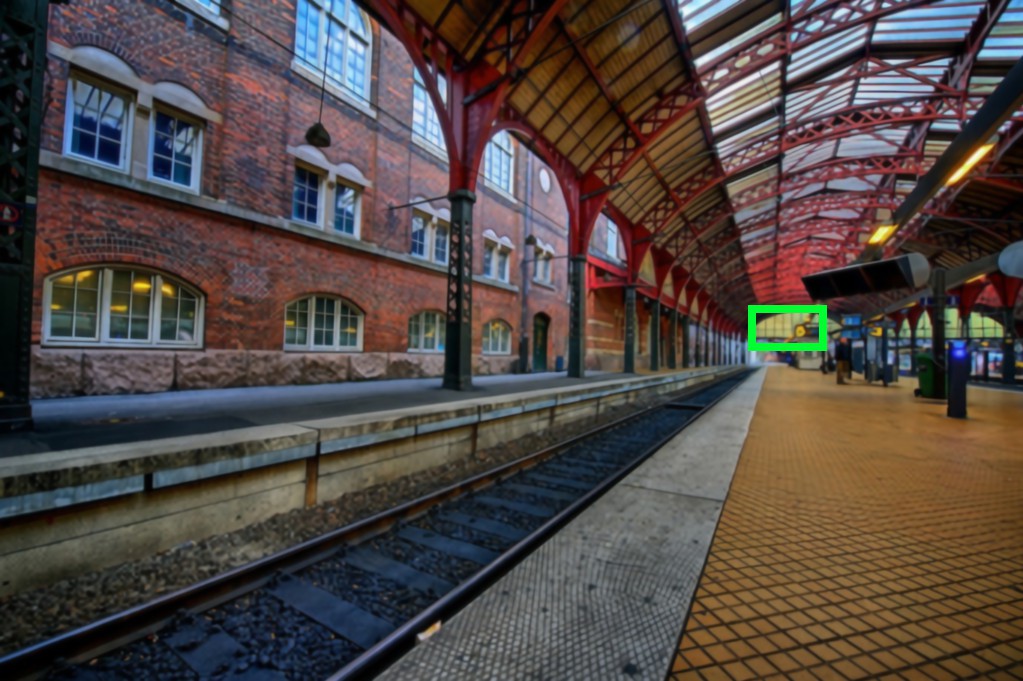} & 
			\includegraphics[width = 0.1\textwidth]{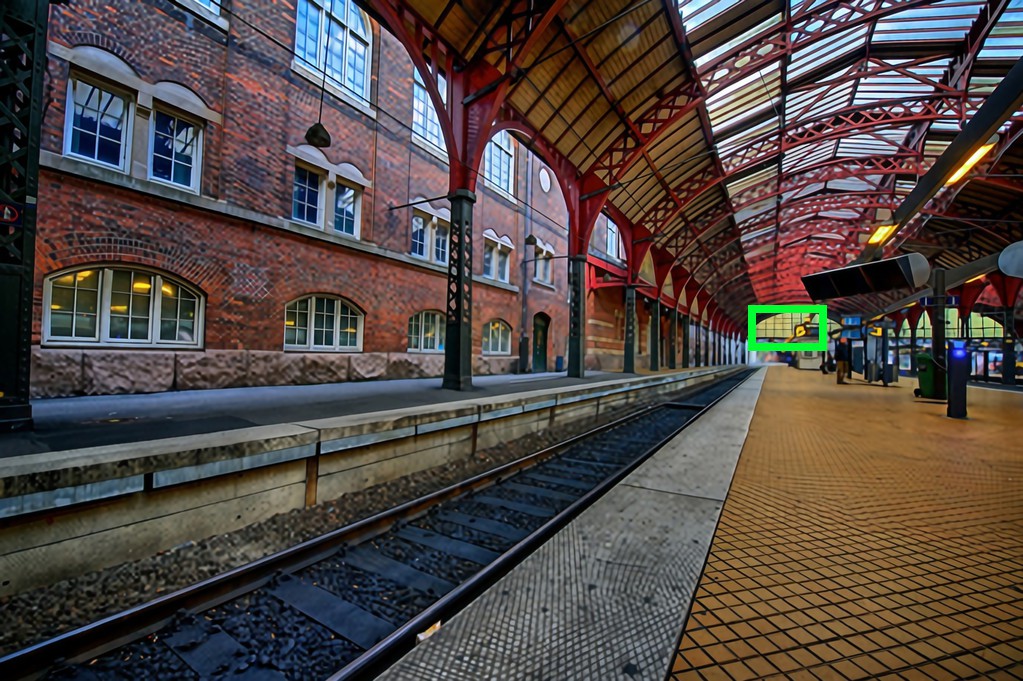} & 
			\includegraphics[width = 0.1\textwidth]{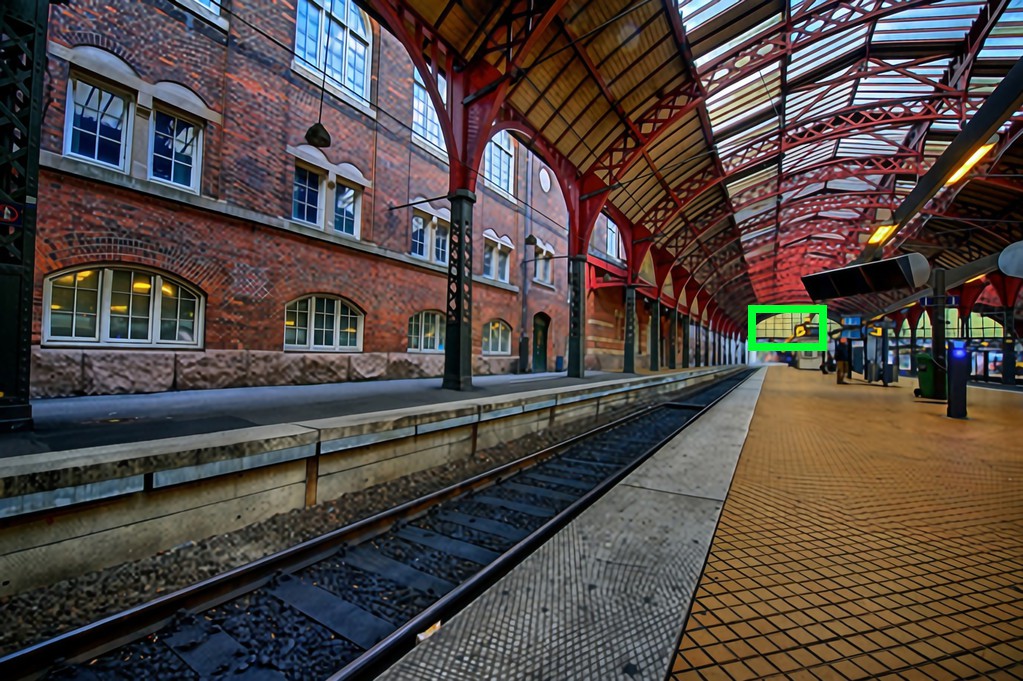} \\
			
			\includegraphics[width = 0.1\textwidth]{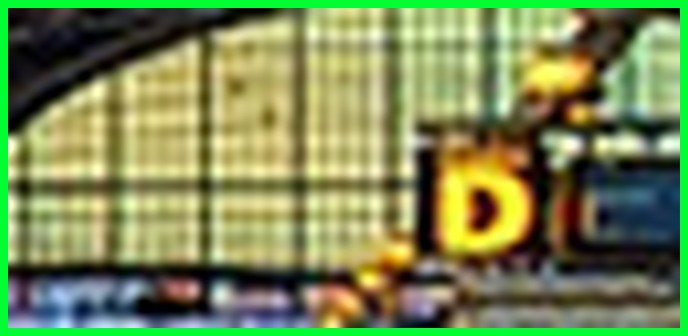}&
			\includegraphics[width = 0.1\textwidth]{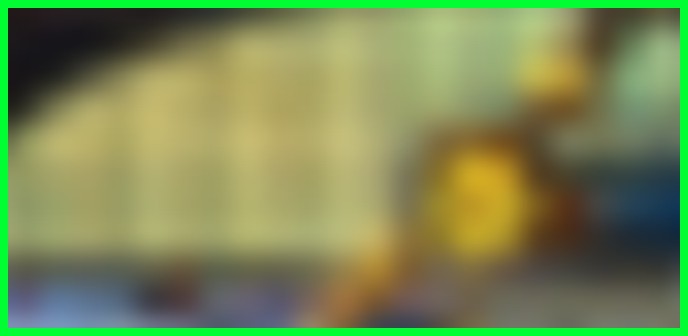} & 
			\includegraphics[width = 0.1\textwidth]{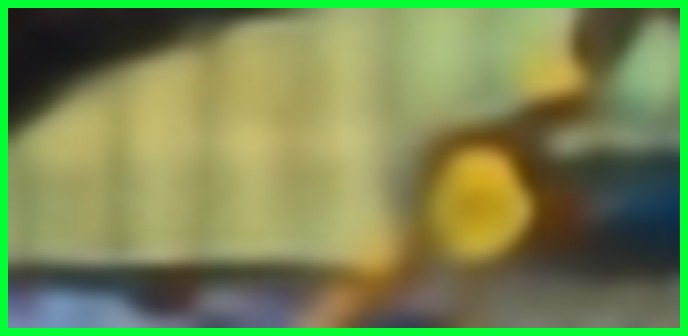} &
			\includegraphics[width = 0.1\textwidth]{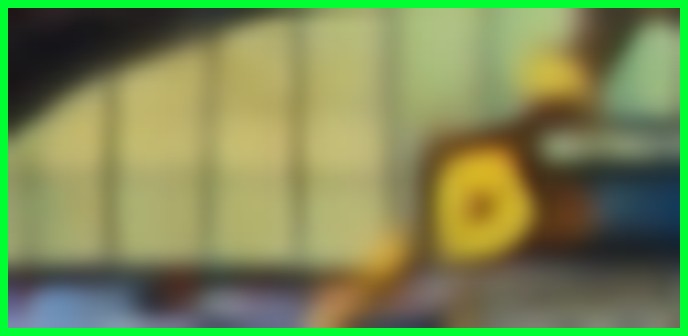} & 
			\includegraphics[width = 0.1\textwidth]{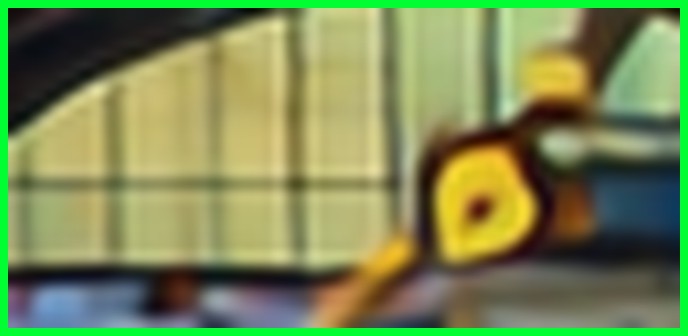} & 
			\includegraphics[width = 0.1\textwidth]{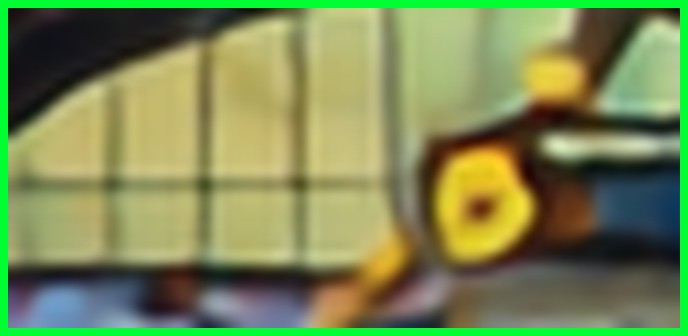}& 
			\includegraphics[width = 0.1\textwidth]{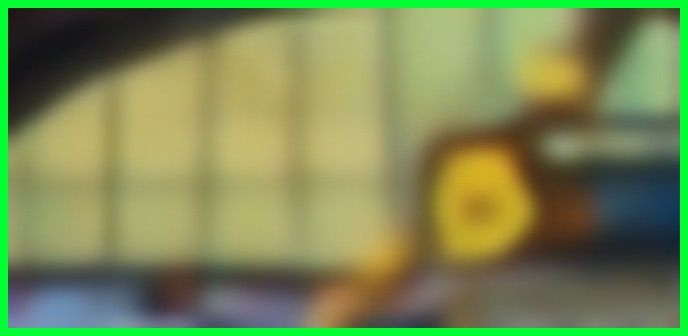} & 
			\includegraphics[width = 0.1\textwidth]{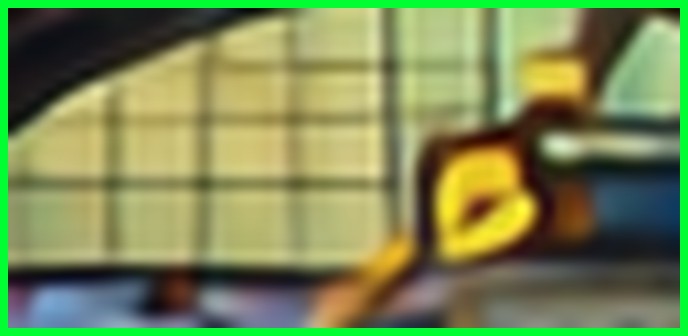} & 
			\includegraphics[width = 0.1\textwidth]{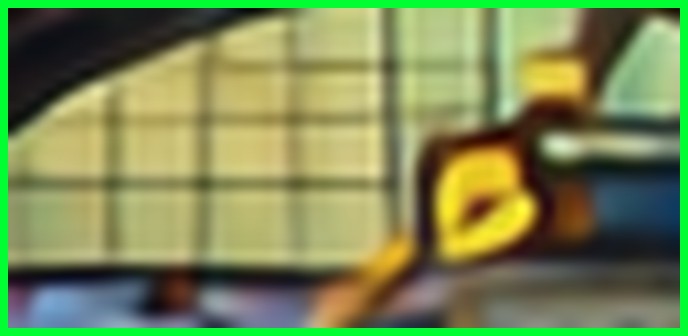} \\

			PSNR/SSIM & 19.93/0.425 &20.66/ 0.508 &20.89/0.531 & 22.34/ 0.675 &20.92/ 0.534 &22.07/0.656 &22.35/.677 &\textbf{22.49}/\textbf{0.681}\\
			
			\includegraphics[width = 0.1\textwidth]{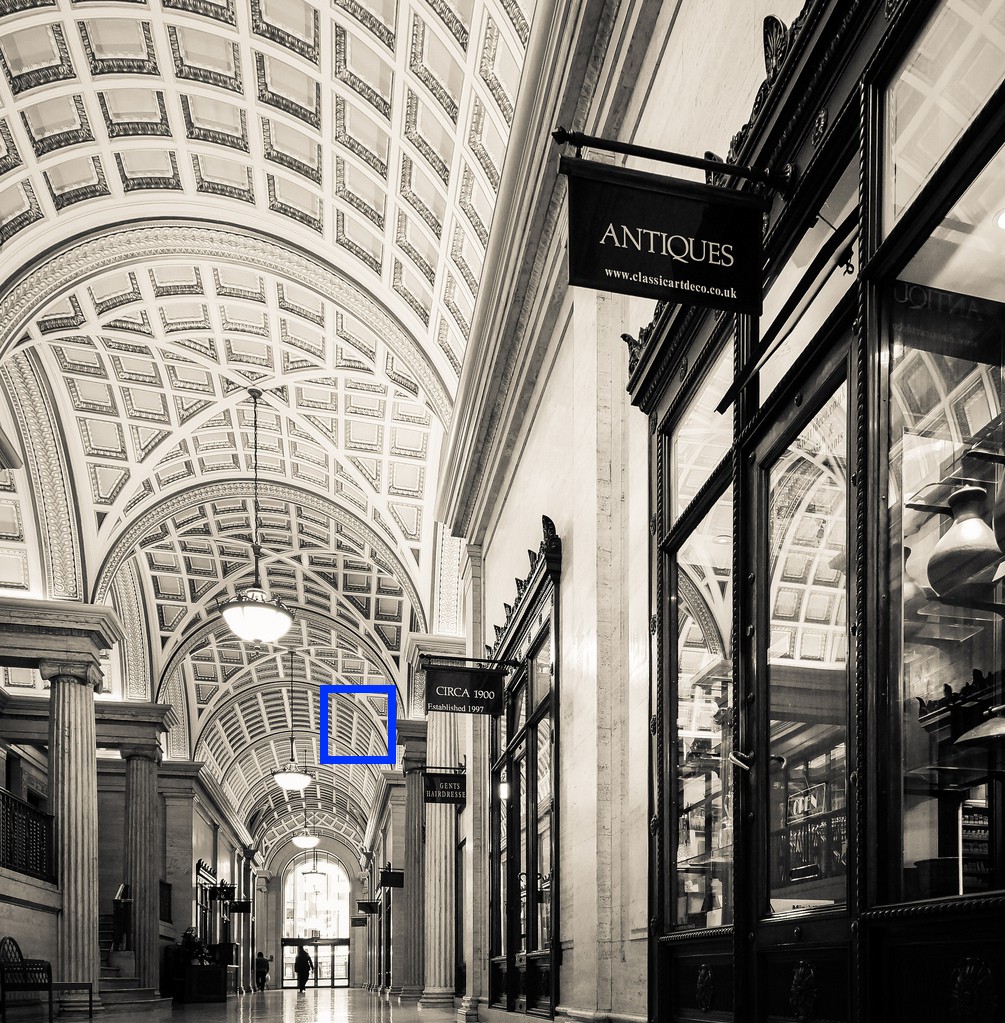}&
			\includegraphics[width = 0.1\textwidth]{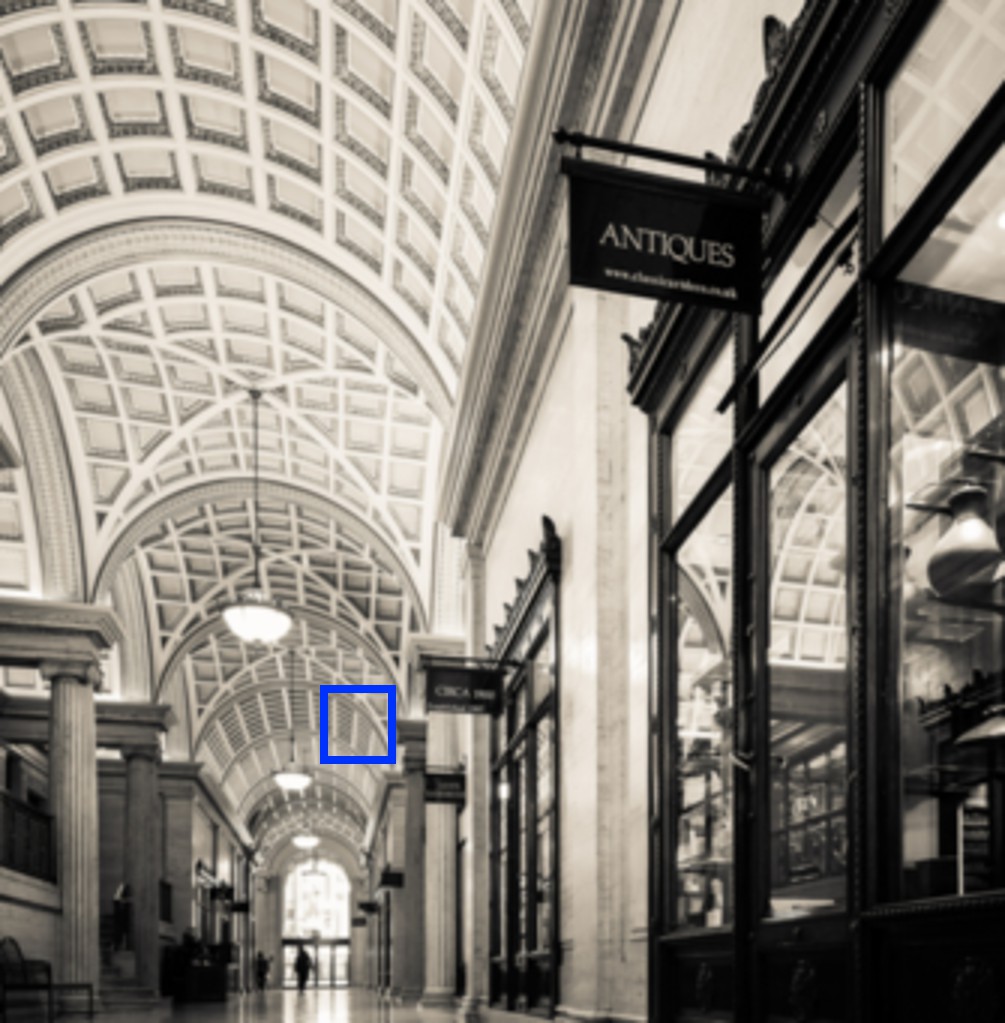} & 
			\includegraphics[width = 0.1\textwidth]{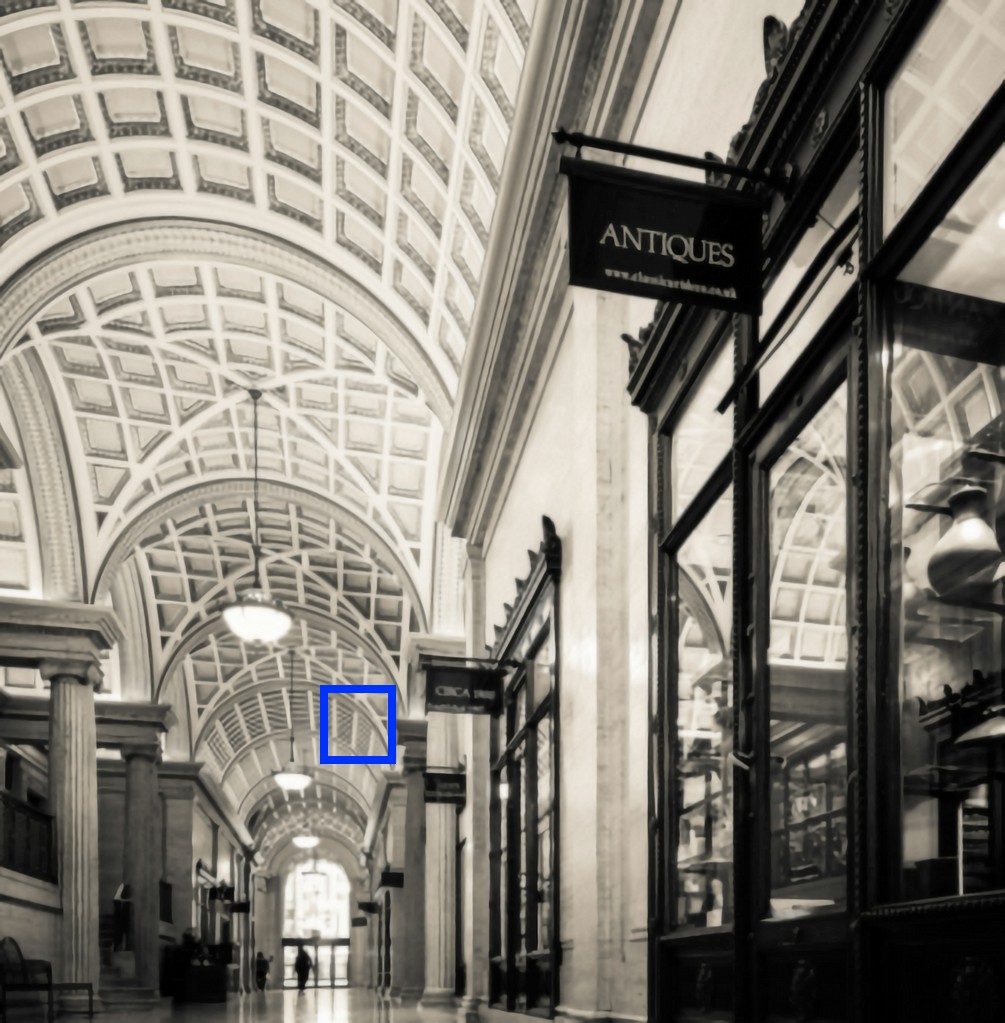} &
			\includegraphics[width = 0.1\textwidth]{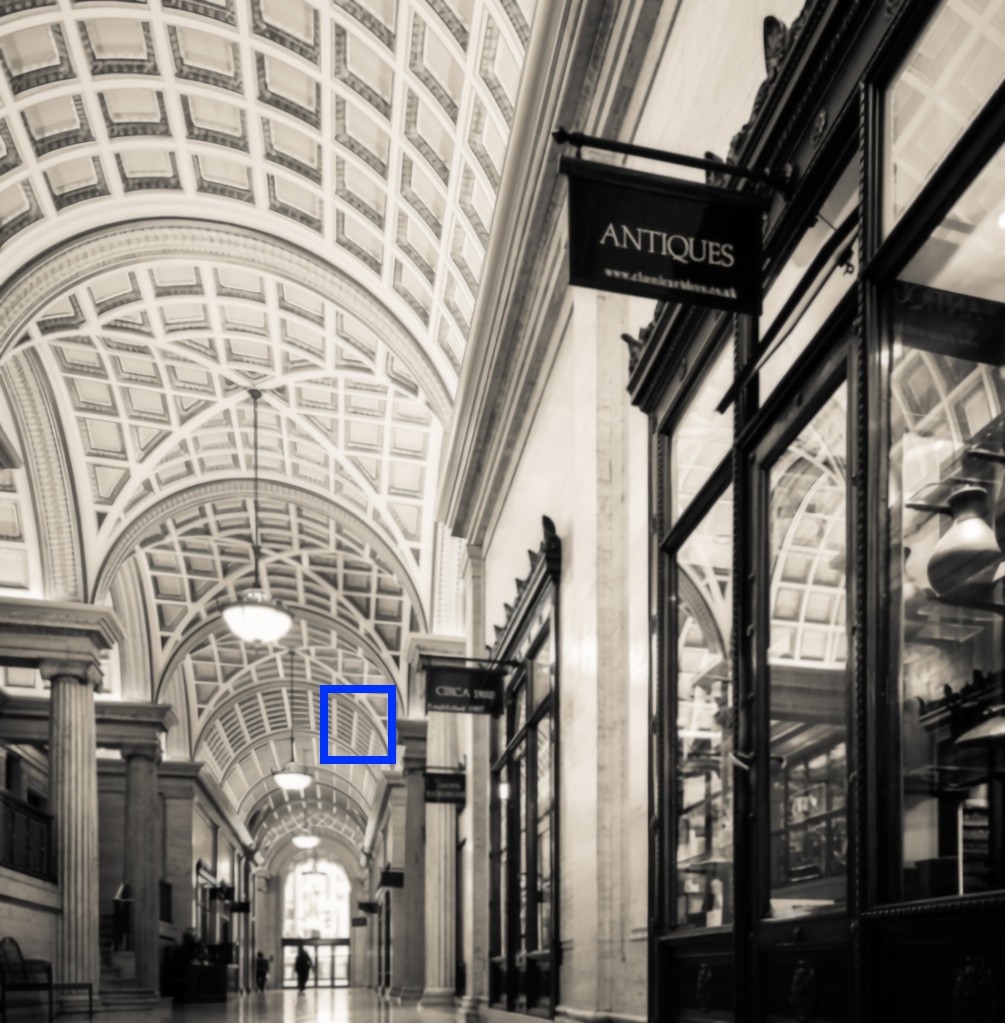} & 
			\includegraphics[width = 0.1\textwidth]{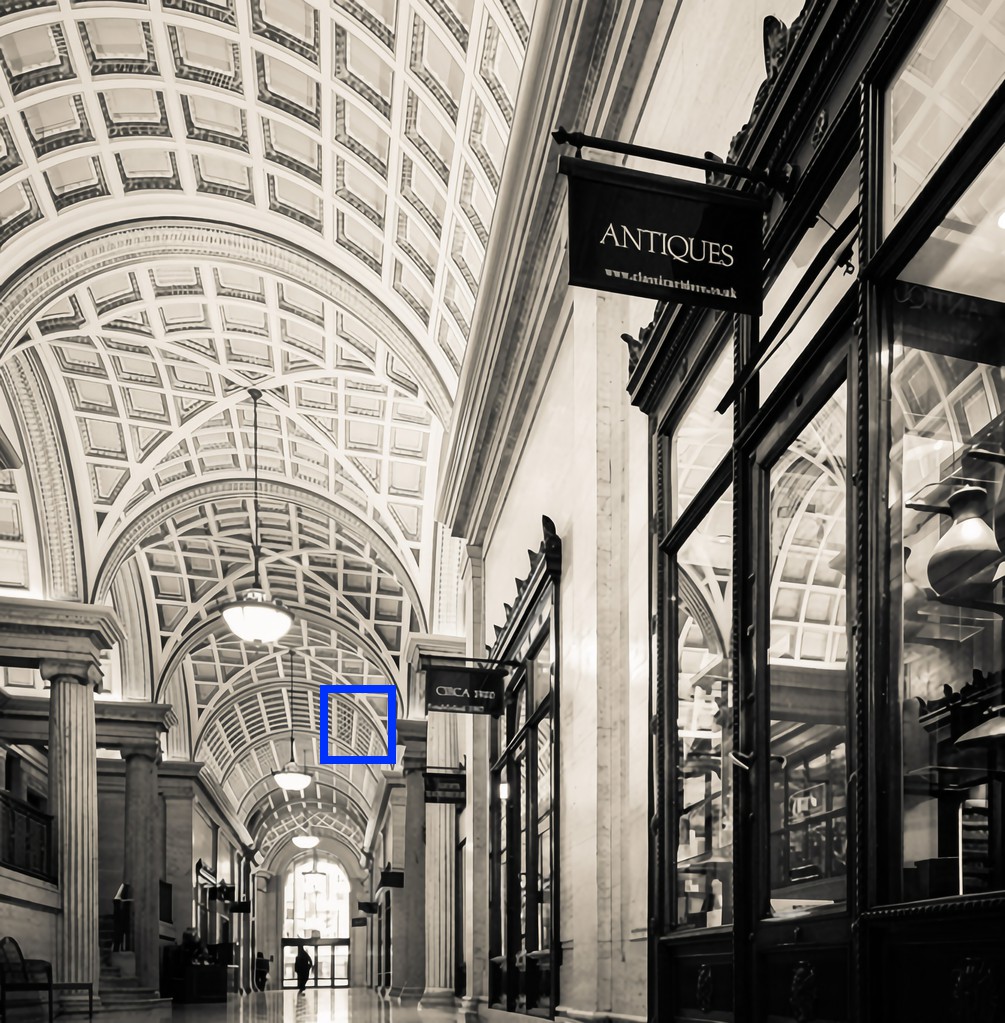} & 
			\includegraphics[width = 0.1\textwidth]{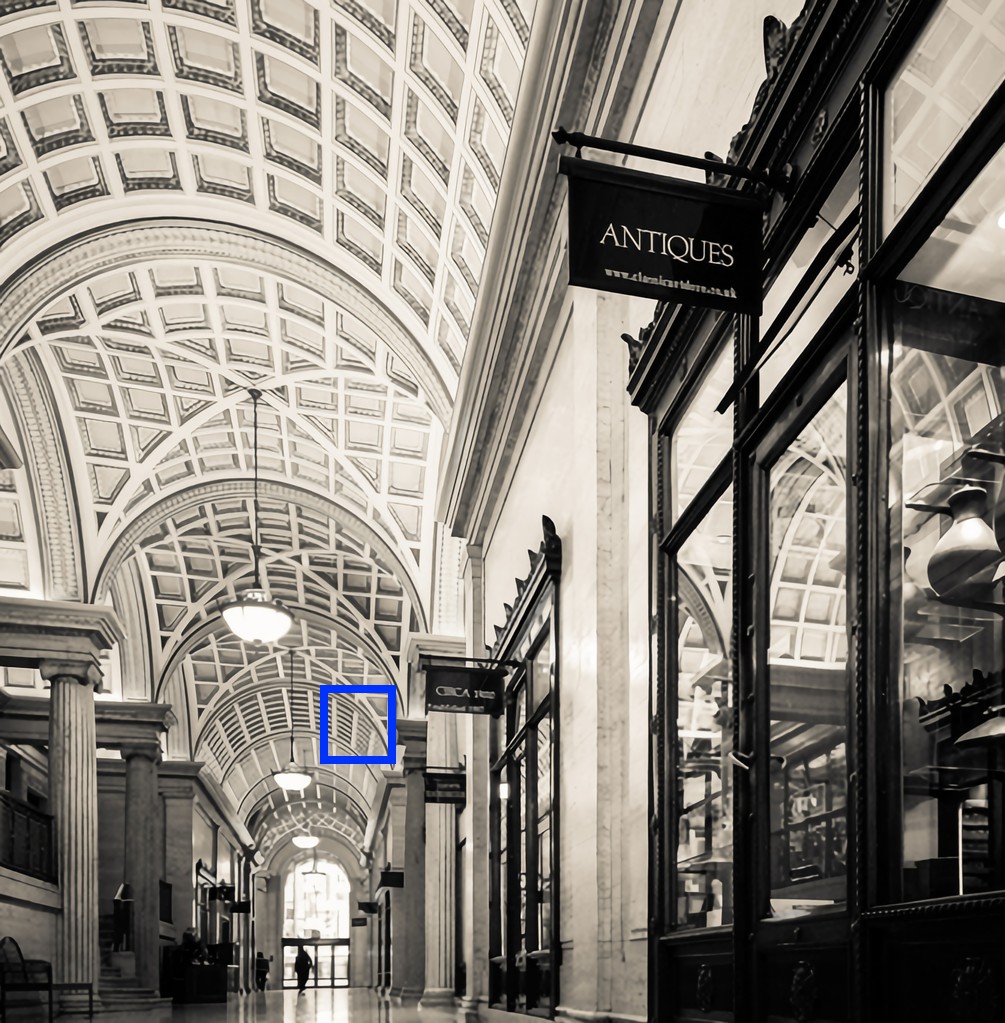}& 
			\includegraphics[width = 0.1\textwidth]{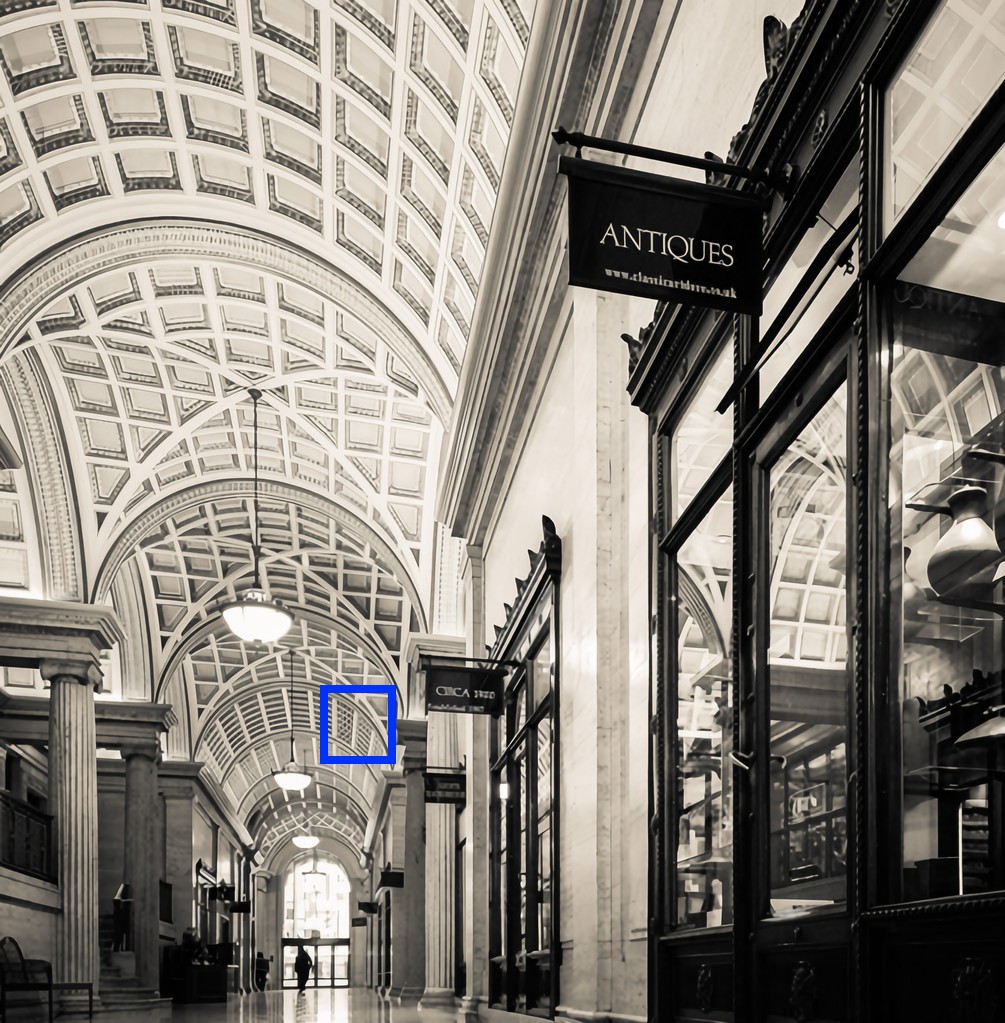} & 
			\includegraphics[width = 0.1\textwidth]{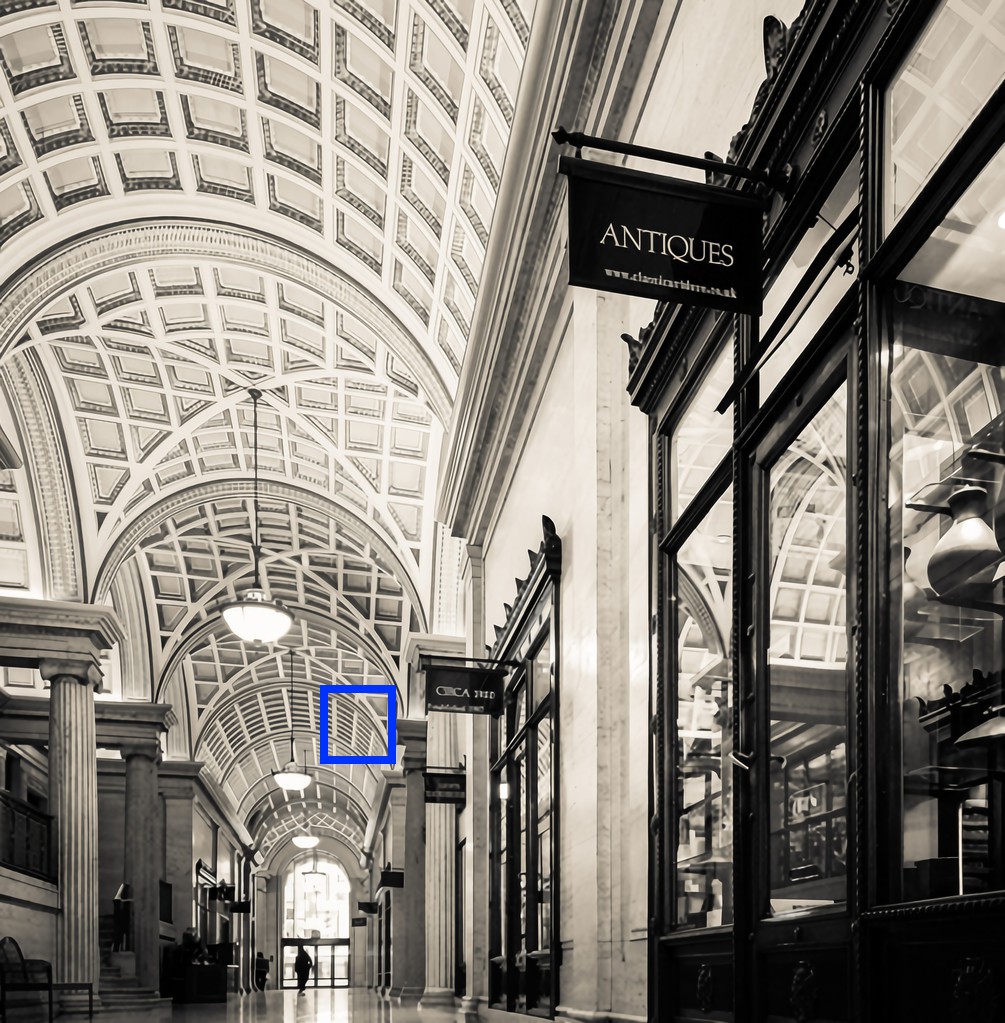} & 
			\includegraphics[width = 0.1\textwidth]{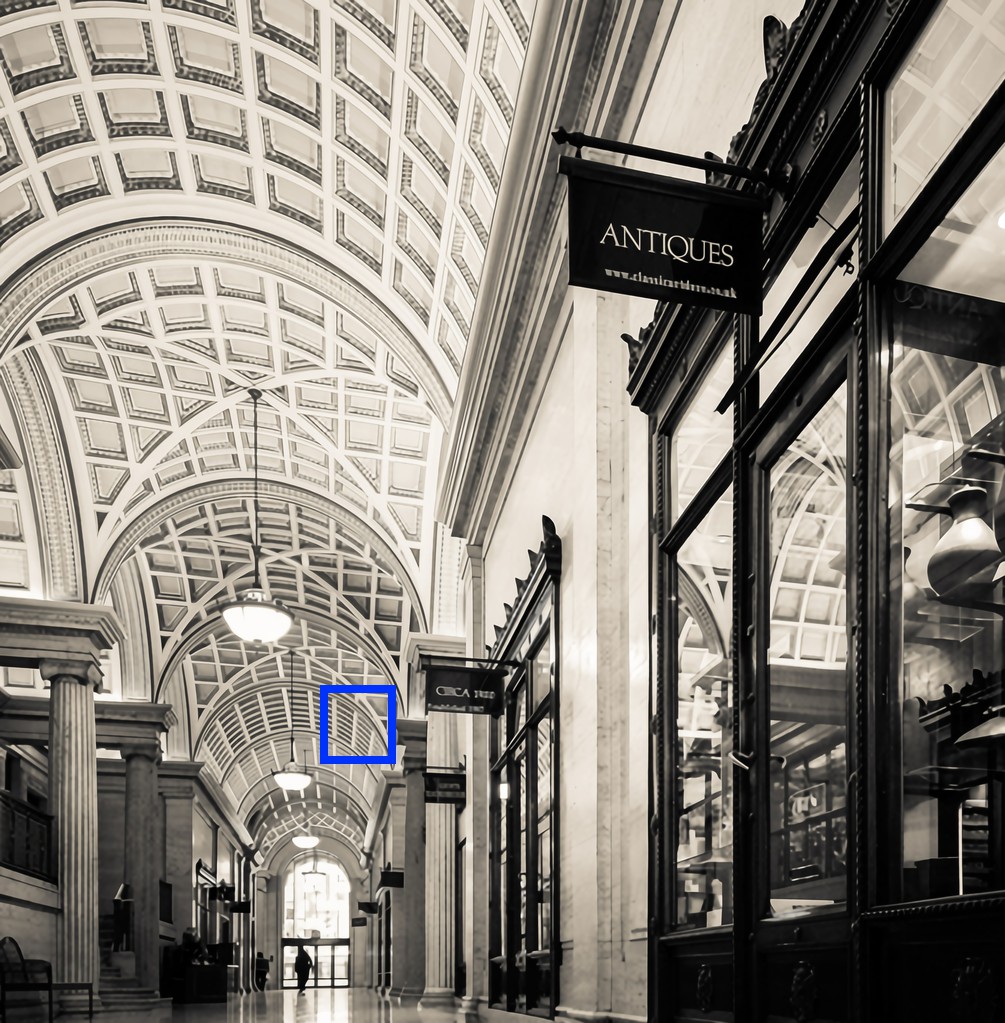} \\
			
			\includegraphics[width = 0.1\textwidth]{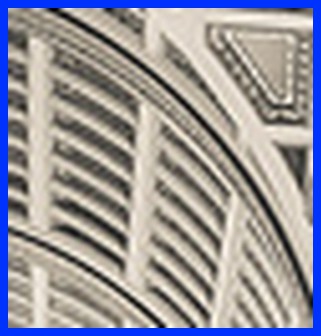}&
			\includegraphics[width = 0.1\textwidth]{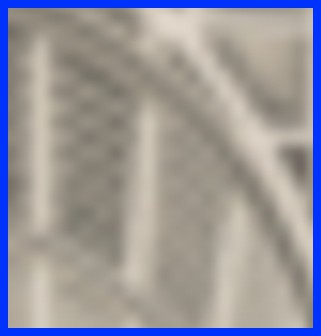} & 
			\includegraphics[width = 0.1\textwidth]{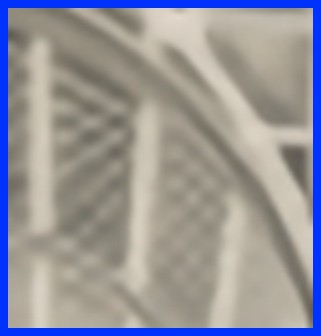} &
			\includegraphics[width = 0.1\textwidth]{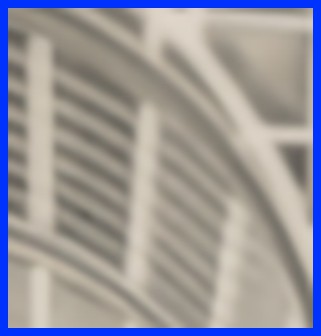} & 
			\includegraphics[width = 0.1\textwidth]{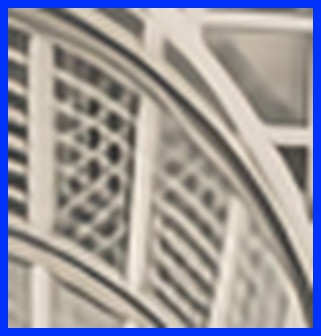} & 
			\includegraphics[width = 0.1\textwidth]{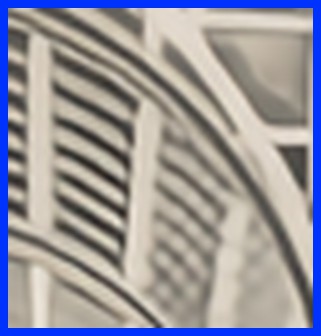}& 
			\includegraphics[width = 0.1\textwidth]{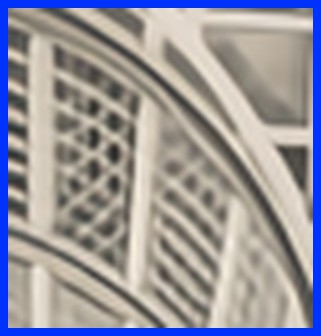} & 
			\includegraphics[width = 0.1\textwidth]{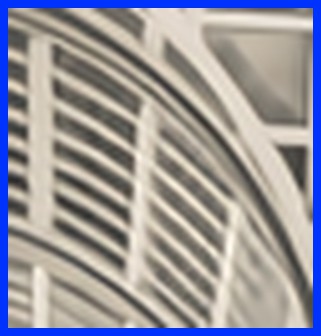} & 
			\includegraphics[width = 0.1\textwidth]{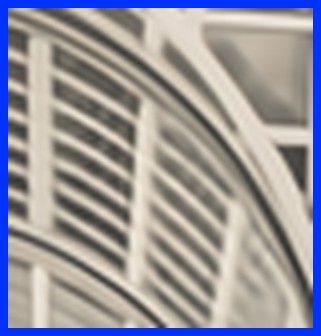} \\
			PSNR/SSIM & 20.85/0.590 & 21.92/0.671 &22.17/ 0.692 & 24.26/ 0.814 &23.98/ 0.802 &24.20/ 0.805 &24.28/0.819 &\textbf{24.65}/\textbf{0.828}\\
			
		\end{tabular}
	\end{center}

	\caption{Visual comparison for 3$\times$ SR with BD model on the Urban100 dataset. The best results are highlighted
	}

	\label{fig-BD}
\end{figure}

%

%\renewcommand\arraystretch{1}
%\begin{table}[!htbp]
%	\centering
%	\caption{Computational and parameter comparison (4$\times$ Manga109).}
%	\begin{tabular}{|p{1cm}<{\centering}|p{1cm}<{\centering}|p{1.2cm}<{\centering}|p{1cm}<{\centering}|p{1cm}<{\centering}|p{1cm}<{\centering}|p{1cm}<{\centering}|p{1.2cm}<{\centering}|p{1cm}<{\centering}|p{1cm}<{\centering}|}
%		\hline
%		& EDSR & MemNet & DBPN & RDN & RCAN & SRFBN & SAN  & HAN \\
%		\hline
%		Para.& 43M & 677k & 10M & 22.3M & 16M & 3M & 15.7M  & 16.1M \\
%		\hline
%		PSNR & 31.02 & 29.42 & 30.91 & 31.00 & 31.22 &31.15 & 31.18  & 31.42 \\
%		\hline
%	\end{tabular}%
%	\label{tab:parameter}%
%\end{table}%	

\textbf{Implementation Details.} We implement the proposed network using PyTorch platform and use the pre-trained RCAN ($ \times 2 $), ($ \times 3 $), ($ \times 4 $), ($ \times 8 $) model to initialize the corresponding holistic attention networks, respectively. 
%We use the pre-trained RCAN x2, x3, x4, x8 model to initialize the corresponding holistic attention networks, respectively.
%The framework of proposed network basing on is selected as Pytorch. 
%
%We select RCAN $ \times 2 $ reconstruction results as the pre-training model
%
In our network, patch size is set as $ 64 \times 64 $. We use
ADAM \cite{kingma2014adam} optimizer with a batch size 16 for training. The
learning rate is set as $ 10^{-5} $.
% and we decrease the learning rate by one half every $2 \times 10^{5} $ iterations.
Default values of $ \beta_{1}$ and $\beta_{2}$ are used, which are 0.9 and 0.999, respectively, and we set $ \epsilon=10^{-8} $.
We do not use any regularization operations such as batch normalization and group normalization in our network. In addition to random rotation and translation, we do not apply other data augmentation methods in the training.
%
%BN, Group Normalization(GN)~\cite{wu2018group} and other regularization operations are not used in our network. 
%
The input of the LAM is selected as the outputs of all residual groups of RCAN, we use $N=10$ residual groups in out network. 
For all the results reported in the paper, we train the network for  
%$2 \times 10^{5} $ iterations
250 epochs, which takes about two days on an Nvidia GTX 1080Ti GPU.
%The convolution kernel size k of the channel-spatial attention mechanism is chosen to be 3. 
%
%Following the setting of RCAN, 
%the input degraded images are all patched, and the patch size is $ 64\times64 $, batch size is set as 16. 
%
%Our model is trained by ADAM optimizer with $ \beta_{1}  = 0.9 $,$\beta_{2} = 0.999 $, and $ \epsilon=10^{8} $. We select RCAN $ \times 2 $ reconstruction results as the pre-training model, learning rate is initialized to $ 10^{-5} $ and training cycle is 100 epochs.
%
%

%\subsection{Ablation Study}
%
%\subsubsection{Ablation study about the proposed LAM and CSAM}
\begin{table}[t]
	\centering
	\scriptsize
	\caption{Quantitative results with BD degradation model. The best and second best results are highlighted in \textbf{bold} and \underline{underlined}}
	\begin{tabular}{|p{6.5em}|p{2.5em}|p{2.5em}|p{3em}|p{2.5em}|p{3em}|p{2.5em}|p{3em}|p{2.5em}|p{3em}|p{2.5em}|p{3em}|}
		
		\hline
		\multicolumn{1}{|c|}{\multirow{2}{*}{{ Method} }} & \multicolumn{1}{c|}{\multirow{2}{*}{Scale}} & \multicolumn{2}{c}{Set5} & \multicolumn{2}{c}{Set14} & \multicolumn{2}{c}{B100} & \multicolumn{2}{c}{Urban100} & \multicolumn{2}{c|}{ Manga109} \\
		\cline{3-12}  &  &  PSNR  & SSIM  & PSNR   & SSIM  & PSNR  & SSIM  & PSNR  & SSIM  & PSNR  & SSIM \\
		\hline
		
		Bicubic \newline{}SPMSR~\cite{peleg2014statistical} \newline{}SRCNN~\cite{dong2014learning}  \newline{}FSRCNN~\cite{dong2016accelerating} \newline{}VDSR~\cite{kim2016accurate} \newline{}IRCNN~\cite{zhang2017learning}  \newline{}SRMDNF~\cite{zhang2018learning} \newline{}RDN~\cite{zhang2018residual} \newline{}RCAN~\cite{zhang2018image} \newline{} SRFBN~\cite{li2019feedback} \newline{} SAN~\cite{dai2019second} \newline{} HAN(ours) \newline{} HAN+(ours)	& 
		$\times3$ \newline{}$\times3$ \newline{}$\times3$ \newline{}$\times3$ \newline{}$\times3$ \newline{}$\times3$ \newline{}$\times3$ \newline{}$\times3$ \newline{}$\times3$ \newline{} $\times3$ \newline{}$\times3$ \newline{}$\times3$ \newline{}$\times3$ 
		%\multirow{15}{|c|}{$\times$2}
		& 28.78 \newline{}32.21 \newline{}32.05 \newline{}26.23 \newline{}33.25 \newline{}33.38 \newline{}34.01 \newline{}34.58 \newline{}34.70 \newline{}34.66 \newline{}34.75 \newline{}\underline{34.76} \newline{}\bfseries{34.85}  & 0.8308 \newline{}0.9001 \newline{}0.8944 \newline{}0.8124 \newline{}0.9150 \newline{}0.9182 \newline{}0.9242 \newline{}0.9280 \newline{}0.9288 \newline{}0.9283 \newline{}0.9290 \newline{}\underline{0.9294} \newline{}\bfseries{0.9300}  & 26.38 \newline{}28.89 \newline{}28.80 \newline{}24.44 \newline{}29.46 \newline{}29.63 \newline{}30.11 \newline{}30.53 \newline{}30.63 \newline{}30.48 \newline{}30.68 \newline{}\underline{30.70} \newline{}\bfseries{30.79}& 0.7271 \newline{}0.8105 \newline{}0.8074 \newline{}0.7106 \newline{}0.8244 \newline{}0.8281 \newline{}0.8364 \newline{}0.8447 \newline{}0.8462 \newline{}0.8439 \newline{}0.8466 \newline{}\underline{0.8475} \newline{}\bfseries{0.8487} & 26.33 \newline{}28.13 \newline{}28.13 \newline{}24.86 \newline{}28.57 \newline{}28.65 \newline{}28.98 \newline{}29.23 \newline{}29.32 \newline{}29.21 \newline{}29.33 \newline{}\underline{29.34} \newline{}\bfseries{29.41} & 0.6918 \newline{}0.7740 \newline{}0.7736 \newline{}0.6832 \newline{}0.7893 \newline{}0.7922 \newline{}0.8009 \newline{}0.8079 \newline{}0.8093 \newline{}0.8069 \newline{}0.8101 \newline{}\underline{0.8106} \newline{}\bfseries{0.8116}& 23.52 \newline{}25.84 \newline{}25.70 \newline{}22.04 \newline{}26.61 \newline{}26.77 \newline{}27.50 \newline{}28.46 \newline{}28.81 \newline{}28.48 \newline{}28.83 \newline{}\underline{28.99} \newline{}\bfseries{29.21}  & 0.6862 \newline{}0.7856 \newline{}0.7770 \newline{}0.6745 \newline{}0.8136 \newline{}0.8154 \newline{}0.8370 \newline{}0.8582 \newline{}0.8647 \newline{}0.8581 \newline{}0.8646 \newline{}\underline{0.8676} \newline{}\bfseries{0.8710}  &25.46 \newline{}29.64 \newline{}29.47 \newline{}23.04 \newline{}31.06 \newline{}31.15 \newline{}32.97 \newline{}33.97 \newline{}34.38 \newline{}34.07 \newline{}34.46 \newline{}\underline{34.56} \newline{}\bfseries{34.87} &0.8149\newline{} 0.9003\newline{} 0.8924\newline{} 0.7927\newline{} 0.9234\newline{} 0.9245\newline{} 0.9391\newline{} 0.9465\newline{} 0.9483 \newline{} 0.9466 \newline{} 0.9487 \newline{}\underline{0.9494} \newline{}\bfseries{0.9509} \\
		\hline
	\end{tabular}%
	\label{tab-BD}%
\end{table}%

\subsection{Ablation Study about the Proposed LAM and CSAM}
The proposed LAM and CSAM ensure that the proposed SR method generate the feature correlations between hierarchical layers, channels, and locations. One may wonder whether the LAM and CSAM help SISR.
To verify the performance of these two attention mechanisms, we compare the method without using LAM and CSAM in Table \ref{tab-psnr-ssim}, where we conduct experiments on the Manga109 dataset with the magnification factor of $\times4$. 
%
%In this work, we propose two novel attention modules to generate the feature correlations between hierarchical layers, channels, and locations.
%information and strengthen the feature accordingly.
%The one is LAM, it can generate a N $\times$ N correlation matrix which can distinguish the importance of feature maps. The other one is CSAM, the self-attention map which obtained by 3D convolution integrates the  channel and spatial dimensions of features. 
%To verify the performance of the two attention mechanisms, we perform experiments with different settings as shown in Table \ref{tab:ablation}. We conduct experiments on the B100 dataset with the magnification factor of $\times4$. 
% represents the application of RCAN network under $ \times4 $ super resolution reconstruction. CSAM stands for channel-spatial attention module and LAM denotes layer attention module.
%Compared with RCAN which is chosen as base model, PSNR of LAM was 27.78 and 0.02 higher than RCAN.

Table \ref{tab-psnr-ssim} shows the quantitative evaluations. Compared with the baseline method which is identical to the proposed network except for the absence of these two modules LAM and CSAM. CSAM achieves better results by up to 0.06 dB in terms of PSNR, while LAM promotes 0.16 dB on the test dataset. In addition, the improvement of using both LAM and CSAM is significant as the proposed algorithm improves 0.2 dB, which demonstrates the effectiveness of the proposed layer attention and channel-spatial attention blocks. 
Figure \ref{fig-BI} further shows that using the LAM and CSAM is able to
generate the results with clearer structures and details.
% Compared with the baseline method of without using both LAM and CSAM, LAM Meanwhile, the performance of the CSAM alone was 0.01 higher than base model. Results from the second and third comparative experiment verify the effectiveness of our individual module, since the module employed alone improves the performance over RCAN. However, we combined the results of the two attention modules in parallel and improved the performance to 27.80. These results show that our two attention modules bring great benefits to the SR reconstruction task.

%\renewcommand\arraystretch{2}
%\begin{table}[h]
%	\centering
%	\caption{Ablation experiments for attention modules. We observe the best PSNR (dB) values on B100 (4$\times$) 5$\times$ $10^{4} $ iterations}
%	\label{tab:performance_comparison}  
%	\begin{tabular}{|p{4.5cm}<{\centering}|p{1cm}<{\centering}|p{1cm}<{\centering}|p{1cm}<{\centering}|p{1cm}<{\centering}|}
%		\hline
%		\bf Global attention(GA) & $\times$ & $\times$ &$\surd $ & $\surd $\\
%		\hline
%		\bf Local attention(LA) & $\times$ & $\surd $ & $\times$ & $\surd $\\
%		\hline
%		\bf PSNR on B100 ($\times$4) & 27.77 & 27.78 & 27.79 &27.80\\
%		\hline
%	\end{tabular}%
%	\label{tab:ablation}%
%\end{table}%
\begin{table}[!t]
	\tabcolsep 8pt
	\caption{Ablation study about using different numbers of CSAMs}
	
	\begin{center}\scriptsize{
			\begin{tabular}{cccccc}
				\toprule
				&Set5 & Set14 & B100 & Urban100 & Manga100 \\
				\midrule
				HAN(1 CSAM)       &32.64  & 28.90   & 27.80  & 26.85  &31.42   \\
				HAN(3 CSAM)       &32.67 & 28.91   & 27.80  & 26.89  &\textbf{31.46}   \\
				HAN(5 CSAM)       &\textbf{32.69}  & 28.91   & 27.80  & 26.89  &31.43   \\
				HAN(10 CSAM)      &32.67  & \textbf{28.91}   & \textbf{27.80}  & \textbf{26.89} &31.43  \\
				\bottomrule
		\end{tabular}}
		\label{tabR2}
	\end{center}
	
\end{table}

\subsection{Ablation Study about the Number of Residual Group}
%\subsection{Ablation study about the number of residual group}
%
%For a fair comparison, 
%In our work, the number of residual group (RG) is the same as the RCAN model. All of the RGs’ outputs will be fed into the proposed layer attention model. 
%
We conduct an ablation study about feeding different numbers of RGs to the proposed LAM. Specifically, we apply severally three, six, and ten RGs to the LAM, and we evaluate our model on five standard datasets. As shown in Table \ref{tab4}, we compare our three models with RCAN, although using fewer RGs, our algorithm still generates higher PSNR values than the baseline of RCAN. This ablation study demonstrates the effectiveness of the proposed LAM.

%\makeatletter
%\renewcommand{\thetable}{\@arabic\c@table}
%\makeatother

%\renewcommand{\tabcolsep}{8pt}
%\begin{table}[t]\scriptsize
%
%	\tabcolsep 8pt
%	\caption{Ablation study about using different numbers of RGs}
%
%	\begin{center}\scriptsize{
%			\begin{tabular}{cccccc}
%				\toprule
%				&Set5 & Set14 & B100 & Urban100 & Manga100 \\
%				\midrule
%				RCAN       &32.63  & 28.87   & 27.77  & 26.82  &31.22   \\
%				HAN 3RGs      &32.63  & 28.89   & 27.79  & 26.82  &31.40   \\
%				HAN 6RGs      &\textbf{32.64}  & \textbf{28.90}    & 27.79  & 26.84  &\textbf{31.42}   \\
%				HAN 10RGs      &\textbf{32.64}  & \textbf{28.90}   & \textbf{27.80}  & \textbf{26.85}  &\textbf{31.42}   \\
%				\bottomrule
%		\end{tabular}}
%		\label{tab4}
%	\end{center} 
%\end{table}
%%

\subsection{Ablation Study about the Number of CSAM}
In the paper, the channel-spatial attention module (CSAM) can extract powerful representations to describe inter-channel and intra-channel information in continuous channels. We conduct an ablation study about using different numbers of CSAM. We use one, three, five, and ten CSAMs in RGs. As shown in Table\ref{tabR2}, with the increase of CSAM, the values of PSNR are increasing on the testing datasets. 
%But the speed of training will slow down at the same time. The five-CSAMs model and the ten-CSAMs model will consume 1.2 times and 1.5 times training time than the one-CSAM model, respectively. This is because CSAM contains 3D convolution, which will affect the speed of network training and testing. 
This ablation study demonstrates the effectiveness of the proposed CSAM.

\subsection{Results with Bicubic (BI) Degradation Model}
%
%We only use DIV2K dataset to train our model HAN. 
%To test the effectiveness of our proposed model, 
We compare the proposed algorithm with 11 state-of-the-art methods: SRCNN~\cite{dong2014learning}, FSRCNN~\cite{dong2016accelerating}, VDSR~\cite{kim2016accurate}, LapSRN~\cite{lai2017deep}, MemNet~\cite{tai2017memnet}, SRMDNF~\cite{zhang2018learning}, D-DBPN~\cite{haris2018deep}, RDN~\cite{zhang2018residual}, EDSR~\cite{lim2017enhanced}, SRFBN~\cite{li2019feedback} and SAN~\cite{dai2019second}. We provide more comparisons in supplementary material.
Following~\cite{lim2017enhanced,dai2019second,zhang2018image},  we also propose self-ensemble model and donate it as HAN+.
% which can improve our model a lot.
%

\textbf{Quantitative results.} Table \ref{table-BI} shows the comparison of 2$\times$, 3$ \times$, 4$ \times $, and 8$ \times $ SR quantitative results. Compared to existing methods, our HAN+ performs best on all the scales of reconstructed test datasets. Without using self-ensemble, our network HAN still obtains great gain compared with the recent SR methods. 
In particular, our model is much better than SAN which also uses the same backbone network of RCAN and has more computationally intensive attention module.
%
%Although our model only increases two attention modules based on RCAN, our model achieves more performance gain. 
%We compare the reconstruction results of different scales on Set5 and Set14 dataset. 
%
Specifically, when we compare the reconstruction results at $ \times $8 scale on the Set5 dataset, the proposed HAN advances 0.11 dB in terms of PSNR than the competitive SAN.
% the PSNR of SAN is 27.22, and the result of our model HAN on this index is 0.11 higher. However, our backbone is similar to SAN, which demonstrates the effectiveness of our LAM and CSAM modules.
%
%For the Manga109 dataset, the reconstructed results generated by HAN achieve 0.18 dB and 0.24dB higher than SAN \cite{bibid}, and surpasses RCAN \cite{bibid} by up to 0.04 dB and 0.2 dB at $\times$3 and $\times$4 scales, respectively.
%
%0.04 dB higher than RCAN and surpass SAN by up to 0.18 dB.
% On the other hand, RCAN enhanced the reconstruction feature by introducing deeper residual connections and channel activation modules. 
%
%For example, on the Manga109 dataset with a scale of 4, our method achieve 0.2 higher result than RCAN.

% the  It also employed a deeper channel (128) to enhance feature extraction. SRFBN used an intermediate supervisory designer to artificially process data of different degradation types. Compared with the previous approach, we do not adopt any intermediate monitoring loss or feature channel deepening. We introduce two attention modules to employ global dependency and extract detailed textures to achieve better performance.
%

To further evaluate the proposed HAN, we conduct experiments on the large test sets of B100, Urban100, and Manga109. Our algorithm still performs favorably against the state-of-the-art methods. For example, the super-resolved results by the proposed HAN is 0.06 dB and 0.35 dB higher than the very recent work of SAN for the 4$ \times $ and 8$ \times $ scales, respectively.
%
%In the $ \times $4 scale reconstruction results on the Urban100 dataset, the result of state-of-the-art, namely SAN, is 26.79, and the HAN is 0.06 higher than the result. In the x8 reconstruction result of Manga109 test set, SAN result is 24.85, and HAN was 25.20, higher than 0.35.
%

\textbf{Visual results.} We also show visual comparisons of various methods on the Urban100 dataset for 4$\times$ SR in Figure \ref{fig-BI}. As shown, most compared SR networks cannot recover the grids of buildings accurately and suffer from unpleasant blurring artifacts. In contrast, the proposed HAN obtains clearer details and reconstructs sharper high-frequency textures.
% such as the fine lines of the grid above the building. 

Take the first and fourth images in Figure \ref{fig-BI} as example, 
%most comapared methods generate heavy blurring artifacts. Some methods bicubic, 
VDSR and EDSR fail to generate the clear structures. The results generated by the recent work of RCAN, SRFBN, and SAN still contain noticeable artifacts caused by spatial aliasing. 
In contrast, our approach effectively suppresses such artifacts through the proposed two attention modules. As shown, our method accurately reconstructs the grid patterns on windows in the first row and the parallel straight lines on the building in the fourth image.
% and recover the grid lines on windows in the first row image.

For 8$\times$ SR, we also show the super-resolved results by different SR methods in Figure \ref{fig-BD8}. As show, it is challenging to predict HR images from bicubic-upsampled input by VDSR and EDSR. Even the state-of-the-art methods of RCAN and SRFBN cannot super-resolve the fine structures well.
In contrast, our HAN reconstructs high-quality HR images for 8$\times$ results by using cross-scale layer attention and channel-spatial attention modules on the limited information.
%
%our HAN makes full use the  through cross-scale layer attention module and channel-spatial attention module.
%
%These visual comparisons suggest that the proposed HAN based on LAM and CSAM works better on image SR, especially for a large magnification factor (e.g., 8$\times$), than the state-of-the-art approaches.

%
%\begin{figure}[t] 
%	\centering 
%	\subfigure[Results on Manga109 (3$\times$)]{ 
%		\label{fig:subfig:a} %% label for first subfigure 
%		\includegraphics[width=2in]{img/compare/3x.pdf} 
%	} 
%	\subfigure[Results on Manga109 (4$\times$)]{ 
%		\label{fig:subfig:b} %% label for second subfigure 
%		\includegraphics[width=2in]{img/compare/4x.pdf} 
%	} 
%	\caption{Performance and number of parameters. Results are evaluated on Manga109} 
%	\label{fig-compare}  
%\end{figure}

\subsection{Results with Blur-downscale Degradation (BD) Model}
\noindent \textbf{Quantitative results.} Following the protocols of \cite{zhang2018learning,zhang2017learning,zhang2018residual}, we further compare the SR results on images with blur-downscale degradation model. We compare the proposed method with nine state-of-the-art super-resolution methods: SPMSR~\cite{peleg2014statistical}, SRCNN~\cite{dong2014learning}, FSRCNN~\cite{dong2016accelerating}, VDSR~\cite{kim2016accurate}, IRCNN~\cite{zhang2017learning}, SRMD~\cite{zhang2006edge}, RDN~\cite{zhang2018residual}, RCAN~\cite{zhang2018image},SRFBN~\cite{li2019feedback} and SAN~\cite{dai2019second}. Quantitative results on the 3$\times$ SR are reported in Table \ref{tab-BD}. 
As shown, both the proposed HAN and HAN+ perform favorably against existing methods. 
In particular, our HAN+ yields the best quantitative results and HAN obtains the second best scores for all the datasets, 0.06-0.2 dB PSNR better than the attention-based methods of RCAN and SAN and 0.2-0.8 dB better than the recently proposed SRFBN.
%
%SAN has achieved great results. However, our proposed method is 0.04 higher than SAN. 
 
\textbf{Visual quality.} In Figure \ref{fig-BD}, we show visual results on images from the Urban 100 dataset with blur-downscale degradation model by a scale factor of 3. Both the full images and the cropped regions are shown for comparison. 
We find that our proposed HAN is able to recover structured details that were missing in the LR image by properly exploiting the layer, channel, and spatial attention in the feature space.

As shown, VDSR and EDSR suffer from unpleasant blurring artifacts and some results even are out of shape. RCAN alleviate it to a certain extent, but still misses some details and structures.
SRFBN and SAN also fail to recover these structured details. 
In contrast, our proposed HAN effectively suppresses artifacts and exploits the scene details and the internal natural image statistics to super-resolve the high-frequency contents.
%
%Taking the fourth row image in \ref{fig-BD} as example, we can see that the early proposed SR methods bicubic, VDSR, EDSR cannot recover the texture of the building. The recent SR network RCAN generates distorted details in the image. SRFBN and SAN can reduce the ambiguity, but we still can find some blurring artifacts easily. However our HAN can recover more high-frequence texture due to the proposed scale-cross layer module, which can conduct a holistic feature weighting for the preceding features. 

%\subsection{Model Size}
%
%As shown in Figure \ref{fig-compare}, we can make a comparison of model size and performance of different SR reconstruction networks. Among all the networks shown in the figure, our HAN performs best and achieves 31.42 dB on the Manga109 dataset with 4$\times$ SR, which is 0.3 dB higher than the recent work of RCAN, and the parameter amount is only 16.1M.
%%
%The high-quality results of the proposed model are attributed to the global enhanced features obtained by the holistic attention model. Although we use RCAN as the backbone of our model, the parameters do not increase too many as shown in Figure \ref{fig-compare} (a) and (b). These results demonstrate that our proposed HAN can obtain a better trade-off between the performance and model complexity.

\section{Conclusions}
In this paper, we propose a holistic attention network for single image super-resolution, which adaptively learns the global dependencies among different depths, channels, and positions using the self-attention mechanism. 
Specifically, the layer attention module captures the long-distance dependencies among hierarchical layers. Meanwhile, the channel-spatial attention module incorporates the
channel and contextual information in each layer. 
These two attention modules are collaboratively applied to multi-level features and then more informative features can be captured.
% and the fusion features of the deepest feature space and channel, respectively. 
%
%The ablation study and contrastive experiments show that our proposed HAN effectively learns the feature dependence and yields better reconstruction results.
%
Extensive experimental results on benchmark datasets demonstrate that the proposed model performs favorably against the state-of-the-art SR algorithms in terms of accuracy and visual quality.

\subsubsection{Acknowledgements:}
This work is supported by the National Key R\&D Program of China under Grant 2019YFB1406500, National Natural Science Foundation of China (No. 61971016, U1605252, 61771369), Fundamental Research Funds of Central Universities (Grant No. N160504007), Beijing Natural Science Foundation (No. L182057), Peng Cheng Laboratory Project of Guangdong Province PCL2018KP004, and the Shaanxi Provincial Natural Science Basic Research Plan (2019JM-557). 
\clearpage

\bibliographystyle{splncs04}
\bibliography{sr2020_}

\begin{thebibliography}{10}
\providecommand{\url}[1]{\texttt{#1}}
\providecommand{\urlprefix}{URL }
\providecommand{\doi}[1]{https://doi.org/#1}

\bibitem{bevilacqua2012low}
Bevilacqua, M., Roumy, A., Guillemot, C., Alberi-Morel, M.L.: Low-complexity
  single-image super-resolution based on nonnegative neighbor embedding. In:
  BMVC (2012)

\bibitem{dai2019second}
Dai, T., Cai, J., Zhang, Y., Xia, S.T., Zhang, L.: Second-order attention
  network for single image super-resolution. In: CVPR (2019)

\bibitem{dong2014learning}
Dong, C., Loy, C.C., He, K., Tang, X.: Learning a deep convolutional network
  for image super-resolution. In: ECCV (2014)

\bibitem{dong2016accelerating}
Dong, C., Loy, C.C., Tang, X.: Accelerating the super-resolution convolutional
  neural network. In: ECCV (2016)

\bibitem{NIPS2014_5423}
Goodfellow, I., Pouget-Abadie, J., Mirza, M., Xu, B., Warde-Farley, D., Ozair,
  S., Courville, A., Bengio, Y.: Generative adversarial nets. In: NIPS (2014)

\bibitem{haris2018deep}
Haris, M., Shakhnarovich, G., Ukita, N.: Deep back-projection networks for
  super-resolution. In: CVPR (2018)

\bibitem{he2016deep}
He, K., Zhang, X., Ren, S., Sun, J.: Deep residual learning for image
  recognition. In: CVPR (2016)

\bibitem{hu2018squeeze}
Hu, J., Shen, L., Sun, G.: Squeeze-and-excitation networks. In: CVPR (2018)

\bibitem{hu2019channel}
Hu, Y., Li, J., Huang, Y., Gao, X.: Channel-wise and spatial feature modulation
  network for single image super-resolution. IEEE Transactions on Circuits and
  Systems for Video Technology  (2019)

\bibitem{huang2017densely}
Huang, G., Liu, Z., Van Der~Maaten, L., Weinberger, K.Q.: Densely connected
  convolutional networks. In: CVPR (2017)

\bibitem{huang2015single}
Huang, J.B., Singh, A., Ahuja, N.: Single image super-resolution from
  transformed self-exemplars. In: CVPR (2015)

\bibitem{huang2018robust}
Huang, S., Sun, J., Yang, Y., Fang, Y., Lin, P., Que, Y.: Robust single-image
  super-resolution based on adaptive edge-preserving smoothing regularization.
  TIP  \textbf{27}(6),  2650--2663 (2018)

\bibitem{ji20123d}
Ji, S., Xu, W., Yang, M., Yu, K.: 3d convolutional neural networks for human
  action recognition. TPAMI  \textbf{35}(1),  221--231 (2012)

\bibitem{johnson2016perceptual}
Johnson, J., Alahi, A., Fei-Fei, L.: Perceptual losses for real-time style
  transfer and super-resolution. In: ECCV (2016)

\bibitem{kim2016accurate}
Kim, J., Kwon~Lee, J., Mu~Lee, K.: Accurate image super-resolution using very
  deep convolutional networks. In: CVPR (2016)

\bibitem{kim2016deeply}
Kim, J., Kwon~Lee, J., Mu~Lee, K.: Deeply-recursive convolutional network for
  image super-resolution. In: CVPR (2016)

\bibitem{kim2018ram}
Kim, J.H., Choi, J.H., Cheon, M., Lee, J.S.: Ram: Residual attention module for
  single image super-resolution. arXiv preprint arXiv:1811.12043  (2018)

\bibitem{kingma2014adam}
Kingma, D.P., Ba, J.: Adam: A method for stochastic optimization. arXiv
  preprint arXiv:1412.6980  (2014)

\bibitem{lai2017deep}
Lai, W.S., Huang, J.B., Ahuja, N., Yang, M.H.: Deep laplacian pyramid networks
  for fast and accurate super-resolution. In: CVPR (2017)

\bibitem{ledig2017photo}
Ledig, C., Theis, L., Husz{\'a}r, F., Caballero, J., Cunningham, A., Acosta,
  A., Aitken, A., Tejani, A., Totz, J., Wang, Z., et~al.: Photo-realistic
  single image super-resolution using a generative adversarial network. In:
  CVPR (2017)

\bibitem{li2019feedback}
Li, Z., Yang, J., Liu, Z., Yang, X., Jeon, G., Wu, W.: Feedback network for
  image super-resolution. In: CVPR (2019)

\bibitem{lim2017enhanced}
Lim, B., Son, S., Kim, H., Nah, S., Mu~Lee, K.: Enhanced deep residual networks
  for single image super-resolution. In: CVPR (2017)

\bibitem{martin2001database}
Martin, D., Fowlkes, C., Tal, D., Malik, J.: A database of human segmented
  natural images and its application to evaluating segmentation algorithms and
  measuring ecological statistics. In: ICCV (2001)

\bibitem{matsui2017sketch}
Matsui, Y., Ito, K., Aramaki, Y., Fujimoto, A., Ogawa, T., Yamasaki, T.,
  Aizawa, K.: Sketch-based manga retrieval using manga109 dataset. Multimedia
  Tools and Applications  \textbf{76}(20),  21811--21838 (2017)

\bibitem{peleg2014statistical}
Peleg, T., Elad, M.: A statistical prediction model based on sparse
  representations for single image super-resolution. TIP  \textbf{23}(6),
  2569--2582 (2014)

\bibitem{ren2019face}
Ren, W., Yang, J., Deng, S., Wipf, D., Cao, X., Tong, X.: Face video deblurring
  using 3d facial priors. In: Proceedings of the IEEE International Conference
  on Computer Vision. pp. 9388--9397 (2019)

\bibitem{ren2018deep}
Ren, W., Zhang, J., Ma, L., Pan, J., Cao, X., Zuo, W., Liu, W., Yang, M.H.:
  Deep non-blind deconvolution via generalized low-rank approximation. In:
  Advances in Neural Information Processing Systems. pp. 297--307 (2018)

\bibitem{sajjadi2017enhancenet}
Sajjadi, M.S., Scholkopf, B., Hirsch, M.: Enhancenet: Single image
  super-resolution through automated texture synthesis. In: ICCV (2017)

\bibitem{shi2016real}
Shi, W., Caballero, J., Husz{\'a}r, F., Totz, J., Aitken, A.P., Bishop, R.,
  Rueckert, D., Wang, Z.: Real-time single image and video super-resolution
  using an efficient sub-pixel convolutional neural network. In: CVPR (2016)

\bibitem{tai2017image}
Tai, Y., Yang, J., Liu, X.: Image super-resolution via deep recursive residual
  network. In: CVPR (2017)

\bibitem{tai2017memnet}
Tai, Y., Yang, J., Liu, X., Xu, C.: Memnet: A persistent memory network for
  image restoration. In: ICCV (2017)

\bibitem{timofte2017ntire}
Timofte, R., Agustsson, E., Van~Gool, L., Yang, M.H., Zhang, L.: Ntire 2017
  challenge on single image super-resolution: Methods and results. In: CVPRW
  (2017)

\bibitem{wang2015deep}
Wang, Z., Liu, D., Yang, J., Han, W., Huang, T.: Deep networks for image
  super-resolution with sparse prior. In: ICCV (2015)

\bibitem{woo2018cbam}
Woo, S., Park, J., Lee, J.Y., So~Kweon, I.: Cbam: Convolutional block attention
  module. In: ECCV (2018)

\bibitem{yang2008image}
Yang, J., Wright, J., Huang, T., Ma, Y.: Image super-resolution as sparse
  representation of raw image patches. In: CVPR (2008)

\bibitem{zeyde2010single}
Zeyde, R., Elad, M., Protter, M.: On single image scale-up using
  sparse-representations. In: International conference on curves and surfaces
  (2010)

\bibitem{zhang2017learning}
Zhang, K., Zuo, W., Gu, S., Zhang, L.: Learning deep cnn denoiser prior for
  image restoration. In: CVPR (2017)

\bibitem{zhang2018learning}
Zhang, K., Zuo, W., Zhang, L.: Learning a single convolutional super-resolution
  network for multiple degradations. In: CVPR (2018)

\bibitem{zhang2006edge}
Zhang, L., Wu, X.: An edge-guided image interpolation algorithm via directional
  filtering and data fusion. TIP  \textbf{15}(8),  2226--2238 (2006)

\bibitem{zhang2018image}
Zhang, Y., Li, K., Li, K., Wang, L., Zhong, B., Fu, Y.: Image super-resolution
  using very deep residual channel attention networks. In: ECCV (2018)

\bibitem{zhang2018residual}
Zhang, Y., Tian, Y., Kong, Y., Zhong, B., Fu, Y.: Residual dense network for
  image super-resolution. In: CVPR (2018)

\bibitem{zhang2019image}
Zhang, Z., Wang, Z., Lin, Z., Qi, H.: Image super-resolution by neural texture
  transfer. In: CVPR (2019)

\end{thebibliography}
\end{document}